\pdfoutput=1
\documentclass[12pt]{article}

\usepackage{graphicx,psfrag,epsf,color}
\usepackage{amsmath,amssymb,amsfonts}
\usepackage{array}
\usepackage{bm} 
\usepackage{cite}
\usepackage{multirow}
\usepackage{rotating}
\usepackage{slashed}
\DeclareMathOperator*{\sumint}{%
\mathchoice%
  {\ooalign{$\displaystyle\sum$\cr\hidewidth$\displaystyle\int$\hidewidth\cr}}
  {\ooalign{\raisebox{.14\height}{\scalebox{.7}{$\textstyle\sum$}}\cr\hidewidth$\textstyle\int$\hidewidth\cr}}
  {\ooalign{\raisebox{.2\height}{\scalebox{.6}{$\scriptstyle\sum$}}\cr$\scriptstyle\int$\cr}}
  {\ooalign{\raisebox{.2\height}{\scalebox{.6}{$\scriptstyle\sum$}}\cr$\scriptstyle\int$\cr}}
}

\newcounter{MBQ}

\newcommand{\be}{\begin{equation}}
\newcommand{\ee}{\end{equation}}
\newcommand{\bea}{\begin{eqnarray}}
\newcommand{\eea}{\end{eqnarray}}
\newcommand{\bi}{\begin{itemize}}
\newcommand{\ei}{\end{itemize}}
\newcommand{\ben}{\begin{enumerate}}
\newcommand{\een}{\end{enumerate}}
\newcommand{\bt}{\begin{tabular}}
\newcommand{\et}{\end{tabular}}

\newcommand{\Eres}{E^\gamma_{\rm res}}
\newcommand{\mchi}{m_\chi}
\newcommand{\la}{\lambda}
\newcommand{\np}{n_+}
\newcommand{\nm}{n_-}

\def\lmu{l_{\mu^2}}

\DeclareMathOperator{\dif}{d\!}

\numberwithin{equation}{section}

\begin{document}
\allowdisplaybreaks

\begin{titlepage}

\begin{flushright}
{\small
TUM-HEP-1191/19\\
arXiv:1903.08702 [hep-ph]\\[0.0cm]
December 16, 2020
}
\end{flushright}

\vskip1cm
\begin{center}
{\Large \bf Resummed photon spectrum from 
dark  matter\\[0.0cm] 
annihilation for intermediate and narrow energy\\[0.2cm] 
resolution}\\[0.2cm]
\end{center}

\vspace{0.45cm}
\begin{center}
{\sc M.~Beneke$^{a}$, A.~Broggio$^{b,c}$, C.~Hasner$^{a}$,} \\ 
{\sc K.~Urban$^{a}$,} and  {\sc M.~Vollmann$^{a}$}\\[6mm]
{\it ${}^a$Physik Department T31,\\
James-Franck-Stra\ss e~1, 
Technische Universit\"at M\"unchen,\\
D--85748 Garching, Germany \\[0.3cm]
${}^b$ Universit\`{a} degli Studi di Milano-Bicocca,\\ 
Piazza della Scienza 3,\\ I--20126 Milano, Italy\\[0.3cm]
${}^c$ INFN, Sezione di Milano-Bicocca,\\ 
Piazza della Scienza 3,\\ I--20126 Milano, Italy}
\\[0.3cm]
\end{center}

\vspace{0.55cm}
\begin{abstract}
\vskip0.2cm\noindent
The annihilation cross section of weakly interacting 
TeV scale dark matter particles $\chi^0$ into photons is affected 
by large quantum corrections due to electroweak Sudakov 
logarithms and the Sommerfeld effect. We extend our 
previous work on the resummation of the 
semi-inclusive photon energy spectrum in 
$\chi^0\chi^0\to \gamma+X$ in the vicinity of the maximal photon 
energy $E_\gamma = m_\chi$ with NLL' accuracy from the 
case of narrow photon energy resolution $\Eres$ of 
order $m_W^2/m_\chi$ to intermediate resolution of 
order $\Eres \sim m_W$. We also provide details on the previous narrow resolution calculation. The two calculations, performed 
in different effective field theory set-ups for the wino 
dark matter model, are then shown to match well, providing an 
accurate representation up to energy resolutions of 
about 300~GeV. 
\end{abstract}
\end{titlepage}

\section{Introduction}
\label{sec:introduction}

High-energy photons may constitute an important signal for the 
particle nature of dark matter (DM) through the pair annihilation of
DM particles. In order to distinguish the DM component from the 
astrophysical $\gamma$-ray background, one searches for the 
line signal of the two-body annihilation $\chi^0\chi^0\to\gamma\gamma$ 
(or $\gamma Z$) at (or very close to) $E_\gamma = m_\chi$, where 
$m_\chi$ is the mass of the dark matter particle, to be 
determined.

In particular, the paradigmatic WIMP with mass in the 100~GeV to 10~TeV 
range and electroweak charge is expected to be observed or ruled 
out by the Cherenkov Telescope Array (CTA) \cite{Consortium:2010bc} 
under construction even under conservative assumptions on astrophysical 
uncertainties, especially due to the dark matter density profile near 
the Galactic center. Precise theoretical computations of the photon 
yield from DM annihilation are therefore well motivated. 

Recent theoretical work has focused on two aspects of the problem. First, 
for dark matter annihilation into energetic particles, electroweak Sudakov 
(double) logarithms ${\cal O}((\alpha_2\ln^2 (\mchi/m_W))^n)$ 
are large and should be summed to all 
orders~\cite{Baumgart:2014vma,Bauer:2014ula,Ovanesyan:2014fwa,Ovanesyan:2016vkk}, 
in addition to the summation of ladder diagrams known as the Sommerfeld 
effect. Second, since $\gamma$-ray telescopes do not measure two photons 
from a single annihilation in coincidence, 
the observable is not  $\chi^0\chi^0\to\gamma\gamma$ (or $\gamma Z$) but 
rather the semi-inclusive single-photon 
energy spectrum $\gamma+X$, where $X$ denotes the unidentified other 
final state particles. Although the leading term in the 
perturbative expansion of the semi-inclusive annihilation rate 
arises from the two-body final states 
$\gamma\gamma, \gamma Z$, the logarithmically enhanced terms 
differ in higher orders and this affects their resummation 
\cite{Baumgart:2017nsr,Beneke:2018ssm,Baumgart:2018yed}. It has 
been shown, both for the exclusive $\gamma\gamma$ annihilation 
rate \cite{Ovanesyan:2016vkk}, as well as for the semi-inclusive 
rate at narrow energy resolution (as defined below) 
\cite{Beneke:2018ssm}, that resummation with NLL' accuracy, which 
combines the full one-loop calculations with next-to-leading logarithmic 
resummation provides precise results for the photon rate with 
uncertainties around 1\%.

The resummation of the semi-inclusive spectrum 
is performed for the primary photon energy spectrum 
$d(\sigma v_{\rm rel})/dE_\gamma$ of the DM pair annihilation cross section 
multiplied by the relative velocity of the annihilating particles. 
While in forecasts for the rate observed by a specific telescope, 
the spectrum will have to be smeared with an instrument-specific 
resolution function of some width $\Eres$ in energy, the expected 
impact and accuracy of the theoretical prediction can be equally 
discussed for the spectrum integrated over the energy interval 
$\Eres$ from its kinematic endpoint:
\begin{equation}
\langle \sigma v \rangle (E^\gamma_{\rm res}) = 
\int_{\mchi-E^\gamma_{\rm res}}^{\mchi} 
dE_\gamma\, \frac{d(\sigma v)}{dE_\gamma}\,.
\label{eq:defines}
\end{equation}
The endpoint-integrated spectrum depends on the three scales $\mchi$, 
$m_W$ (representative of electroweak scale masses), and $\Eres$. We consider 
TeV scale dark matter, hence the hierarchy $m_W\ll \mchi$ is 
always assumed. The details of the resummation of electroweak Sudakov 
logarithms near the endpoint, $\Eres\ll m_\chi$, differ according 
to the scaling of $\Eres$ and $m_W$ with respect to each other. We 
distinguish the following three regimes:
\begin{align}
\mbox{narrow}: \quad & \Eres \sim m_W^2/\mchi \nonumber \\
\mbox{intermediate}:  \quad & \Eres \sim m_W \nonumber \\
\mbox{wide}:  \quad & \Eres \gg m_W
\end{align}
The wide resolution regime was considered 
in \cite{Baumgart:2017nsr,Baumgart:2018yed} and resummed at the 
NLL order. Due to the double hierarchy $\mchi\gg \Eres \gg m_W$ a 
two-step procedure applies to simultaneously sum the unrelated large 
logarithms of $\mchi/m_W$ and $\Eres/m_W$. This procedure requires 
large dark matter masses to satisfy both hierarchies. Resummation 
of electroweak Sudakov logarithms for the narrow resolution case 
was accomplished in \cite{Beneke:2018ssm} at the NLL' order. 
The intermediate resolution regime has not been considered up to 
now. 

In the present paper we close this theoretical gap. 
We develop the effective field theory (EFT) for the 
intermediate resolution regime and sum 
the electroweak logarithms at the NLL' order. We show that the result 
can be smoothly joined to the narrow resolution regime to provide a 
precise prediction of the photon energy spectrum near $m_\chi$ in 
the entire region from the line signal ($\Eres =0$) to $\Eres 
\approx 4 m_W$. We also provide details and derivations for the 
narrow resolution regime not given in the 
letter \cite{Beneke:2018ssm}.

\begin{figure}[t]
\centering
\vskip-0.2cm
\hspace*{-0.6cm}
\includegraphics[width=0.67\textwidth]{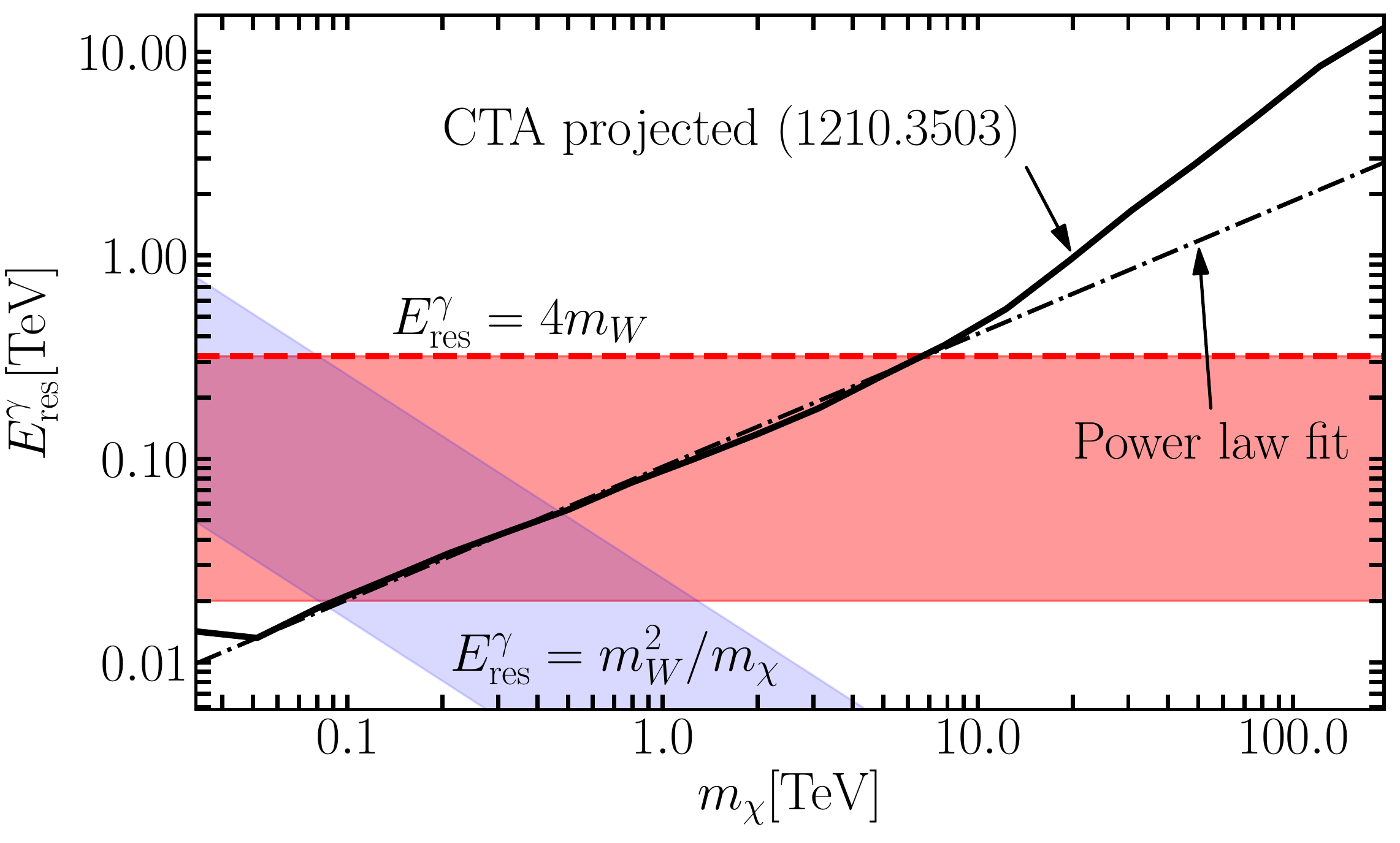} 
\caption{Energy resolution of the CTA experiment (solid black line, 
from \cite{Bernlohr:2012we}), and the power-law fit $\Eres = 
0.0915 \,(E_\gamma/\mbox{TeV})^{0.653}$ (dash-dotted) 
with $E_\gamma = \mchi$. 
The dark-grey (red) and light-grey (blue) bands show where the 
intermediate and narrow resolution 
resummation applies, respectively. The boundaries are defined 
by $m_W \,[1/4,4]$ (intermediate resolution) and $m_W^2/\mchi\, 
[1/4,4]$ (narrow resolution).
\label{fig:CTA}}
\end{figure}

The intermediate resolution regime is relevant to present and upcoming 
DM searches. For example, assuming the regime to apply to 
$\Eres$ in $[m_W/4,4 m_W]$ the energy resolution 
of the H.E.S.S. experiment $\Eres/E_\gamma \approx 10\%$ 
\cite{deNaurois:2009ud} implies that dark matter masses in the 
range 200~GeV to 3.2~TeV are covered by the intermediate resolution 
calculation. For the CTA experiment, we obtain the power-law fit 
$\Eres/E_\gamma = 0.0915\,(E_\gamma/\mbox{TeV})^{-0.347}$ 
from Figure~11 of \cite{Bernlohr:2012we}
in the range of photon energies of interest, which is shown as the 
dash-dotted line in Figure~\ref{fig:CTA} together with the 
unapproximated resolution (solid line). The horizontal band (dark-grey/red)  
represents the region of applicability of the intermediate 
resolution regime, which extends to 6.8~TeV for the CTA experiment. 
Thus, the intermediate resolution calculation covers the mass values 
where the thermal relic density of the pure Higgsino (electroweak doublet) 
and pure wino (triplet) models agrees with the observed relic density.

The outline of the paper is as follows. In Section~\ref{sec:theory} 
we discuss the momentum modes and effective Lagrangians relevant to the 
problem and derive a factorization formula for the photon energy 
spectrum valid when $\mchi-E_\gamma ={\cal O}(m_W)$, which corresponds 
to the intermediate resolution regime. We also discuss the modifications 
that apply to narrow resolution
\cite{Beneke:2018ssm}. In Section~\ref{sec:nllprime} we calculate 
the hard, jet and soft functions that appear in the factorization 
formula at the one-loop order, as required for NLL' resummation, 
provide the renormalization group equations (RGEs), 
and solve them with corresponding 
accuracy. These calculations are performed for the specific 
case of the pure wino model, which corresponds to the Standard Model 
extended by an SU(2) triplet with zero hypercharge, of which 
the electrically neutral member is the dark matter particle.
In the subsequent Section~\ref{sec:results} we show our 
main result, the resummed endpoint-integrated photon spectrum 
in the range of dark matter masses of interest and for various 
$\Eres$. We match the intermediate resolution calculation to the 
narrow resolution one from \cite{Beneke:2018ssm} and find very good 
agreement. We pursue and explain this numerical observation in 
Section~\ref{sec:logexpansions} by expanding analytically the resummed 
expressions to the two-loop order and comparing the logarithmic and 
constant terms. We conclude in Section~\ref{sec:conclusion}. A series 
of appendices collects additional technical details on soft and jet 
function integrals, including the narrow resolution case and the treatment
of the Z-boson resonance, 
the RGE invariance check for the narrow resolution 
case, and the complete analytic expressions for the expansion of 
the resummation formula to the two-loop order.


\section{Factorization of the energy spectrum}
\label{sec:theory}

The annihilation cross section of TeV scale dark matter with electroweak 
charges can be strongly modified by the Sommerfeld effect \cite{Hisano:2004ds} 
due to non-relativistic scattering of the dark matter particles before 
they annihilate. This effect is well understood. Our concern are the 
large electroweak logarithms in the annihilation rate $\chi^0\chi^0\to\gamma +X$,
when the photon energy is close to maximal, $E_\gamma \sim \mchi 
\gg m_W$, more precisely $m_\chi-E_\gamma \leq \Eres\ll \mchi$. The observation 
of a photon with this energy implies that the unobserved final state 
$X$ is ``jet-like'' with small invariant mass $m_X=\sqrt{4\mchi\Eres}$. 
The logarithmic enhancements of such final states are caused by soft 
and collinear physics relative to the large scale $\mchi$. In a systematic 
expansion in $m_W/\mchi$, where $\Eres$ is parametrically related to $m_W$, 
$\mchi$ as above, the $\chi^0\chi^0\to\gamma +X$ process is separated into a 
hard annihilation process and the low-scale initial- and final-state 
dynamics, which is described by suitable effective Lagrangians valid at scales 
$\mu\ll \mchi$.

We assume that the DM particle is the component of an SU(2) 
multiplet of the electroweak interaction, which remains electrically 
neutral after electroweak symmetry breaking. Since $\mchi\gg m_W$, this is always 
a good approximation at leading order in an expansion in $m_W/\mchi$ unless 
there are two nearly degenerate heavy multiplets, such that large 
mixing can occur. For definiteness, we assume (as in \cite{Beneke:2018ssm}) 
that $\chi_a$, $a=1,\ldots, 2j+1$, is a $2j+1$ dimensional isospin-$j$ SU(2) 
multiplet of Majorana fermions with integer $j$ 
(thus hypercharge vanishes). The essence of 
the derivation of the factorization formula below does not rely on 
these assumptions.

\subsection{Effective Lagrangians and annihilation operators}

After integrating out virtualities of order $\mchi^2$, the short-distance 
part of the annihilation process is represented by an operator that destroys
the two DM particles at a single point, and a set of collinear and 
anti-collinear fields along opposite light-like directions starting 
from this point, which describe the energetic particles in $X$ and those 
that convert to the observed photon. We refer to the direction $\nm^\mu$ of the 
jet $X$ as ``collinear''. The direction of the photon momentum defines 
the ``anti-collinear'' direction, $p_\gamma^\mu = E_\gamma \np^\mu$. 
The reference vectors satisfy $\nm^2=\np^2=0$, $\np\cdot\nm=2$. A general 
momentum is written in components as $k^\mu =
(\np k,\nm k,\,\mathbf{k}_\perp)$, such that for 
collinear momenta $\np\!\cdot k \gg \nm\!\cdot k$, vice-versa for anti-collinear 
momenta. 

The low-energy dynamics of the intermediate resolution case 
is described by non-relativistic effective field 
theory~\cite{Bodwin:1994jh} and soft-collinear effective field theory 
(SCET) \cite{Bauer:2000yr,Bauer:2001yt,Beneke:2002ph}. The kinematics 
of the annihilation process considered here is a mixture of an inclusive 
process in the collinear direction, also called a SCET$_{\rm I}$ problem, 
and an exclusive final state of the  SCET$_{\rm II}$ type in the other 
direction with the added complication of electroweak symmetry breaking 
and gauge boson masses. The effective Lagrangian must describe the 
interactions of the relevant modes with momentum scaling
\begin{align}
\text{hard-collinear}\,(hc):\quad & k^\mu \sim  \mchi (1,\la,\sqrt{\la})
\nonumber \\
\text{collinear}\,(c): \quad & k^\mu \sim  \mchi (1,\la^2,\la)
\nonumber \\
\text{anti-collinear}\,(\bar{c}): \quad & k^\mu \sim  \mchi (\la^2,1,\la)
\nonumber \\
\text{soft}\,(s): \quad & k^\mu \sim  \mchi (\la,\la,\la) \nonumber
\\
\text{potential}\, (p) :\quad & k^0\sim m^2_W/\mchi,\, \mathbf{k}\sim m_W
\nonumber\\
\text{ultrasoft}\,(s): \quad & k^\mu \sim  \mchi (\la^2,\la^2,\la^2)
\end{align}
Hard modes with momentum $ k^\mu \sim  \mchi (1,1,1)$ are integrated out into 
matching coefficients and are no longer part of 
the effective Lagrangian by construction. 
The power counting parameter is the small ratio $\lambda = m_W/\mchi$. 
Compared to the narrow resolution case \cite{Beneke:2018ssm}, an additional 
hard-collinear mode is needed to describe the unobserved final state $X$ 
with hard-collinear virtuality of order $\mchi^2 \lambda =\mchi m_W$. 
On the other hand, the effective theory for the wide resolution 
case \cite{Baumgart:2017nsr,Baumgart:2018yed}
requires a yet more numerous set of modes to account for the independent 
scales $\Eres$ and $m_W$. This set collapses to the one above when 
$\Eres$ is set parametrically to $m_W$. 

The leading hard annihilation processes are those into two energetic 
final-state particles. Adding another collinear or anti-collinear 
field to the primary annihilation vertex, implies a suppression by at 
least one power in $\lambda$ due to the scaling of the fields in the 
effective Lagrangian. In this work as well as in all previous works 
on electroweak Sudakov resummation for dark matter annihilation, the 
aim is to sum logarithms of $m_W/\mchi$ at leading power in the expansion 
in $\lambda$. Power-suppressed effects in $\lambda=m_W/\mchi$ are 
systematically neglected in this treatment. 

A consequence of neglecting power corrections is that 
(anti-) collinear fields must preserve their identity while emitting 
soft radiation. Since the energetic final state in the anti-collinear 
direction consists of a single photon with nearly maximal energy, which 
hence cannot be generated from an energetic fermion or Higgs boson, the 
leading-power operators for the hard annihilation process contain a single 
anti-collinear SU(2) or U(1)${}_Y$ gauge field. The collinear part 
of the operator must then also be an SU(2) or U(1)${}_Y$ 
gauge field by gauge invariance, because the
non-relativistic initial state consists of a DM two-particle 
state with vanishing hypercharge and colour, and integer weak isospin. 
It is therefore not possible to combine the anti-collinear SU(2) (U(1)${}_Y$) 
gauge field with any other single Standard Model (SM) field to form these 
quantum numbers, except with another SU(2)  (U(1)${}_Y$) gauge field. The 
hard annihilation 
process is reproduced by the effective Lagrangian 
\begin{equation} 
\label{eq:lann}
{\cal L}_{\rm ann} = \frac{1}{2 m_\chi} \sum_{i} 
\int dsdt\,\hat{C}_i(s,t,\mu) \mathcal{O}_i
\end{equation}
with operators of the form
\begin{equation}
\label{eq:opdef}
\mathcal{O}_i = \chi_v^{c\dagger}\Gamma^{\mu\nu}_i T^{AB}_i \chi_v\,
\mathcal{A}^A_{\perp c,\mu}(sn_+) \mathcal{A}^B_{\perp \bar{c},\nu}(tn_-)\,.
\end{equation}
Here $\chi_v$ is a two-component non-relativistic spinor field in the 
SU(2) weak isospin-$j$ representation, $\chi_v^c = -i\sigma^2 \chi_v^* 
=-\epsilon \chi_v^*$ (with $\epsilon$ the antisymmetric 
2~x~2 matrix with $\epsilon_{12}=1$)
the charge-conjugated field, and $\mathcal{A}^A_{\perp c,\mu}$ 
($\mathcal{A}^B_{\perp \bar{c},\nu}$) the collinear (anti-collinear) 
SU(2) gauge field of soft-collinear effective theory. (For the case of 
U(1)${}_Y$, replace $\mathcal{A}^A_{\perp,\mu}$ by 
$\mathcal{B}_{\perp,\mu}$ and omit $T^{AB}_i$.) The definitions 
will be given below. Fields without position arguments are 
evaluated at $x=0$. The operator is non-local, since (anti-) collinear field 
operators are integrated along the light-cone of the respective direction 
with the coefficient function $\hat{C}_i$.\footnote{In momentum space, this simply 
implies that the hard coefficient depends on the large (anti-) collinear momentum 
component.}  The spin matrix  
$\Gamma^{\mu\nu}_i$ is contracted with the two-spinor indices of the 
DM fields (not written explicitly) and the Lorentz index of the 
gauge fields. Similarly, the SU(2) group index matrix $T^{AB}_i$ is contracted with 
the two isospin-$j$ representation indices of the 
DM fields (not written explicitly) and the adjoint index of the 
gauge fields. 
The operator basis is given by the list of distinct spin- and group-matrix 
structures.

\subsubsection{Non-relativistic dynamics}

For energies below $\mchi$ but above $m_W$ the DM interactions are described by 
the standard non-relativistic Lagrangian 
\begin{equation}
\label{eq:LNRDM}
\mathcal{L}_{\rm NRDM} = \chi_v^\dagger(x) \left(i D^0 + 
\frac{\mathbf{D}^2}{2\mchi}\right)\chi_v(x)
\end{equation}
with $D_\mu = \partial_{\mu} - i g_2 A^C_\mu\, T^C$  the SU(2) covariant 
derivative, $T^C$, $C=1,2,3$, the SU(2) generators in the isospin-$j$ 
representation of the DM field, and $g_2$ the SU(2) gauge coupling. 
The Lagrangian can be extended to 
include interactions suppressed by powers of $\mathbf{p}/\mchi$. Since 
the largest non-relativistic momentum scale is $m_W$, they correspond 
to power corrections, which are neglected.

The non-relativistic Lagrangian describes the soft, potential and ultrasoft 
modes (see, for example, \cite{Beneke:1997zp}) of the non-relativistic DM and 
light SM fields. The soft modes can be integrated out from the non-relativistic 
Lagrangian in straightforward analogy with heavy quark anti-quark systems 
in non-relativistic QCD. Together with the potential 
modes of the light particles, they generate instantaneous but spatially 
non-local interactions between the DM fields, that is, DM potentials. 
The effective Lagrangian for the remaining potential modes 
of the DM field and the ultrasoft modes of the light fields, is the 
potential-non-relativistic dark matter Lagrangian~\cite{Beneke:2012tg}. 
At leading power, 
\begin{eqnarray}
\label{eq:LPNRDM}
\mathcal{L}_{\rm PNRDM} &=& \sum_i \chi_{vi}^\dagger(x) \left(i D^0(t,\mathbf{0}) - 
\delta m_i+ \frac{\bm{\partial}^2}{2\mchi}\right)\chi_{vi}(x) 
\nonumber\\
&&-\,\sum_{\{i,j\},\{k,l\}}
\int d^3\mathbf{r}\, V_{\{ij\}\{kl\}}(r)\,
\chi_{vk}^\dagger(t,\mathbf{x})\chi_{vl}^\dagger(t,\mathbf{x}+\mathbf{r}) 
\chi_{vi}(t,\mathbf{x})\chi_{vj}(t,\mathbf{x}+\mathbf{r}) \,.
\quad
\end{eqnarray}
We indicated explicitly the space-time arguments of the fields to highlight 
the non-locality of the potential interaction and the fact that the ultrasoft 
gauge field in the covariant derivative $D^0$ is multipole-expanded around 
$\mathbf{x}=0$.\footnote{The multipole expansion must be done around 
the center-of-mass point, which here is assumed to be zero. If the 
center-of-mass were at some $\vec{a}$, the matrix elements would acquire an 
irrelevant phase due to translation invariance of the 
center-of-mass. However, the multipole expansion breaks 
translation invariance in the relative coordinate by the explicit 
appearance of the special point $\vec{x}=0$ (and explicit factors of 
$\vec{x}$ in higher-order terms in the Lagrangian). The breaking 
is always of higher-order in the EFT expansion than the one of 
consideration, and is reduced successively with order of accuracy. 
This can best be seen in momentum space, where setting $\vec{x}=0$ 
corresponds to neglecting subleading ultrasoft momentum 
components in the interactions with heavy particles, and the 
explicit appearance of $\vec{x}$ to an expansion in these 
components. The same 
issue arises in the standard quantum mechanics problem of the interaction of 
light with atoms, and in the multipole-expansion of 
soft-collinear effective theory below.} 
Note that the covariant derivative is now the one with respect 
to the unbroken electromagnetic gauge symmetry, since ultrasoft light fields 
with momentum $k\sim \mchi \lambda^2 \sim m_W^2/\mchi$ exist only for fields 
with masses much smaller than $m_W$. The electroweak gauge bosons no longer 
appear as dynamical fields in PNRDM effective theory.

The soft modes of the light particles have virtuality of order $m_W^2$. 
Hence, in matching NRDM EFT to PNRDM EFT, the masses of 
the electroweak gauge bosons and of the top quark and Higgs boson cannot 
be neglected. The potential $V_{\{ij\}\{kl\}}(r)$ depends on these masses, 
resulting in Yukawa (electroweak gauge bosons, Higgs bosons) and Coulomb 
potentials (photons). Furthermore, the components of the original isospin-$j$ 
DM multiplet acquire slightly different masses after electroweak 
symmetry breaking. The Lagrangian above uses $\delta m_i = m_i - m_{\chi^0} 
\geq 0$, where $m_i$ is the mass of eigenstates labelled by $i$. Since 
$\mathcal{L}_{\rm PNRDM}$ is no longer invariant under the SU(2)$_L\times$U(1)$_Y$ 
gauge symmetry and calculations are carried out in broken theory, we 
express it in terms of mass eigenstate fields $\chi_{vi}$ rather than 
the fields $\chi_{va}$ of the SU(2) multiplet. 

While soft subgraphs not connected to the annihilation vertex generate 
potential interactions, soft momentum running through the annihilation 
vertex, dresses the non-relativistic fields in the operators (\ref{eq:opdef}). 
Since the leading soft interaction is of the eikonal type, this 
dressing takes the form of a Wilson line. This is seen most easily by 
noting that the temporal soft gauge-field coupling in the covariant derivative 
$D^0$ in (\ref{eq:LNRDM}) can be removed by the field redefinition
\begin{equation}
\label{eq:Sw}
\chi_{va}(x) = Y_{v,a b}(x_0) \chi^{(0)}_{vb}(x)\,,
\end{equation}
where the soft Wilson line $Y_v(x)$ is defined as the path-ordered 
exponential 
\begin{equation}
Y_{v}(x) = \mbox{P} \exp \left[i g_2 \int^0_{-\infty} d t \, 
v\cdot A^C_s(x+v t)  \,T^{C} \right].
\end{equation}
with $T^C$ the SU(2) generators in the spin-$j$ representation and 
$v^\mu=(1,\mathbf{0})$.\footnote{\label{footnote1}
The field redefinition is analogous to but not the same as the field 
redefinition discussed in \cite{Beneke:2010da}. There ultrasoft gauge bosons 
are decoupled from potential fields in {\it potential} non-relativistic EFT. 
Here the decoupling refers to soft gluons in NRDM EFT. In both cases the soft 
Wilson lines can and must be evaluated at the multipole-expanded position 
$x_0 = (t,\mathbf{0})$, since the three-momentum of the gauge boson does not enter 
the virtual heavy particle propagator in the leading-power approximation.}
Dropping the superscript (0), the non-relativistic part of the operators 
(\ref{eq:opdef}) takes the form
\begin{equation}
\label{eq:chipart}
\chi_v^{c\dagger}\Gamma^{\mu\nu}_i [Y_v^\dagger T^{AB}_i Y_v] \chi_v
\end{equation} 
after the field redefinition.  The coupling of soft 
electroweak gauge bosons to the DM field is now (at leading power) 
fully accounted by the Wilson lines $Y_v$ in the operator 
and the soft gauge bosons are decoupled from the Lagrangian (\ref{eq:LPNRDM}).

The main use of the PNRDM Lagrangian after the field redefinition is related 
to the computation of the Sommerfeld effect. In this context it is convenient 
to write the sums over the field indices in terms of a sum over the composite 
indices $I=\{ij\}$ and $K=\{kl\}$ of two-particle states according to the 
bound- and scattering-states of the Schr\"odinger problem for the relative 
coordinate. For example, for the triplet wino model, the index $i$ takes the 
values $0,+,-$ corresponding to the electric charge of the DM mass eigenstates. 
The index $I$ runs over the nine two-particle values $00,+-,-+,0+,+0,0-,-0,
++,--$, ordered by the modulus of the electric charge. Since electric charge 
is conserved, it is sufficient for the computation of the $\chi^0\chi^0$ 
annihilation rate to solve the Sommerfeld problem in the charge-0 sector of 
the two-particle states. For the SU(2) $j=1$ triplet the fields are related to the 
mass basis by $\chi^\pm = (\chi_1 \mp i \chi_2)/\sqrt{2}$, $\chi^0 = \chi_3$. 
The two-particle states are related by  
\begin{equation}
\chi_{va}^{c\dagger} \chi_{vb} = K_{ab,I} [\chi_{v}^{c\dagger} \chi_{v}]_I\,,
\end{equation}
where the $9\times9$ matrix $K_{ab,I}$ can be read off from 
\begin{equation}
\begin{pmatrix}
\chi_1^T\epsilon\chi_1\\
\chi_1^T\epsilon\chi_2\\
\chi_1^T\epsilon\chi_3\\
\chi_2^T\epsilon\chi_1\\
\chi_2^T\epsilon\chi_2\\
\chi_2^T\epsilon\chi_3\\
\chi_3^T\epsilon\chi_1\\
\chi_3^T\epsilon\chi_2\\
\chi_3^T\epsilon\chi_3
\end{pmatrix}_{\!\!ab}
\!=
\begin{pmatrix}
 0 & \frac{1}{2} & \frac{1}{2} & 0 & 0 & 0 & 0 & \frac{1}{2} & \frac{1}{2} \\
 0 & -\frac{i}{2} & \frac{i}{2} & 0 & 0 & 0 & 0 & \frac{i}{2} & -\frac{i}{2} \\
 0 & 0 & 0 & 0 & \frac{1}{\sqrt{2}} & 0 & \frac{1}{\sqrt{2}} & 0 & 0 \\
 0 & \frac{i}{2} & -\frac{i}{2} & 0 & 0 & 0 & 0 & \frac{i}{2} & -\frac{i}{2} \\
 0 & \frac{1}{2} & \frac{1}{2} & 0 & 0 & 0 & 0 & -\frac{1}{2} & -\frac{1}{2} \\
 0 & 0 & 0 & 0 & \frac{i}{\sqrt{2}} & 0 & -\frac{i}{\sqrt{2}} & 0 & 0 \\
 0 & 0 & 0 & \frac{1}{\sqrt{2}} & 0 & \frac{1}{\sqrt{2}} & 0 & 0 & 0 \\
 0 & 0 & 0 & \frac{i}{\sqrt{2}} & 0 & -\frac{i}{\sqrt{2}} & 0 & 0 & 0 \\
 1 & 0 & 0 & 0 & 0 & 0 & 0 & 0 & 0 
\end{pmatrix}
\begin{pmatrix}
\chi^{0T}\epsilon \chi^0 \\
\chi^{+T}\epsilon \chi^- \\
\chi^{-T}\epsilon \chi^+ \\
\chi^{0T}\epsilon \chi^+ \\
\chi^{+T}\epsilon \chi^0 \\
\chi^{0T}\epsilon \chi^- \\
\chi^{-T}\epsilon \chi^0 \\
\chi^{+T}\epsilon \chi^+ \\
\chi^{-T}\epsilon \chi^-
\end{pmatrix}_{\!\!I} \, .
\end{equation}
While the specific form of the $K$-matrix depends on the SU(2) representation 
of the DM field, the formalism is general. 

From this point on the non-relativistic part of the problem follows the 
discussion of the computation of the Sommerfeld effect for an arbitrary 
set of heavy fermions nearly degenerate with the DM particle, developed 
for the general minimal supersymmetric 
SM in~\cite{Beneke:2012tg, Hellmann:2013jxa,Beneke:2014gja,Beneke:2014hja}. 
The framework described in these papers in turn 
generalizes the original work~\cite{Hisano:2004ds,Hisano:2006nn} 
to mixed DM states and reformulates it in the DM EFT context. For the 
case of the pure wino triplet, the result is identical to the original 
treatment, but can in principle be extended to systematically include 
radiative and velocity corrections to the Sommerfeld effect. In this paper, 
however, as in all previous studies, the Sommerfeld effect is computed 
only at leading order in the PNRDM EFT. 

What will appear in the factorization formula is the matrix element of 
non-relativistic annihilation operators of the form 
\begin{equation}
\chi^\dagger_{ve_4} \Gamma\chi^c_{ve_3}
\, \chi^{c \dagger}_{ve_2}\Gamma\chi^{}_{ve_1}\,,
\end{equation}
which arise from (\ref{eq:chipart}) after squaring the amplitude. The 
matrix $\Gamma = 1$ or $\sigma^i$ (Pauli matrix) operates on the spinor 
index of $\chi_{v}$, depending on whether the fermion bilinear destroys or 
creates a spin-0 or spin-1 state. By assuming that the spin matrices 
in the two bilinears are the same, we implicitly make use of the fact that 
the potential $V_{\{ij\}\{kl\}}(r)$, while being spin-dependent, does not 
change the spin of the incoming two-particle state before it annihilates. 
The NRDM EFT matrix element of the above operator in an incoming 
$\chi_{i} \chi_{j}$ DM state with some relative velocity $v_{\rm rel}$, 
orbital quantum number $L=0$ (S-wave) and total spin $S$ is given by 
(no sum over $i,j$) \cite{Beneke:2014gja}
\begin{eqnarray}
&& 
\langle \chi_i \chi_j | \, \chi^\dagger_{ve_4} \Gamma\chi^c_{ve_3}
\, \chi^{c \dagger}_{ve_2}\Gamma\chi^{}_{ve_1} \, | \chi_i \chi_j \rangle = 
\langle \chi_i \chi_j | \, 
\chi^\dagger_{ve_4} \Gamma \chi^c_{ve_3} |0 \rangle \, 
\langle 0| \chi^{c \dagger}_{ve_2}\Gamma\chi^{}_{ve_1}\, | \chi_i \chi_j \rangle
\nonumber\\
&&  
= \,\Big[\,\langle \xi^{c \dagger}_{j} \Gamma \xi_{i} \rangle \, 
\big( \psi^{(0,S)}_{e_4 e_3,\,ij} \, + \,  (-1)^S 
\psi^{(0,S)}_{e_3 e_4,\,ij} \big) 
\Big]^*
\,\langle \xi^{c \dagger}_{j}\Gamma \xi_{i} \rangle \,
\big( \psi^{(0,S)}_{e_1 e_2, \,ij} \, + \,  (-1)^S \psi^{(0,S)}_{e_2 e_1, \,ij} 
\big)\, ,\quad
\label{eq:wave}
\end{eqnarray}
where $\psi^{(L,S)}_{e_1 e_2,\,ij}$ is the $\chi_{e_1} \chi_{e_2}$-component 
of the scattering wave function for the incoming $\chi_{i} \chi_{j}$ state, 
evaluated for zero relative distance and
normalized to the free scattering solution, that is 
$\psi^{(L,S)}_{e_a e_b,\,ij}\to \delta_{e_a i}\,\delta_{e_b j}$ in the 
absence of interactions.\footnote{The first equality in (\ref{eq:wave}) 
holds, since the PNRDM Lagrangian (\ref{eq:LPNRDM}) conserves particle 
number. Hence, 
inserting a complete set of states between the two 
DM bilinears, only the vacuum state contributes.} 
The symbols $\xi_{i},\,\xi_j$ in the second
line of~(\ref{eq:wave}) denote the Pauli spinor of the incoming particles 
$\chi_{i}$ and $\chi_j$, and $\langle \dots \rangle$ stands for the trace in 
spin space (spin sum). The multi-component wave function 
$\psi^{\,(L,S)}_{e_2 e_1,ij}$ 
accounts for the potential interactions of the incoming 
$\chi_{i} \chi_{j}$ state, which couple it to all possible intermediate
two-body states $e_2 e_1$ with the same charge, spin and orbital angular momentum. 
Both wave-function components $e_1 e_2$ and $e_2 e_1$ contribute to the 
matrix-element of the operator  $\chi^{c \dagger}_{e_2}\chi^{}_{e_1}$. For an operator 
with quantum numbers $L$ and $S$, there is a 
relative sign $(-1)^{L+S}$ between the two components. 
The above-defined $\psi^{(0,S)}_{e_1 e_2,\,ij}$ is related via 
\begin{eqnarray}
\psi^{(0,S)}_{e_1 e_2, \,ij} = [\psi_{E}(0)]_{e_1 e_2, \,ij}^*
\label{eq:psiwf0}
\end{eqnarray}
to the coordinate-space scattering wave-function $[\psi_E(\mathbf{r}\,)]_{I,ij}$ 
at the origin, which in turn can be obtained directly from the matrix-Schr\"odinger 
equation
\begin{equation}
\left(\left[-\frac{\bm{\nabla}^{\,2}}{2\mu_{I}} - E\right] \delta_{IK}
+ V_{IK}(r)\right) [\psi_E(\mathbf{r}\,)]_{K,ij} =0
\label{eq:schroedinger1}
\end{equation}
with the potential (\ref{eq:LPNRDM}), now including the mass splitting between 
the mass $M_I$ of the two-particle state $I$ and the mass of the $\chi^0\chi^0$ state 
via $V_{IK} \to V_{IK} +\delta_{IK} (M_I - 2\mchi)$. 
The energy $E$ is fixed through the 
relative velocity of the initial state, and the label $ij$ refers to the fact 
that this equation should be solved with the initial condition corresponding 
to the particular incoming two-particle state $ij$. $\mu_I$ is the reduced mass 
of the two-particle state $I$, which can be set to $\mchi/2$ in the leading-order 
treatment of the Sommerfeld effect. We refer to \cite{Beneke:2014gja} 
for further details and the methods employed to solve this equation. 

For the task at hand, we focus on the initial state $I=ij=00$. The other 
two-particle states appear only as virtual states in the ladder diagrams 
summed by the Schr\"odinger equation.\footnote{They also appear as convenient 
external states for the computation of the matching coefficients. Since 
the potential interaction can convert the $00$ into the $+-$ state, 
which then annihilates at short distances, the short-distance coefficients 
have to be computed for all two-particle states, including off-diagonal 
terms, see \cite{Beneke:2012tg,Beneke:2014gja}.} Due to electric charge conservation, 
the potential is block-diagonal and it is sufficient to solve 
(\ref{eq:schroedinger1}) in the charge-0 sector, which, for the wino 
example, consists of $I=00,+-,-+$. The description in terms of three 
two-particle states is convenient, since the framework can be formulated 
without additional rules for the construction of the potential for different 
$S$ and $L$. The (anti-) symmetrization is encoded automatically in the 
(anti-) symmetry of the operator and its short-distance coefficient. However, 
the description is redundant, since the fermion bilinear with $+-$ fields 
is identical up to a possible sign to the one with $-+$. It is customary 
in the discussion of the Sommerfeld effect to reduce the basis of two-particle 
states to non-identical ones (six instead of nine, for the triplet model, 
and two instead of three for the charge-0 sector). 
In the following we adopt this convention. 
Specifically, for the wino (triplet) model, $I$ shall then refer to $00,+-$
only. The precise relation between the two formulations, referred to as 
method-1 and method-2, respectively, is explained in \cite{Beneke:2014gja}, 
including the explicit forms of the potential and tree-level short-distance 
annihilation coefficients in both methods. Irrespective of the method, 
the Sommerfeld factors are defined as 
\begin{equation}
S_{IJ} = \left[\psi^{(0,S)}_{J,\,00}\right]^* \psi^{(0,S)}_{I, \,00}\,.
\label{eq:SFdef}
\end{equation}

The discussion up to now ignored the coupling of ultrasoft photons to 
the charged members of the DM multiplet through the electromagnetic 
covariant derivative in (\ref{eq:LPNRDM}). This is justified, since the 
field redefinition mentioned in footnote~\ref{footnote1} removes the 
ultrasoft photon field from the Lagrangian at the expense of modifying 
the DM fermion bilinear as 
\begin{equation}
[\chi_{v}^{c\dagger} \chi_{v}]_I \to S_{vi} S_{vj} 
[\chi_{v}^{c\dagger} \chi_{v}]_I\,
\end{equation}
where $S_{vi}$ is an electromagnetic time-like Wilson line corresponding to the 
charge of the field $\chi_{vi}$ in $I=\{ij\}$. In the charge-0 sector, 
the charges of the fields $i,j$ add to zero, which implies $S_{vi} S_{vj} = 
S_{vi} S^\dagger_{vi}=1$, reflecting the well-known fact that photons 
with wave-length much larger than the size of the system only couple to 
the total charge (here, zero) of the system.

Finally we note that the factorization of non-relativistic dynamics from 
the soft and collinear dynamics of the final state is independent of the 
photon energy resolution as defined above, at least to the accuracy 
considered here. The key requirement is the decoupling of 
soft and ultrasoft interactions 
from the ladder diagrams that build up the Sommerfeld effect, which holds 
because soft gauge bosons throw potential DM propagators off-shell, and 
because ultrasoft photons do not interact with the electrically neutral 
two-particle state.

\subsubsection{Soft-collinear  dynamics}
\label{sec:scetdynamics}

The characteristics of the final state is an energetic photon, whose 
momentum is balanced by a jet of unobserved particles. The low-energy 
physics of energetic objects with small invariant mass interacting 
with soft modes is described by SCET. Its application 
to electroweak Sudakov situations was first discussed in 
\cite{Chiu:2007yn,Chiu:2009mg} for the production of two particles 
with electroweak charges in a high-energy collision. As in the 
non-relativistic sector, one needs two different effective Lagrangians 
depending on whether the virtuality is much larger than or of order $m_W^2$. 

Although in higher orders all SM fields are present in the collinear 
and soft interactions, we restrict ourselves to the gauge boson 
Lagrangian, since the gauge boson SCET fields appear directly in the 
annihilation operators, and since the discussion of fermions is fairly 
standard from QCD applications of SCET. The SCET Lagrangian consists 
of 
\begin{equation}
{\cal L}_{\rm SCET-I} = \mathcal{L}_c + \mathcal{L}_{\bar c} + 
\mathcal{L}_{\rm soft}\,,
\end{equation}
where $\mathcal{L}_{\rm soft}$ is the purely soft field Lagrangian 
that takes the same form as the corresponding SM Lagrangian except that 
all fields are assumed to be soft. The 
collinear Lagrangian at leading power is
\bea
{\cal L}_{c} &=& 
-\frac12\,{\rm tr}\left(F_c^{\mu\nu} 
F^c_{\mu\nu}\right) 
+(D^\mu \varphi_c)^\dagger D_{\mu}\varphi_c \,,
\label{eq:SCETLagrangian}
\eea
where $g_2 F_c^{\mu\nu}=i \,[D^\mu,D^\nu]$ as usual, but the collinear 
SU(2) covariant derivative is given by 
\bea
D^\mu &=& \partial^\mu -ig_2A_c^\mu(x)-ig_2\nm A_s(x_-+x_\perp)\frac{\np^\mu}{2} \,.
\label{eq:collinearcovariantderivative}
\eea
We included the 
collinear Higgs doublet field $\varphi_c$ for later purposes.
At leading power, soft-collinear interactions involve only the $\nm A_s$ 
projection of the soft gauge field. Moreover, the soft gauge field is evaluated 
at the multipole-expanded position $x_-^\mu+x_\perp^\mu$ with 
$x_-^\mu = (\np x) \,\nm^\mu/2$, reflecting the fact that the $\np k$ 
component of the soft momentum can be neglected relative to the corresponding 
large component of hard-collinear and collinear momentum.
The Lagrangian above accounts for collinear modes of both, the hard-collinear 
and collinear type and is formulated in the unbroken phase of SU(2) gauge symmetry, 
since this is relevant to the hard-collinear fields of virtuality $\mchi m_W$. 
The anti-collinear Lagrangian ${\cal L}_{c}$ is the 
same up to the interchange of $\np\leftrightarrow \nm$. The above expressions 
should be amended in an obvious manner to include the gauge 
field for the U(1) hypercharge interaction and its coupling to the 
Higgs field.

The SCET Lagrangian enjoys separate collinear, anti-collinear and soft gauge 
symmetries \cite{Beneke:2002ni}. 
It is convenient to express the collinear Lagrangian in terms of manifestly 
collinear-gauge invariant collinear fields\footnote{The following 
construction can be extended to the hypercharge gauge field by 
introducing an abelian U(1) collinear Wilson line. Since the Higgs 
field carries hypercharge, the collinear-invariant field includes the 
hypercharge Wilson line.} $\Phi_c(x) = 
W_c^\dagger(x) \varphi_c(x)$ and
 \begin{equation}
\mathcal{A}^{B,\mu}_c(x) T^B = 
\mathcal{A}^\mu_c(x) = \frac{1}{g_2} W_c^\dagger [i D^\mu W_c](x) 
= \int_{-\infty}^0 ds\,n_{+\nu} [W_c^\dagger F^{\nu\mu}_c W_c](x+s \np)\,,
\label{eq:defAc}
\end{equation}
where the collinear Wilson line is defined as 
\begin{equation}
W_{c}(x) = \mbox{P} \exp \left[i g_2 \int^0_{-\infty} d s \, 
\np\cdot A_c^C(x+s \np)  \,T^{C} \right].
\label{eq:defWc}
\end{equation}
In terms of these, together with
\begin{equation}
i \mathcal{D}^\mu \equiv W_c^\dagger i D^\mu W_c 
= i\partial^\mu+g_2\mathcal{A}_c^\mu,
\qquad \mathcal{F}^{B}_{c,\mu\nu} T^B 
= \frac{i}{g_2} [\mathcal{D}_\mu,\mathcal{D}_\nu],
\end{equation}
the collinear Lagrangian is expressed as 
\bea
{\cal L}_{c} &=& 
-\frac12\,{\rm tr}\left(\mathcal{F}_c^{\mu\nu} 
\mathcal{F}^c_{\mu\nu}\right) 
+(\mathcal{D}^\mu \Phi_c)^\dagger \mathcal{D}_{\mu}\Phi_c \,.
\label{eq:SCETLagrangian2}
\eea
In this form it is apparent that the collinear gauge field degrees of freedom 
are represented by the transverse fields, since $\np \mathcal{A}_c=0 $ 
from (\ref{eq:defAc}),  while $\nm \mathcal{A}_c$ can be eliminated using the 
gauge-field equation of motion (see, for instance \cite{Beneke:2017ztn}, 
Appendix~B for the operator equation in QCD).

At scales $\mu\ll \mchi$ there are no interactions between collinear modes 
of different directions, as well as between collinear and non-relativistic 
DM modes, since this would result in hard virtualities, which are already 
integrated out into the short distance coefficients of annihilation operators.  
However, they all interact with each other through the soft gauge fields. 
As was the case for the non-relativistic DM field, the soft gauge field 
can be decoupled from the hard-(anti-)collinear fields through the 
field redefinition\footnote{In the coupling to hard-collinear (as opposed to 
collinear) fields, the argument of the soft field in the covariant 
derivative (\ref{eq:collinearcovariantderivative}) can be set to 
$x_-$ at leading-power accuracy, since the transverse momentum of the soft 
mode is negligible compared to the hard-collinear one. This simplification 
is required, so that the field redefinition in the following equation 
removes the soft field from the covariant derivative on hard-collinear 
fields. Also note the order of adjoint indices in (\ref{eq:Yn}). With the 
alternative definition $\mathcal{A}^B_{c}(x) = \mathcal{A}^{C (0)}_{c}(x)
Y_+^{CB}(x_-)$, the sign in the exponent of (\ref{eq:Ypmdef}) is $+i g_2$.} 
\begin{equation}
\label{eq:Yn}
\mathcal{A}^B_{c}(x) = Y_+^{BC}(x_-)\mathcal{A}^{C (0)}_{c}(x)
\qquad
\mathcal{A}^B_{\bar c}(x) = Y_-^{BC}(x_+)\mathcal{A}^{C (0)}_{\bar c}(x)\,,
\end{equation}
with \cite{Bauer:2001yt}
\begin{equation}
Y_{\pm}(x) = \mbox{P} \exp \left[-i g_2 \int_0^{\infty} d s \, 
n_\mp\cdot A_s^D(x+s n_\mp)  \,T^{D} \right].
\label{eq:Ypmdef}
\end{equation}
Here the SU(2) generator $T^D$ refers to the adjoint representation, 
$(T^D)_{BC} = -i \epsilon^{DBC}$, in case of the adjoint gauge field, 
and the fundamental representation for the analogous decoupling transformation 
of the (anti-) collinear Higgs fields. 
As a result of this field redefinition, soft Wilson lines 
appear in the annihilation operator. 

In the intermediate resolution case, the virtuality of the unobserved jet 
is not resolved by the measurement below the hard-collinear scale $\mchi m_W$. 
The dynamics of this jet is described by hard-collinear modes. On the 
other hand the scale for the anti-collinear direction of the observed photon 
is set by the virtuality $m_W^2$ of the collinear electroweak gauge bosons, 
whose masses cannot be neglected. The photon ``jet-function'' as well as 
possible mass effects within the hard-collinear jet must be computed with the SCET 
Lagrangian for the (anti-) collinear modes of the massive electroweak 
gauge bosons and the photon after electroweak symmetry breaking. 
The gauge boson mass term follows from the Higgs covariant kinetic 
term in (\ref{eq:SCETLagrangian2}) from
\begin{eqnarray}
&& (\mathcal{D}^\mu \Phi_c)^\dagger \mathcal{D}_{\mu}\Phi_c = 
(\np\partial \Phi_c)^\dagger \nm \mathcal{D}\Phi_c +
(\nm\mathcal{D} \Phi_c)^\dagger \np \partial\Phi_c +
(\mathcal{D}^\mu_\perp \Phi_c)^\dagger \mathcal{D}_{\perp,\mu}\Phi_c
\nonumber\\
&& \stackrel{\Phi_c = (0,v/\sqrt{2})}{\longrightarrow} \;
\frac{g_2^2 v^2}{8} \,\mathcal{A}^{B,\mu}_{\perp c}\mathcal{A}^{B}_{\perp c,\mu}\,,
\end{eqnarray}
which shows explicitly that the mass term arises only for the transverse field. 
The collinear gauge field Lagrangian for virtualities of order $m_W^2$, ignoring 
now the Higgs field,  
reads
 \bea
{\cal L}_{c} &=& 
-\frac{1}{4}\,\mathcal{F}_c^{B,\mu\nu} 
\mathcal{F}^{c,B}_{\mu\nu} + \frac{m_W^2}{2} 
\,\mathcal{A}^{B,\mu}_{\perp c}\mathcal{A}^{B}_{\perp c,\mu}\,.
\label{eq:SCETLagrangian3}
\eea
The hypercharge interaction should be added in the standard way. It is then 
convenient to express the Lagrangian in terms of mass eigenstates fields. 
The collinear fields no longer interact with soft fields, as discussed above, 
but the interaction with ultrasoft fields is still present. However, only 
fields with masses much smaller than $m_W$ can be ultrasoft, and the 
leading-power ultrasoft interactions are included through electromagnetic 
covariant derivatives acting on the electrically charged electroweak 
gauge fields with covariant derivatives defined as in 
(\ref{eq:collinearcovariantderivative}), except that now $\nm A_s(x_-)$ 
refers the ultrasoft photon field only.

\subsubsection{Annihilation operator basis}

The annihilation operators (\ref{eq:opdef}) and their short-distance 
coefficients do not depend on the photon energy resolution as long as 
$\Eres\ll \mchi$. Two relevant operators have been identified in 
previous work \cite{Ovanesyan:2014fwa,Ovanesyan:2016vkk,Beneke:2018ssm}, 
but the arguments 
that these two operators form a complete basis to all orders in 
perturbation theory have not been explicitly provided.
In the following we use symmetries to reduce the possible operators to 
\begin{eqnarray}
\mathcal O_1 &=& \chi_v^{c\dagger}\Gamma^{\mu\nu}\chi_v\,
\mathcal{A}^B_{\perp c,\mu}(sn_+) \mathcal{A}^B_{\perp \bar{c},\nu}(tn_-)\,,
\label{eq:opbasis1}
\\
\mathcal O_2 &=& \frac{1}{2} \,\chi_v^{c\dagger}\Gamma^{\mu\nu}
\{T^A,T^B\}\chi_v\,
\mathcal{A}^A_{\perp c,\mu}(sn_+) \mathcal{A}^B_{\perp \bar{c},\nu}(tn_-)\,,
\label{eq:opbasis2}
\\
\mathcal O_3 &=& \chi_v^{c\dagger}\sigma^\rho(n_{-\rho}-n_{+\rho})
T^C \chi_v\,
\epsilon^{CAB} \mathcal{A}^A_{\perp c,\mu}(sn_+) 
\mathcal{A}^{B,\mu}_{\perp \bar{c}}(tn_-)
\label{eq:opbasis3}
\end{eqnarray}
with the spin matrix in $d$ space-time dimensions given by 
\begin{equation}
\Gamma^{\mu\nu} = \frac{i}{4}\,[\sigma^\mu,\sigma^\nu] \,\sigma^\alpha 
(n_{-\alpha}-n_{+\alpha}) 
= \frac{1}{2 i}\,[\sigma^m,\sigma^n] \,\bm{\sigma}
\cdot\mathbf{n} \stackrel{d=4 \;\mbox{\tiny only}} 
= \frac{1}{2} \epsilon^{\mu\nu\alpha\beta} n_{+\alpha} n_{-\beta} 
\equiv \epsilon_{\perp}^{\mu\nu}\,.
\label{eq:spinmatrix}
\end{equation}
(Conventions  $v^\mu = (1,0,0,0)$, $n^\mu_\pm = (1,0,0,\mp 1)$, 
$\mathbf{n}=(0,0,1)$, $m,n=1,2,3$,  
$\epsilon^{0123}=-1$ are used.) We then show that the third operator does 
not contribute when the detected gauge boson is a photon.

We have already shown that the collinear and anti-collinear field must 
each consist of a single SU(2) or U(1)${}_Y$ gauge field. 
Since the wino does not carry hypercharge, operators with 
$\mathcal{B}_{\perp c}$ ($\mathcal{B}_{\perp \bar{c}}$) fields 
cannot be generated at tree level. Their matching coefficients 
can be non-zero starting from the two-loop order through closed 
loops of particles that carry SU(2) {\em and} hypercharge, 
such as the Higgs boson and the fermions of the SM. Since 
two-loop matching coefficients are needed only for 
resummation with NNLL' or 
higher accuracy, we drop these operators here. 
We therefore start from the 
general form (\ref{eq:opdef})
\begin{equation}
\label{eq:opdef2}
\mathcal{O}_i = \chi_v^{c\dagger} \Gamma^{\mu\nu}_i T^{AB}_i \chi_v\,
\mathcal{A}^A_{\perp c,\mu}(sn_+) \mathcal{A}^B_{\perp \bar{c},\nu}(tn_-)\,,
\end{equation}
and note that the two DM fields must couple to an operator with 
SU(2) isospin 0, 1 or 2. Thus, the group-index matrix must be from 
\begin{equation}
T_1^{AB} = \delta^{AB},\quad
T_2^{AB} = \frac{1}{2}\, \{T^A,T^B\},\quad
T_3^{AB} = \epsilon^{CAB} T^C\,,
\end{equation}
where $T^A$ are the SU(2) generators in the isospin-$j$ representation.

Turning to the spinor and Lorentz indices, the two spin-1/2 DM fields 
can couple to spin-0 or spin-1. In the first case the implicit pair of two-spinor 
indices of $\Gamma^{\mu\nu}$ must be of the form $\delta_{\alpha\beta}$. The  
spin-1 structure is the vector of Pauli matrices $(0,\bm{\sigma})$ or 
$[\sigma^\rho- (v\cdot \sigma) v^\rho]_{\alpha\beta}$.
For the spin-0 case, noting that $\mu$, $\nu$ are transverse indices, 
we obtain two different $\Gamma^{\mu\nu}$ by multiplying with 
\begin{equation}
g_\perp^{\mu\nu} = g^{\mu\nu} - \frac{\np^\mu \nm^\nu+\nm^\mu \np^\nu}{2}
\qquad\mbox{or}\qquad
\epsilon_\perp^{\mu\nu},
\label{eq:symstructures}
\end{equation}
defined in (\ref{eq:spinmatrix}). For spin-1, the three independent combinations 
\begin{equation}
(n_{-\rho}-n_{+\rho}) \,g_\perp^{\mu\nu},\quad
(n_{-\rho}-n_{+\rho}) \,\epsilon_\perp^{\mu\nu},\quad
g_{\rho\lambda}v_\kappa\epsilon^{\lambda\kappa\mu\nu}
\label{eq:antisymstructures}
\end{equation}
can be formed. Here $v_\rho \,(\sigma^\rho- (v\cdot \sigma) v^\rho) =0$ 
was used to reduce a number of further structures to the given ones. Together 
with the three independent SU(2) structures, this results in six spin-0 and 
nine spin-1 operators.

The final state of two gauge bosons must respect Bose symmetry, hence the operator 
has to be symmetric under the simultaneous exchange of all labels, 
$c\leftrightarrow \bar{c}$, $\np\leftrightarrow \nm$, $A\leftrightarrow B$, 
$\mu \leftrightarrow \nu$. The admissible structures are therefore the 
product of $T_1^{AB}$, $T_2^{AB}$ ($T^{AB}_3$) with the symmetric (antisymmetric) 
tensors from (\ref{eq:symstructures}), (\ref{eq:antisymstructures}), 
which leaves four spin-0 and three spin-1 operators.

The DM gauge interaction conserves CP symmetry and consequently parity for 
Majorana fermions. Since $\chi_v^{c\dagger}\chi_v$  
($\chi_v^{c\dagger}\bm{\sigma}\chi_v$) has negative (positive) parity, 
this excludes $g_\perp^{\mu\nu}$ in (\ref{eq:symstructures}) and 
all except the first structure in (\ref{eq:antisymstructures}), resulting 
in the two spin-0 operators $\mathcal{O}_{1,2}$ and one spin-1 operator 
$\mathcal{O}_3$ as given above.

In the process $\chi^0\chi^0\to\gamma +X$, the assumption of a single photon 
in the anti-collinear final state implies that the SU(2) index $B$ in  
$\mathcal{O}_i$ must necessarily be $B=3$. For the operator $\mathcal{O}_3$, 
the index $C$ in $\chi_v^{c\dagger}T^C \chi_v$ must then be $C=1$ or 2, 
in which case the bilinear cannot annihilate an electrically neutral 
two-particle state. Thus, $\mathcal{O}_3$ does not contribute to the annihilation 
into $\gamma+X$ when the photon is required to have nearly 
maximal energy. We note that the remaining two operators are both spin-singlet, 
so the dominant short-distance annihilation process occurs in the ${}^1S_0$ 
configuration. 

When the matching calculations are done with dimensional regularization, the 
question arises whether the above operator basis is complete in $d$ 
dimensions or whether it has to be complemented by evanescent operators, 
which vanish in $d=4$. We find that no evanescent operators arise. 
The $d$-dimensional basis is given by the same $\mathcal{O}_{1,2}$ 
except that the first form on the right-hand side of (\ref{eq:spinmatrix}) 
should be used for the spin matrix rather than the four-dimensional 
expression $\epsilon_\perp^{\mu\nu}$. To see this we note that an arbitrary 
full theory diagram in the calculation of the hard matching coefficients 
contains a single string of Dirac matrices of the form\footnote{Closed 
loops of DM lines do not contain $\gamma_5$ and the corresponding trace 
is well-defined in $d$ dimensions. Loops of SM chiral fermions require the 
same treatment as in the SM.} 
\begin{equation}
\bar{v}(\mchi v)\gamma^{\mu_1}\gamma^{\mu_2}\ldots \gamma^{\mu_N} u(\mchi v)
\end{equation}
with indices $\mu_i$ that can be contracted among each other, with the 
vectors $v$ or $n_\pm$ and with the two transverse polarization vectors  
$\varepsilon_{c\perp}$, $\varepsilon_{\bar{c}\perp}$ 
of the external gauge boson lines. To obtain an S-wave annihilation operator 
in the non-relativistic EFT, $N$ must be odd. By systematically exploiting 
the on-shell condition $\slashed{v}u(p)=u(p)$, the relations 
$\nm=2 v-\np$ and $\np\cdot\varepsilon_{c\perp} = 
\np\cdot\varepsilon_{\bar{c}\perp}=0$,
which imply $\{\slashed{n}_+,\slashed{\varepsilon}_{c\perp}\} = 
\{\slashed{n}_+,\slashed{\varepsilon}_{\bar{c}\perp}\} =0$
the string can be reduced to the two structures
\begin{equation}
\varepsilon_{c\perp}\cdot \varepsilon_{\bar{c}\perp}
\,\bar{v}(\mchi v)\slashed{n}_+ u(p),\qquad 
\bar{v}(\mchi v)[\slashed{\varepsilon}_{c\perp},
\slashed{\varepsilon}_{\bar{c}\perp}]
\,\slashed{n}_+ u(p)\,.
\end{equation}
After expressing the Dirac spinors in terms of non-relativistic two-spinors, 
the first structure corresponds to the spin matrix of $\mathcal{O}_3$, 
and the second to $\Gamma^{\mu\nu}$. 

The operator basis holds for any integer isospin-$j$ DM multiplet with vanishing 
hypercharge. The coefficient functions $C_{1,2}$ of $\mathcal{O}_{1,2}$ 
and their renormalization group evolution (RGE) to scales 
$\mu\ll \mchi$ can be found in \cite{Beneke:2018ssm} and we refer to this 
paper and Section~\ref{sec:hardfn} below for the detailed expressions.

\subsection{Factorization}

In this section we derive the factorization formula for the photon energy 
spectrum for intermediate photon resolution. We also comment on the modifications 
for the narrow-resolution case, in this way providing the derivation of 
this case omitted in \cite{Beneke:2018ssm}. To guide the reader let us 
preview here the main result of this section by giving the equation that 
will subsequently be used for the calculation of the resummed spectrum 
at the NLL' order.

Independent of the resolution we can represent the energy spectrum 
in the form 
\begin{equation}
\frac{d (\sigma v_{\text{rel}})}{d E_{\gamma}} 
= 2 \, \sum_{I,J} S_{IJ} \,\Gamma_{IJ}(E_\gamma) 
= 2\, \sum_{I,J} S_{IJ} \sum_{i,j=1,2} C_i(\mu) C_j^*(\mu) 
\,\gamma_{IJ}^{ij}(E_\gamma,\mu)\,,
\label{eq:SIJGIJ}
\end{equation}
where the sums over $I,J$ run over all electrically neutral two-particle 
states that can be formed from the $2j+1$ single-particle states of the 
electroweak DM multiplet, and the sums over $i,j$ refer to the two operators 
$\mathcal{O}_{1,2}$. The expression after the first equality expresses the 
factorization of the Sommerfeld enhancement factor from the remainder of the 
process. $S_{IJ}$ is the same 
Sommerfeld factor as usual, except that the tree-level short-distance 
annihilation matrices are replaced by matrices $\Gamma_{IJ}(E_\gamma)$, 
which include electroweak Sudakov resummation and other 
radiative corrections up to the specified accuracy. The expression after 
the second equality factors the hard matching coefficients, evolved to the scale 
$\mu$ with their RGE equation, which are 
also universal. The quantity $\gamma_{IJ}^{ij}(E_\gamma,\mu)$ is 
therefore related to the square of matrix elements of $\mathcal{O}_i$ 
in the state $\langle \gamma X| \ldots| [\chi\chi]_I\rangle$, summed and integrated 
over the phase-space of the final-state particles. For intermediate resolution, 
we shall derive
\begin{eqnarray}
\gamma^{ij}_{IJ}(E_\gamma,\mu) &=&
\frac{1}{(\sqrt{2})^{n_{id}}}\,
\frac{1}{4} \,\frac{2}{\pi\mchi} 
\,Z_{\gamma}^{33}(\mu,\nu)
\nonumber\\
&&\times\,
\int d\omega \, J_{\rm int}(4\mchi (\mchi-E_\gamma-\omega/2),\mu)
\, W^{ij}_{IJ}(\omega,\mu,\nu)\,.
\label{eq:factformulaint}
\end{eqnarray}
This equation as well as the corresponding 
equation~\eqref{eq:factformulanarrow} below refer to the method-2 
computation of the Sommerfeld effect~\cite{Beneke:2014gja}. 
The overall factor of $2$ in~\eqref{eq:SIJGIJ} is required to  
compensate the method-2 factor $1/(\sqrt{2})^{n_{id}}$
 for the annihilation of two identical particles 
in~\eqref{eq:factformulaint} and in~\eqref{eq:factformulanarrow}.
The various functions 
appearing in (\ref{eq:factformulaint})
will be defined below. Some of them require rapidity 
regularization in addition to the conventional dimensional regularization, 
resulting in a dependence of the renormalized function on the rapidity 
factorization scale $\nu$ in addition to the dimensional regularization 
scale $\mu$. We note the convolution of 
the jet function $J_{\text{int}}$ for the unobserved final state $X$ with a soft 
function $W$, which accounts for radiation of soft electroweak gauge bosons 
and other soft particles into the final state, and virtual corrections. 
We can compare this to the corresponding formula for narrow resolution,
\begin{eqnarray}
\gamma^{ij}_{IJ}(E_\gamma,\mu)&=&
\frac{1}{(\sqrt{2})^{n_{id}}}\,
\frac{1}{4} \,\frac{2}{\pi\mchi} 
\,Z_{\gamma}^{33}(\mu,\nu)\,
\nonumber\\
&&\times\,
D_{I,33}^{i}(\mu,\nu)D_{J,33}^{j \, *}(\mu,\nu) 
J_{\rm nrw}^{33}(4\mchi (\mchi-E_\gamma),\mu,\nu)\,.
\label{eq:factformulanarrow}
\end{eqnarray}
The main difference is that the smaller invariant mass of the final 
state $X$ forbids soft real radiation. The soft effects are purely 
virtual, and appear at the amplitude level in the factors $D$.

The starting point for the derivation is the general expression for 
the initial-state spin-averaged and final-state spin-summed 
annihilation cross section
\begin{eqnarray}
\frac{d (\sigma v_{\text{rel}})}{d E_{\gamma}} &=&
\frac{1}{4} \frac{1}{4\mchi^2}
\hspace*{0.2cm}\int\limits_X\hspace*{-0.6cm}\sum 
\!\int\!\frac{d^3\mathbf{p}_\gamma}{(2\pi)^3 2 p_\gamma^0} \,
(2\pi)^4\delta^{(4)}(p_{\chi\bar{\chi}}-p_\gamma-p_X) 
\delta(E_\gamma-|\mathbf{p}_\gamma|)\left|T_{\chi^0\chi^0\to \gamma X}\right|^2
\!.
\nonumber
\\[-0.6cm]&&
\label{eq:annrate1}
\end{eqnarray}
The sum-integral symbol implies a sum over all kinematically allowed final 
states $X$ with total momentum $p_X$ and the phase-space integral over the 
final-state momenta. Summation over spins is understood for the initial and 
final state and the overall factor $1/4$ accounts for the 
initial-state spin average. In the center-of-mass frame the initial-state 
momentum is $p_{\chi\bar{\chi}} = (2\mchi +E) v$. 
$T_{i\to f}$ is the T-matrix element for the transition, and 
$E$ denotes the small kinetic energy of the DM two-particle state.  
After integrating out the hard momentum modes, the T-matrix 
element is non-vanishing only if it involves the effective interaction
(\ref{eq:lann}), and we can write
\begin{equation}
T_{\chi^0\chi^0\to \gamma X} = \frac{1}{2\mchi}\sum_{i=1,2} 
\,\int ds dt \,\hat{C}_i(s,t,\mu)\,2\mchi
\langle \gamma(p_\gamma) X_c X_s|\,\mathcal{O}_i \,|[\chi\chi]_{00}\rangle\,.
\label{eq:tmatrix}
\end{equation}
We have split the sum over $X$ into a sum over (hard-) collinear particles 
$X_c$ and soft particles $X_s$. The matrix element is to be evaluated in 
non-relativistic and soft-collinear EFT. The factor $2\mchi$ arises from 
the non-relativistic normalization of external DM state.

After the field redefinitions (\ref{eq:Sw}), (\ref{eq:Yn}) 
that decouple soft gauge bosons from the 
(hard-) collinear, (hard-) anti-collinear and non-relativistic fields 
the operators are
\begin{equation}
\label{eq:opdecoupled}
\mathcal{O}_i = \chi_v^{c\dagger}\Gamma^{\mu\nu}_i 
[Y_v^\dagger T^{AB}_i Y_v] \chi_v\,
\mathcal{Y}_+^{AV}
\mathcal{Y}_-^{BW}
\mathcal{A}^{V}_{\perp c,\mu}(sn_+) \mathcal{A}^{W}_{\perp \bar{c},\nu}(tn_-)
\,.
\end{equation}
We now use the symbol $\mathcal{Y}_\pm(x)$ to denote SU(2) Wilson lines 
in the adjoint representation and recall that fields and Wilson lines 
without space-time argument are evaluated at $x=0$. Because the different 
types of fields no longer interact, we can factorize the matrix 
element into
\begin{eqnarray}
\langle \gamma(p_\gamma) X_c X_s|\,\mathcal{O}_i \,|[\chi\chi]_{00}\rangle 
&=& 
\langle \gamma(p_\gamma)|\,\mathcal{A}^{W}_{\perp \bar{c},\nu}(tn_-) \,|0\rangle 
\langle X_c|\,\mathcal{A}^{V}_{\perp c,\mu}(sn_+)\,|0\rangle 
\nonumber\\
&&\hspace*{-3cm}\times\,
\langle X_s|\,[Y_v^\dagger T^{AB}_i Y_v]_{ab}\,\mathcal{Y}_+^{AV}
\mathcal{Y}_-^{BW}\,|0\rangle 
\,K_{ab,I}\,\langle 0|\, [\chi_{v}^{c\dagger}\Gamma^{\mu\nu}_i 
\chi_{v}]_I\,|[\chi\chi]_{00}\rangle \,.
\end{eqnarray}
Translation invariance implies
\begin{eqnarray}
&& \langle \gamma(p_\gamma)|\,\mathcal{A}^{W}_{\perp \bar{c},\nu}(tn_-) \,|0\rangle 
= e^{i t \nm\cdot p_\gamma}\,
\langle \gamma(p_\gamma)|\,\mathcal{A}^{W}_{\perp \bar{c},\nu}(0) \,|0\rangle 
\,,
\nonumber\\
&& \langle X_c|\,\mathcal{A}^{V}_{\perp c,\mu}(sn_+)\,|0\rangle = 
e^{i s \np\cdot p_{X_c}}\,
\langle X_c|\,\mathcal{A}^{V}_{\perp c,\mu}(0)\,|0\rangle \,,
\label{eq:fact1}
\end{eqnarray}
where $p_{X_c}$ is the total four-momentum of the collinear final state, 
which allows us to perform the $s$, $t$ integrations in (\ref{eq:tmatrix}) 
and express them in terms of the momentum-space coefficient function
\begin{equation}
C_i(\np p_X,\nm p_\gamma,\mu) = 
\int ds dt \,e^{i s \np\cdot p_{X_c}+i t \nm\cdot p_\gamma}\,
\hat{C}_i(s,t,\mu)\,.
\end{equation}
Up to power-suppressed corrections $\nm p_\gamma =2 E_\gamma \approx 2 \mchi$, 
$\np\cdot p_X\approx 2\mchi$. We therefore define 
\begin{equation}
C_i(\mu) = C_i(2\mchi,2\mchi,\mu)\,,
\end{equation}
and these are given in \cite{Beneke:2018ssm} and below 
in the one-loop approximation required for NLL' accuracy.

We write the four-momentum conservation delta-function in 
(\ref{eq:annrate1}) as the space-time integral of the exponential,  
insert the factorized matrix element 
$\langle \gamma(p_\gamma) X_c X_s|\,\mathcal{O}_i \,|[\chi\chi]_{00}\rangle$ 
into (\ref{eq:tmatrix}), and the square of the resulting expression 
for the T-matrix element in (\ref{eq:annrate1}). In this way, we 
obtain 
\begin{eqnarray}
\frac{d (\sigma v_{\text{rel}})}{d E_{\gamma}} &=&
\sum_{i.j=1,2} C_i(\mu) C_j^*(\mu)\,\sum_{I,J}\,
\frac{1}{4} \frac{1}{4\mchi^2}
\!\int\!\frac{d^3\mathbf{p}_\gamma}{(2\pi)^3 2 p_\gamma^0} \,
\delta(E_\gamma-|\mathbf{p}_\gamma|)
\nonumber\\
&&
\times \int d^4 x\,e^{i (p_{\chi\chi}-p_\gamma)\cdot x}\,
\langle [\chi\chi]_{00}(p_{\chi\chi})|\, [\chi_{v}^{c\dagger}\Gamma^{\mu'\nu'}_j 
\chi_{v}]_J^\dagger\,|0\rangle\,
\langle 0|\, [\chi_{v}^{c\dagger}\Gamma^{\mu\nu}_i 
\chi_{v}]_I\,|[\chi\chi]_{00}(p_{\chi\chi})\rangle
\nonumber\\
&&
\times \,
\langle 0 |\,\mathcal{A}^{Y}_{\perp \bar{c},\nu'} \,|\gamma(p_\gamma)\rangle 
\langle \gamma(p_\gamma)|\,\mathcal{A}^{W}_{\perp \bar{c},\nu} \,|0\rangle 
\hspace*{0.2cm}\int\limits_{X_c}\hspace*{-0.6cm}\sum 
e^{-i p_{X_c}\cdot x}\,\langle 0|\,\mathcal{A}^{X}_{\perp c,\mu'}\,|X_c\rangle 
\langle X_c|\,\mathcal{A}^{V}_{\perp c,\mu}\,|0\rangle 
\nonumber\\[-0.1cm]
&&
\times \hspace*{0.2cm}\int\limits_{X_s}\hspace*{-0.6cm}\sum 
e^{-i p_{X_s}\cdot x}\,K_{ab,I} K^\dagger_{a'b',J}\,
\langle 0|\,\mathcal{Y}_+^{\dagger A'X}
\mathcal{Y}_-^{\dagger B'Y}\,
[Y_v^\dagger T^{A'B'}_j Y_v]_{a'b'}^\dagger\,
\,|X_s\rangle 
\nonumber\\
&&\times\hspace*{0.2cm}\, 
\langle X_s|\,[Y_v^\dagger T^{AB}_i Y_v]_{ab}\,\mathcal{Y}_+^{AV}
\mathcal{Y}_-^{BW}\,|0\rangle \,.
\end{eqnarray}
We use translation invariance again to absorb the exponentials 
$e^{-i p_{X_c}\cdot x}$, $e^{-i p_{X_s}\cdot x}$ into a shift of position 
0 to $x$ in the first halves of the collinear and soft matrix elements, 
after which the sums over complete sets of collinear and soft intermediate 
states can be done. We also use (\ref{eq:wave}), (\ref{eq:SFdef}) 
to express the non-relativistic matrix element in the 
form
\begin{eqnarray}
&&\langle [\chi\chi]_{00}(p_{\chi\chi})|\, [\chi_{v}^{c\dagger}\Gamma^{\mu'\nu'}_j 
\chi_{v}]_J^\dagger\,|0\rangle\,
\langle 0|\, [\chi_{v}^{c\dagger}\Gamma^{\mu\nu}_i 
\chi_{v}]_I\,|[\chi\chi]_{00}(p_{\chi\chi})\rangle
\nonumber\\
&&=\,4\,
\langle \xi^{c \dagger}_{0} \Gamma^{\mu'\nu'}_j  \xi_{0} \rangle^*
\,\langle \xi^{c \dagger}_{0}\Gamma^{\mu\nu}_i \xi_{0} \rangle \,S_{IJ}
\end{eqnarray}
with $\xi_0$ the spinor of an external $\chi^0$ field (with the two 
orientations $\uparrow$, $\downarrow$). The Sommerfeld factor is a function 
of the small kinetic energy $E$ of the DM two-particle state. For 
annihilation in the present Universe $E$ is much smaller than any 
other energy scale in the problem. After factoring the Sommerfeld effect, 
$E$ can be neglected in the other parts of the calculation, that is, 
we set $p_{\chi\chi} = 2\mchi$.

After these manipulations in (\ref{eq:fact1}), by comparing 
to (\ref{eq:SIJGIJ}) we can read off the 
quantity $\gamma_{IJ}^{ij}(E_\gamma)$:
\begin{eqnarray}
\gamma_{IJ}^{ij}(E_\gamma) &=&
\frac{1}{4} \frac{1}{4\mchi^2}
\!\int\!\frac{d^3\mathbf{p}_\gamma}{(2\pi)^3 2 p_\gamma^0} \,
\delta(E_\gamma-|\mathbf{p}_\gamma|)
\,4\,
\langle \xi^{c \dagger}_{0} \Gamma_j^{\mu'\nu'} \xi_{0} \rangle^*
\,\langle \xi^{c \dagger}_{0}\Gamma_i^{\mu\nu} \xi_{0} \rangle
\nonumber\\
&&
\times\,
\langle 0 |\,\mathcal{A}^{Y}_{\perp \bar{c},\nu'}(x) \,|\gamma(p_\gamma)\rangle 
\langle \gamma(p_\gamma)|\,\mathcal{A}^{W}_{\perp \bar{c},\nu}(0)\,|0\rangle \nonumber \\
&& \times \int d^4 x\,e^{i (p_{\chi\chi}-p_\gamma)\cdot x}\,
\langle 0|\,\mathcal{A}^{X}_{\perp c,\mu'}(x)\,\mathcal{A}^{V}_{\perp c,\mu}(0)\,|0\rangle 
\nonumber\\[-0.1cm]
&&
\times \hspace*{0.2cm}\int\limits_{X_s}\hspace*{-0.6cm}\sum \,
\langle 0|\,[\mathcal{S}^{\dagger}]^{j}_{J,XY}(x)\,|X_s\rangle 
\, \langle X_s|\,\mathcal{S}^i_{I,VW}(0)\,|0\rangle \,,
\label{eq:gamma1}
\end{eqnarray}
introducing the soft operator
\begin{equation}
\mathcal{S}^i_{I,VW}(x) = K_{ab,I}\,[Y_v^\dagger T^{AB}_i Y_v]_{ab}(x)
\,\mathcal{Y}_+^{AV}(x)
\mathcal{Y}_-^{BW}(x)\,.
\label{eq:softop}
\end{equation}
The last three factors in the above equation define, in order, 
an anti-collinear, hard-collinear and soft function, as follows.

\subsubsection{Definitions for the intermediate resolution case}

\paragraph{Photon collinear function}

The ``jet'' function for the exclusive anti-collinear photon state 
is defined by the squared matrix element 
\begin{equation}
-g^{\perp}_{\nu\nu'} \,Z_\gamma^{YW} = 
\sum_{\lambda} \,
\langle 0|\mathcal A^Y_{\perp\bar{c},\nu'}(0)|\gamma(p_\gamma,\lambda) 
\rangle\langle \gamma(p_\gamma,\lambda)
|\mathcal A^W_{\perp\bar{c}\nu}(0)|0\rangle\,.
\label{eq:photJetDef}
\end{equation}
We have made the sum over photon polarizations explicit. 
Obviously, only $Z_\gamma^{33}$ is non-vanishing. From 
(\ref{eq:defAc}) and (\ref{eq:defWc}) is follows that 
$Z^{33}_\gamma/\hat s_W^2$ can be interpreted as the on-shell photon 
field renormalization constant in anti-collinear light-cone gauge 
$\nm\cdot A_{\bar{c}}=0$. $Z_\gamma^{33}$ depends on the electroweak scale 
masses $m_W$, $m_Z$, $m_H$ and $m_t$ of the SM particles, the dimensional 
regularization scale $\mu$ and a rapidity regulator scale $\nu$, since 
the factorization formula involves the separation of regions 
(here anti-collinear and soft) with equal virtuality 
but parametrically different $n_\pm$ momentum components.

\paragraph{Unobserved-jet collinear function}

The jet function pertaining to the inclusive (unobserved) collinear 
final state is defined as 
\begin{eqnarray}
&& -g^{\perp}_{\mu\mu'} \,J^{XV}(p^2,m_W) =
\frac{1}{\pi}\,\text{Im}\big[
-g^{\perp}_{\mu\mu'}\,i  \mathcal{J}^{XV}(p^2,m_W) \big]
\nonumber\\
&&\hspace*{1cm} \equiv \, \frac{1}{\pi}\,\text{Im}\big[\, 
i \int d^4x \,e^{ip\cdot x} 
\langle 0|\mathbf{T}\big\{\mathcal{A}^X_{\perp  c,\mu'}(x) 
\,\mathcal{A}_{\perp c,\mu}^V(0)\big\} |0\rangle\,\big]
\nonumber\\
&& \hspace*{1cm} =\,\frac1{2\pi}\,
\int d^4 x\,e^{i p\cdot x}\,
\langle 0|\,\mathcal{A}^{X}_{\perp c,\mu'}(x)\,\mathcal{A}^{V}_{\perp c,\mu}(0)\,|0
\rangle\,. 
\label{eq:jetfndef}
\end{eqnarray}
The jet function is defined in SCET$_{\rm I}$ in terms of the hard-collinear 
gauge field. It depends on the hard-collinear scale through the 
invariant mass squared $p^2$ of the final state $X$, but also on 
the scale $m_W$ through the electroweak scale masses of the particles 
inside the jet. The jet function as defined above is therefore still 
a two-scale object, which can be further factorized into 
a hard-collinear and collinear function \cite{Baumgart:2017nsr}. 
Up to power corrections of order $m_W^2/p^2\sim m_W/\mchi$, 
\begin{equation}
J^{XV}(p^2,m_W) = J_{\rm int}(p^2) \,J_m^{XV}(m_W) 
+ \mathcal{O}(m_W^2/p^2)\,.
\label{eq:JJ}
\end{equation}
The hard-collinear matching coefficient $J_{\rm int}(p^2)$ 
can be computed in the theory with unbroken electroweak gauge symmetry 
in close analogy with the standard gluon jet function in QCD. 
It depends on the renormalization scale $\mu$, but does not 
require rapidity regularization. The collinear 
mass-jet function $J_m^{XV}(m_W)$ is momentum-independent, but 
can depend on both $\mu$ and the rapidity regulator. However, we 
find that at tree-level and at the one-loop order, the collinear 
mass function is trivial, that is 
\begin{equation}
J_m^{XV}(m_W) = \delta^{XV} + \mathcal{O}(\alpha_2^2)\,.
\label{eq:Jmapprox}
\end{equation}
It is plausible that this result holds to any order in the coupling, 
since the observable 
is not sensitive to the internal jet structure.\footnote{
Naively calculating the expression 
(\ref{eq:jetfndef}) with massive gauge boson propagators reveals a 
leading-power sensitivity to $m_W$. However, this arises from the soft 
region, which must be discarded, since it is already accounted for in 
the soft function defined below. Some technical details on this point 
are given in Appendix~\ref{app:rapreg}. The mass-sensitive collinear mode with 
transverse momentum of order $m_W$ does not 
appear in leading power, at least at the one-loop order.}
We shall make use of this simplification in 
deriving (\ref{eq:factformulaint}).

\paragraph{Soft function}

The sum over the soft final state in (\ref{eq:gamma1}) is the unit 
operator, which allows us to define the soft function in 
momentum space, 
\begin{equation}
\langle 0|\,\mathbf{\bar{T}}[[\mathcal{S}^{\dagger}]^{j}_{J,XY}(x)]\,
\mathbf{T}[\mathcal{S}^i_{I,VW}(0)]\,|0\rangle
\equiv \int \frac{d^4 k}{(2\pi)^4}\,e^{-i k\cdot x}\,
\bm{W}^{ij}_{IJ,VWXY}(k)\,.
\end{equation}
We also define the integrated soft function
\begin{eqnarray}
W^{ij}_{IJ,VWXY}(\omega) &=& 
\frac{1}{2}\int\frac{d(\np k)d^2k_\perp}{(2\pi)^4}\,
\bm{W}^{ij}_{IJ,VWXY}(k)
\nonumber\\ 
&=& \frac{1}{4\pi} 
\int d(\np y)\,e^{i\omega \np\cdot y/2}\,
\langle 0|\,\mathbf{\bar{T}}[[\mathcal{S}^{\dagger}]^{j}_{J,XY}(y_-)]\,
\mathbf{T}[\mathcal{S}^i_{I,VW}(0)]\,|0\rangle\,,\qquad
\label{eq:softfnintdef}
\end{eqnarray}
where $\omega = \nm\cdot k$, and then the 
SU(2) index-contracted soft function
\begin{equation}
W^{ij}_{IJ}(\omega) = W^{ij}_{IJ,V3V3}(\omega)\,.
\label{eq:softtwoindexfn}
\end{equation}
The soft functions must be calculated in the broken SU(2) theory and depend 
on the electroweak masses of the SM particles. They also depend on the 
renormalization scale $\mu$ and the rapidity regularization scale $\nu$.

\subsubsection{Derivation of the final formula}

With the above definitions of the collinear and soft function, we
can rewrite the corresponding terms in (\ref{eq:gamma1}),
\begin{eqnarray}
&&\int d^4 x\,e^{i (p_{\chi\chi}-p_\gamma)\cdot x}\,
\langle 0|\,\mathcal{A}^{X}_{\perp c,\mu'}(x)\,\mathcal{A}^{V}_{\perp c,\mu}\,|0\rangle 
\times \hspace*{0.2cm}\int\limits_{X_s}\hspace*{-0.6cm}\sum \,
\langle 0|\,[\mathcal{S}^{\dagger}]^{j}_{J,XY}(x)\,|X_s\rangle 
\, \langle X_s|\,\mathcal{S}^i_{I,VW}(0)\,|0\rangle 
\nonumber\\[-0.2cm]
&&\hspace*{1cm}
=\,-2\pi\,g^\perp_{\mu\mu'}
\int d^4x\int\frac{d^4 p}{(2\pi)^4}\int \frac{d^4 k}{(2\pi)^4}\,
e^{i (p_{\chi\chi}-p_\gamma-p-k)\cdot x}\,
J^{XV}(p^2,m_W)\,\bm{W}^{ij}_{IJ,VWXY}(k)
\nonumber\\
&&\hspace*{1cm}
=\,-2\pi\,g^\perp_{\mu\mu'}\int \frac{d^4 k}{(2\pi)^4}\,
J^{XV}(4\mchi (\mchi-E_\gamma-\nm k/2),m_W)\,
\bm{W}^{ij}_{IJ,VWXY}(k) 
\nonumber\\
&&\hspace*{1cm}
=\,-2\pi\,g^\perp_{\mu\mu'} \int d\omega 
J^{XV}(4\mchi (\mchi-E_\gamma-\omega/2),m_W)\,
W^{ij}_{IJ,VWXY}(\omega)\,,
\end{eqnarray}
where in passing from the second to the third line we used $p^2 \to (p_{\chi \chi} - p_\gamma-k)^2\approx 4 m_\chi (m_\chi - E_\gamma - \nm k/2)$.
There is no dependence on the direction of the photon momentum, hence we 
can perform the photon phase-space integral in (\ref{eq:gamma1}),
\begin{equation}
\int\!\frac{d^3\mathbf{p}_\gamma}{(2\pi)^3 2 p_\gamma^0} \,
\delta(E_\gamma-|\mathbf{p}_\gamma|) = \frac{E_\gamma}{4\pi^2}\,,
\end{equation}
to obtain 
\begin{eqnarray}
\gamma_{IJ}^{ij}(E_\gamma) &=&
\frac{1}{4} \frac{1}{2\pi\mchi}\,
\langle \xi^{c \dagger}_{0} \Gamma_j^{\mu\nu} \xi_{0} \rangle^*
\,\langle \xi^{c \dagger}_{0}\Gamma_{i,\mu\nu} \xi_{0} \rangle
\nonumber\\
&&
\times\,Z_\gamma^{33}\,
\int d\omega 
J^{XV}(4\mchi (\mchi-E_\gamma-\omega/2),m_W)\,
W^{ij}_{IJ,V3X3}(\omega)\,.
\label{eq:gamma2}
\end{eqnarray}
This equation represents the factorization formula for the intermediate 
resolution case in its general form. To obtain (\ref{eq:factformulaint}), 
we note that both operators involve the same spin matrix (\ref{eq:spinmatrix}), 
that is $\Gamma_1^{\mu\nu} = \Gamma_2^{\mu\nu} = \epsilon_\perp^{\mu\nu}$, 
which implies
\begin{equation}
\langle \xi^{c \dagger}_{0} \Gamma_j^{\mu\nu} \xi_{0} \rangle^*
\,\langle \xi^{c \dagger}_{0}\Gamma_{i,\mu\nu} \xi_{0} \rangle
= \epsilon_\perp^{\mu\nu}\epsilon_{\perp,\mu\nu} 
\langle \xi^{c \dagger}_{0} \xi_{0} \rangle^*
\,\langle \xi^{c \dagger}_{0} \xi_{0} \rangle =4\,.
\end{equation}
We then use the property (\ref{eq:Jmapprox}), which allows us to replace 
\begin{equation}
J^{XV} W^{ij}_{IJ,V3X3} \to J_{\rm int} W^{ij}_{IJ}\,.
\end{equation}
Finally, we switch from method-1 to method-2 (see the discussion before 
(\ref{eq:SFdef})) and sum only over distinguishable 
two-particle states $I,J$. As discussed in \cite{Beneke:2014gja}, this 
implies certain replacement rules for the potential used in the computation 
of the Sommerfeld effect and the annihilation matrix $\Gamma_{IJ}$, 
which introduces the factor $1/(\sqrt{2})^{n_{id}}$ in (\ref{eq:factformulaint}), 
where $n_{id}=0,1,2$ depending on how often the two-particle state 00 
appears in the index pair $IJ$. The objects in the factorization 
formula are assumed to be evolved from their natural scales, where they 
exhibit no large logarithms, to a common scale in $\mu$ and in $\nu$ 
with the renormalization group equations discussed in the following section. 
The evolution factors accomplish the desired resummation of large logarithms.

Let us comment on the treatment of ultrasoft modes that we did not 
mention in the derivation. After the decoupling of soft modes from 
the (anti-) collinear and non-relativistic fields, all of them still 
interact with ultrasoft modes. In writing the various sectors 
in a factorized form, we implicitly made use of the fact that 
an ultrasoft Wilson line field redefinition decouples 
ultrasoft interactions from (anti-) collinear modes 
at leading power. This introduces multiple 
convolutions from the different sectors with an ultrasoft function. 
We omitted this ultrasoft function in the above discussion, since
it is actually absent due to the electric charge neutrality of the 
initial state and the anti-collinear photon final state. The soft and 
the hard-collinear final state, however, are not necessarily electrically 
neutral. However, all momentum components of an ultrasoft mode are 
small compared to the corresponding momentum component of a soft or 
hard-collinear mode, such that  in leading power, the ultrasoft momentum 
transfer to the soft or hard-collinear function can be neglected. 
This eliminates the possibility of a non-trivial convolution and 
allows us to ignore the ultrasoft mode.

\subsubsection{Modifications for the narrow resolution case}

Following the above line of argument, we derive 
the factorization formula for the narrow resolution 
case stated in \cite{Beneke:2018ssm} and written in 
(\ref{eq:factformulanarrow}) in present notation. There is no 
change to the discussion of the non-relativistic and photon 
jet function, but for the unobserved jet function and soft 
function, the following differences need to be noted.

The narrow resolution jet function has the same definition 
as (\ref{eq:jetfndef}), except that now the gauge field is 
collinear rather than hard-collinear. Consequently, there is no 
further factorization. Since there is no soft radiation into 
the final state, the collinear function must be charge-neutral 
which selects the $X=V=3$ component of the jet function. 
The narrow resolution jet function depends 
on the collinear invariant mass squared $p^2\sim m_W^2$ 
and the electroweak scale particle masses of the SM. It further depends 
on the renormalization scale $\mu$ and, contrary to the intermediate 
resolution hard-collinear jet function, also on  the rapidity scale $\nu$. 
The different structure of rapidity logarithms for the 
two cases is matched by different rapidity logarithms 
in the soft function. Details on the calculation of the narrow resolution jet 
function are provided in Appendix~\ref{app:recjetfnnarrow}.

The small energy resolution $\Eres\sim m_W^2/\mchi$ forbids 
soft radiation into the final state, hence the soft factors are 
defined at the amplitude level, rather than for the square of the 
amplitude as above. Technically, the sum over the soft final state 
in (\ref{eq:gamma1}) is empty, such that 
\begin{eqnarray}
&&\int\limits_{X_s}\hspace*{-0.6cm}\sum \,
\langle 0|\,[\mathcal{S}^{\dagger}]^{j}_{J,XY}(x)\,|X_s\rangle 
\, \langle X_s|\,\mathcal{S}^i_{I,VW}(0)\,|0\rangle
\nonumber\\[-0.3cm]
&&\,\rightarrow\quad 
\langle 0|\,[\mathcal{S}^{\dagger}]^{j}_{J,XY}(x)\,|0\rangle 
\, \langle 0|\,\mathcal{S}^i_{I,VW}(0)\,|0\rangle 
\equiv D_{I,VW}^{i}\,D_{J,XY}^{j \, *}\,
\end{eqnarray}
where $D_{I,VW}^{i}$ is defined as the vacuum  matrix element of the 
soft operator (\ref{eq:softop}). The photon jet function selects 
$Y=W=3$, and the unobserved jet function selects $X=V=3$, which 
implies that only the single SU(2) component $D^i_{I,33}$ of the soft amplitude 
is needed. As in the intermediate resolution case the soft 
function must be computed in the broken theory. The inclusive nature of 
the soft function for intermediate resolution entails a partial 
cancellation of infrared singularities between virtual and real 
contributions, which does not happen for the narrow resolution case, 
where real contributions are absent. This also changes the 
structure of the rapidity evolution factor, since the 
narrow resolution soft function couples to the rapidity evolution 
of the collinear and anti-collinear sector, while only the latter 
has rapidity divergences for intermediate resolution.
This explains the differences between the two factorization formulas 
(\ref{eq:factformulanarrow}) and (\ref{eq:factformulaint}).\footnote{
In \cite{Beneke:2018ssm} the rapidity evolution factor $V (\mu_W;\nu_s;\nu_j)$ 
has been made explicit, while in the present notation 
(\ref{eq:factformulanarrow}), the rapidity 
evolution of every factor is implied to be contained in that factor.}

\section{NLL' resummation}
\label{sec:nllprime}

In this section we collect the one-loop results as well as the NLL' 
resummation formulas for the hard, soft and jet functions. These 
functions are the ingredients of the factorization theorems for the 
semi-inclusive photon spectrum in DM annihilation derived in the 
previous section. Furthermore we will show the consistency of the 
renormalization group and discuss different resummation 
schemes. 

The hard functions have been computed for an electroweak DM with 
any integer isospin $j$. The (anti-) collinear functions for the photon and 
for the unobserved jet triggered by an electroweak gauge boson are 
universal. The soft function given below is specific to the triplet 
(wino, $j=1$) DM model, which is the focus of this work.

\subsection{Hard function}
\label{sec:hardfn}
The hard matching coefficients for the annihilation of dark-matter 
particles in an integer isospin-$j$ multiplet were previously 
computed in \cite{Beneke:2018ssm}. For the operators 
$\mathcal{O}_{1,2}$ defined in (\ref{eq:opbasis1}), 
(\ref{eq:opbasis2}), they read
\begin{align}
C_1(\mu) =&\, \frac{\hat{g}^4_2(\mu)}{16 \pi^2}\, c_2(j)\, 
\Big[ (2-2 i \pi) \ln\frac{\mu^2}{4 m^2_\chi}
- \Big(4 - \frac{\pi^2}{2}\Big)\Big]\,,
\label{eq:hmatch1}\\
C_2(\mu) =&\, \hat{g}^2_2(\mu) + \frac{\hat{g}^4_2(\mu)}{16 \pi^2}\Big[16 - \frac{\pi^2}{6} - c_2(j) \Big(10 - \frac{\pi^2}{2}\Big) - 6 \ln\frac{\mu^2}{4 m^2_\chi}
\nonumber \\
&+ 2 i \pi \ln\frac{\mu^2}{4 m^2_\chi}-2 \ln^2\frac{\mu^2}{4 m^2_\chi}\Big]  \, ,
\label{eq:hmatch2}
\end{align}
where $c_2(j)=j \left(j+1\right)$ is the SU(2) Casimir of the isospin 
representation $j$, and $\hat{g}_2(\mu)$ denotes the SU(2) gauge coupling 
in the $\overline{\rm MS}$ scheme at the scale $\mu$.\footnote{
When the argument $\mu$ is omitted in the following, it is implied. 
Similarly for $\hat\alpha_2 = \hat{g}_2^2/(4\pi)$.} They satisfy the RGE equation
\begin{align}
\frac{d}{d \ln \mu}C_i(\mu) = 
(\Gamma^T)_{ij}(\mu) \,C_j(\mu)\,.
\label{eq:rge}
\end{align}
The one-loop anomalous dimension matrix takes the form
\begin{align}
\mathbf{\Gamma} &= \frac{\hat{\alpha}_2}{4\pi}\begin{pmatrix}
\displaystyle
8\ln\frac{4m_\chi^2}{\mu^2} -8i\pi -\frac{43}{3} +\frac{8}{3}n_G  & 0 \\
(4-4 i\pi )c_2(j)  & 
\displaystyle
8\ln\frac{4m_\chi^2}{\mu^2} +4i\pi -\frac{79}{3} +\frac{8}{3}n_G
\end{pmatrix} \,.
\label{eq:anDimWilson}
\end{align}
$n_G=3$ is the number of SM fermion generations. For NLL resummation 
we also include the two-loop cusp anomalous dimension. 
A detailed discussion of the evolution of the Wilson 
coefficients after diagonalization of the anomalous dimension 
can be found in \cite{Beneke:2018ssm}, and will not be repeated here. 
Details on the calculation of the hard functions and anomalous 
dimensions are provided in Appendix~\ref{app:hardFunctions}.

It is convenient to define the vector
\begin{align}
\vec{H} = \big( C_1^* C_1, C_2^* C_1, C_1^* C_2, C_2^* C_2 \big)^{T},
\end{align}
of hard functions which will be used below to demonstrate the scale 
invariance of the annihilation rate. The RGE for $\vec{H}$ reads 
\begin{equation}
\frac{d}{d \ln \mu} \vec{H}(\mu) = 
\mathbf{\Gamma}_{H}^T(\mu) \,\vec{H}(\mu)\,,
\label{eq:rgeH}
\end{equation}
with \begin{align}
\label{eq:anomDimH}
\mathbf{\Gamma}_{H} &= \begin{pmatrix}
2\, \text{Re} \,\Gamma_{11} & 0 & 0 & 0 \\
\Gamma_{21}^* & \Gamma_{11} + \Gamma_{22}^* & 0 & 0 \\
\Gamma_{21} & 0 & \Gamma^*_{11} + \Gamma_{22} & 0 \\
0 & \Gamma_{21} & \Gamma^*_{21} & 2\, \text{Re} 
\,\Gamma_{22}
\end{pmatrix} \,,
\end{align}
as follows from (\ref{eq:rge}).

\subsection{Photon jet function}
\label{sec:photonjet}

The anti-collinear photon jet function is the same as for the 
narrow resolution case and its definition is given in (\ref{eq:photJetDef}). Since 
the photon jet function and the soft function have the same invariant mass squared
of order $m_W^2$, they are defined in SCET$_{\text{II}}$ and require an 
additional rapidity regulator. We chose to use the rapidity regulator 
introduced in \cite{Chiu:2011qc,Chiu:2012ir}. Details on the implementation 
of this regulator can be found in Appendix \ref{app:rapreg}. 
For completeness we report the result for $Z_\gamma \equiv 
Z_\gamma^{33}$, already given in \cite{Beneke:2018ssm}:
\begin{eqnarray}
Z_\gamma(\mu,\nu) &=& \hat s_W^2(\mu) \Bigg\{1
-\frac{\hat{\alpha}_2(\mu)}{4\pi}\,\bigg[
- 16\ln\frac{m_W}{\mu}\ln\frac{2m_\chi}{\nu}
+8\ln \frac{m_W}{\mu}
\nonumber\\
&&   
- \,\hat s^2_W(\mu)\frac{80}{9}\bigg(\ln \frac{m_Z^2}{\mu^2}-\frac{5}{3}\bigg)
- \hat s^2_W(\mu)  \frac{16}{9} \ln \frac{m_t^2}{\mu^2}\nonumber\\
&& +\,\hat s^2_W(\mu)\bigg(3\ln \frac{m_W^2}{\mu^2} -\frac23\bigg) 
- 4\frac{m_W^2}{m_Z^2}\ln\frac{m_W^2}{\mu^2}\bigg]
-\Delta\alpha \Bigg\} \,,
\label{eq:photnJet}
\end{eqnarray}
where $\nu$ is the scale associated with the rapidity regulator and 
$\hat s_W(\mu)$ is the sine of the weak mixing angle in the 
$\overline{\text{MS}}$ scheme. $\Delta \alpha$ is the difference between 
the fine structure constant $\alpha=1/137.036$ and $\alpha_{\rm OS}
(m_Z)=\alpha/(1-\Delta\alpha)$.

Since $Z_\gamma$ depends both on $\mu$ and $\nu$, we need to resum the 
photon jet function in virtuality and rapidity. We will first discuss the resummation in $\mu$ and then in $\nu$. The RG equation is
\begin{align}
\frac{d}{d \ln \mu}\, Z_\gamma(\mu,\nu) 
&= \gamma^{\mu}_{Z_\gamma} Z_\gamma(\mu,\nu) \label{eq:photnRRE} \,,
\end{align}
with anomalous dimension
\begin{align}
\gamma^{\mu}_{Z_\gamma} &= 
4 \gamma_{\text{cusp}} \ln \frac{\nu}{2 m_\chi} 
+ 2 \gamma_{Z_\gamma} \label{eq:photnAnDimVirt} \, .
\end{align}
The anomalous dimensions can be expanded perturbatively in the 
form\footnote{In general, starting from the two-loop order, 
$\gamma_i^{(1)}$, second-order terms involving several SM couplings 
can appear. However, this is not the case for the cusp anomalous 
dimension, which is the only two-loop anomalous dimension needed 
at NLL'.}
\begin{align}
\gamma_i &= \frac{\hat{\alpha}_2}{4 \pi} \gamma_i^{(0)} + 
\left(\frac{\hat{\alpha}_2}{4 \pi}\right)^2 \gamma_i^{(1)} 
+ \mathcal{O}\left(\hat{\alpha}_2^3\right) \,,
\label{eq:expAnDim}
\end{align}
The cusp anomalous dimension coefficients up to the two-loop order 
are given by 
\begin{align}
\gamma_{\text{cusp}}^{(0)} &= 4 \,, \quad
\gamma_{\text{cusp}}^{(1)} = \left(\frac{268}{9}-\frac{4 \pi^2}{3}\right) c_2(\text{ad}) - \frac{80}{9} n_G - \frac{16}{9}
\label{eq:gamcusp}
\end{align}
with $c_2(\text{ad})=2$ and $n_G=3$. The one-loop coefficient
$\gamma^{(0)}_{Z_\gamma}$ can be obtained 
from its definition. Calculating the derivative in $\mu$ of~\eqref{eq:photnJet} using the 
beta-function of $\hat{s}_W^2$, which can be inferred from (\ref{eq:sWrunning}), yields 
\begin{equation}
\gamma^{(0)}_{Z_{\gamma}} 
= \beta_{0,\text{SU}(2)} = \left(\frac{43}{6} - \frac{4}{3}n_G\right) \, .
\label{eq:gam0Zgam}
\end{equation}
In the computation of~\eqref{eq:photnAnDimVirt}, we used the fact that the cusp anomalous dimension appears in the same way at all orders \cite{Korchemsky:1987wg}, 
so only a one-loop calculation is necessary to determine the  
prefactor of the cusp piece. Eq.~\eqref{eq:photnRRE} can 
easily be solved, which results in the following expression for the 
virtuality evolution factor
\begin{align}
Z_\gamma (\mu_f,\nu) &= U(\mu_i,\mu_f,\nu) Z_\gamma (\mu_i,\nu) 
\nonumber \\
&= \exp\left[\int_{\ln \mu_i}^{\ln \mu_f} d \ln \mu 
\left(4 \gamma_{\text{cusp}} \ln \frac{\nu}{2 m_\chi} 
+ 2\, \gamma_{Z_\gamma}\right)\right]Z_\gamma (\mu_i,\nu) \, ,
\label{eq:virtevolfactor}
\end{align}
where $\mu_i$ and $\mu_f$ denote the initial and final 
virtuality scales before and after evolution, respectively. Note that~\eqref{eq:virtevolfactor} is a general solution to the RGE~\eqref{eq:photnRRE}, valid to all orders. The integral in the exponent in~\eqref{eq:virtevolfactor} has to be computed numerically due to the appearance of other Standard Model couplings in the $\beta$-function beyond one-loop. This is also true for the virtuality evolution factors of the other functions in the factorization theorem. 

More care has to be taken when performing the resummation in rapidity. 
The rapidity renormalization group (RRG) equation is given by
\begin{align}
\frac{d}{d \ln \nu}\, Z_\gamma(\mu,\nu) 
&= \gamma_{Z_\gamma}^\nu Z_\gamma(\mu,\nu) 
\label{eq:photnRRG}
\end{align}
with the fixed-order one-loop anomalous dimension
\begin{align}
\gamma_{Z_\gamma}^\nu &= \frac{\hat{\alpha}_2}{4\pi}\, 
4 \gamma_{\text{cusp}}^{(0)} \,\ln\frac{\mu}{m_W}\, .
\label{eq:photnAnDimRap}
\end{align}
One could now use~\eqref{eq:photnAnDimRap} to solve the RRG. This 
procedure imposes that one first evolves in rapidity and only afterwards 
in virtuality, because in higher orders  $\gamma_{Z_\gamma}^\nu$ 
contains terms of the form $\alpha_2^n \ln^m(\mu/m_W)$ with $m \leq n$. 
If the virtuality evolution is done first, these logarithms become large 
and require themselves resummation. To avoid this issue, we note that the independence of any observable of the scales $\mu$ and $\nu$ gives the condition
\begin{align}
\left[\frac{d}{d \ln \mu},\frac{d}{d \ln \nu}\right] = 0 \,.
\label{eq:muNuComm}
\end{align}
From~\eqref{eq:photnRRE},~\eqref{eq:photnRRG} and~\eqref{eq:muNuComm} we deduce the constraint
\begin{align}
\frac{d}{d \ln \mu} \gamma_{Z_\gamma}^\nu = \frac{d}{d \ln \nu} \gamma_{Z_\gamma}^\mu = 4 \gamma_{\text{cusp}} \, .
\label{eq:derivAnDim}
\end{align}
(A similar constraint also applies for the soft function, discussed in 
Section~\ref{sec:intressoftfn} below.) We can now solve~\eqref{eq:derivAnDim} to obtain the integrated form of the rapidity anomalous dimension
\begin{align}
\gamma_{Z_\gamma}^\nu(\mu) = \int^{\ln \mu} d \ln \mu^{\prime} 
\frac{d}{d \ln \nu} \gamma_{Z_\gamma}^\mu (\mu^{\prime}) + \text{const.} \, ,
\label{eq:intPhotnAnDimRap}
\end{align}
where the constant is determined such that one obtains the fixed-order 
non-cusp piece of the rapidity anomalous dimension, which is zero at 
the one-loop 
order~\eqref{eq:photnAnDimRap}. The logarithms  $\ln(\mu/m_W)$  are 
summed by~\eqref{eq:intPhotnAnDimRap} to all orders in perturbation 
theory. Using the integrated form~\eqref{eq:intPhotnAnDimRap} of the 
rapidity anomalous dimension, we solve the RRG~\eqref{eq:photnRRG} to obtain the resummed rapidity evolution factor
\begin{align}
Z_\gamma(\mu,\nu_f)&= V(\mu,\nu_i,\nu_f) Z_\gamma (\mu,\nu_i) = \exp\left[\gamma_{Z_\gamma}^\nu (\mu)\,
\ln \frac{\nu_f}{\nu_i} \right] Z_\gamma(\mu,\nu_i) \, ,
\label{eq:rapevolfactor}
\end{align}
where $\nu_i$ and $\nu_f$ denote the initial and final scales of the 
rapidity evolution, respectively. Expanding the argument in the exponent of $V(\mu,\nu_i,\nu_f)$ in $\hat{\alpha}_2$ to order $\mathcal{O} (\hat{\alpha}_2)$, one would recover the rapidity evolution factor that can be computed from the fixed-order expression for $\gamma_{Z_\gamma}^\nu$ in~\eqref{eq:photnAnDimRap}. Note that in order to confirm the $\mu$-independence of the cross section, which will be discussed below in Section~\ref{sec:RGI}, it suffices to use this fixed-order expression. For more details on the rapidity evolution factor we refer to \cite{Chiu:2012ir}.

Depending on which resummation path is chosen, the anomalous dimensions in both evolution factors~\eqref{eq:virtevolfactor} and~\eqref{eq:rapevolfactor} are required at different order. If we first evolve in rapidity and only afterwards in virtuality, the $\mu$-dependent logarithm in $V(\mu,\nu_i,\nu_f)$ is never large and we only need $\gamma_{\rm cusp}$ at the one-loop order to achieve NLL' accuracy. At the same time, the $\nu$-dependent logarithm in $U(\mu_i,\mu_f,\nu)$ will be large and thus the virtuality evolution factor requires $\gamma_{\rm cusp}$ at two loops. If we first resum in virtuality and then in rapidity, the situation in reversed and we need $\gamma_{\rm cusp}$ at the two-loop order for $V(\mu,\nu_i,\nu_f)$ and at one-loop 
for $U(\mu_i,\mu_f,\nu)$.

Using the resummed expression for $V(\mu,\nu_i,\nu_f)$ and keeping in mind which order of the anomalous dimensions needs to be included ensures path independence for the $\mu - \nu$ resummation, which implies the relation
\begin{align}
V(\mu_f,\nu_i,\nu_f) U(\mu_i,\mu_f,\nu_i) = 
U(\mu_i,\mu_f,\nu_f) V(\mu_i,\nu_i,\nu_f) \, .
\label{eq:resumPath}
\end{align}
For the resummation of the photon jet function, we chose to resum first in rapidity and then in virtuality. As discussed above, for NLL' accuracy, this requires $\gamma_{\rm cusp}$ in the one-loop approximation for $V(\mu,\nu_i,\nu_f)$ and we use (\ref{eq:rapevolfactor}) in the form
\begin{align}
Z_\gamma(\mu,\nu_f)&= \exp\left[\frac{\gamma_{\rm cusp}^{(0)}}{\beta_{0,\text{SU}(2)}} 
\ln \left(\frac{\hat{\alpha}_2(\mu)}{\hat{\alpha}_2(m_W)}\right) 
\ln \frac{\nu_i^2}{\nu_f^2} \right] Z_\gamma(\mu,\nu_i) \, .
\label{eq:rapevolfactorLL}
\end{align}
The virtuality evolution factor $U(\mu_i,\mu_f,\nu)$ is computed 
with the two-loop cusp and one-loop non-cusp anomalous dimension 
from (\ref{eq:virtevolfactor}). The resummed photon jet function reads
\begin{align}
Z_\gamma(\mu_f,\nu_f) &= U(\mu_i,\mu_f,\nu_f) \, V(\mu_i,\nu_i,\nu_f) \, Z_\gamma (\mu_i,\nu_i)\,.
\label{eq:resumPhotJet}
\end{align}
Hence, the rapidity scale 
appearing in the virtuality RGE~\eqref{eq:virtevolfactor} is to be 
understood as the endpoint $\nu_f\sim m_W$ of the rapidity evolution.


\subsection{Jet function for intermediate 
resolution}
\label{sec:intresolutionjetfn}

The jet function (\ref{eq:jetfndef}) in the intermediate energy 
resolution case describes the unobserved hard-collinear final state 
with virtuality 
$m_\chi m_W \gg m_W^2$. It is therefore justified to neglect 
the masses of the electroweak gauge bosons, the fermions, and the 
Higgs boson, and to calculate the jet function in the unbroken 
regime of the SU(2)$_L\times$U(1)$_Y$ gauge symmetry. This implies that 
no additional rapidity regulator is needed (contrary to the case of 
the narrow resolution jet function, which is further discussed 
in Appendices~\ref{app:rapreg} and~\ref{app:recjetfnnarrow}). 
We separately give the results of the Wilson line contribution and of the self-energy contribution, in order to better identify the origin of the different terms, and hence write
\begin{equation}
i \mathcal{J}^{XV}(p^2, \mu) = i \mathcal{J}_{\text{Wilson}}^{XV}(p^2, \mu)
+ i \mathcal{J}_{\text{se}}^{XV}(p^2, \mu)\,.
\end{equation}
The one-loop results for the unrenormalized jet function terms read
\begin{eqnarray}
i \mathcal{J}_{\text{Wilson}}^{XV}(p^2, \mu) &=& 
\frac{\delta^{XV}}{-p^2 - i \epsilon}
\left\{1 +\left(\frac{\mu^2}{-p^2-i \epsilon}
\right)^{\!\epsilon} \,\frac{\hat{g}_2^2(\mu)}{16\pi^2} c_2({\rm ad}) 
\left(\frac{4}{\epsilon^2}
+\frac{2}{\epsilon}+4 - \frac{\pi^2}{3}\right)\right\},\qquad
\label{eq:recoilJetWilson}\\[0.2cm]
i \mathcal{J}_{\text{se}}^{XV}(p^2, \mu)
&=& \frac{\delta^{XV}}{-p^2 - i \epsilon} 
\left(\frac{\mu^2}{-p^2 -i \epsilon}\right)^{\!\epsilon} 
\,\frac{\hat{g}_2^2(\mu)}{16\pi^2} 
\nonumber \\[0.2cm]
&&\hspace*{-2cm} 
\times \left\{\frac{1}{\epsilon}\left(\frac{5}{3} c_2({\rm ad}) 
- \frac{8}{3}T_F n_G-\frac{1}{3}T_s n_s\right) + \frac{31}{9} c_2({\rm ad}) 
- \frac{40}{9}T_F n_G-\frac{8}{9}T_s n_s \right\} \, , 
\label{eq:recoilJetSE}
\end{eqnarray}
where $T_F = T_s = 1/2$ and $n_s = 1$. 
The jet function follows after taking the 
imaginary part and expanding in terms of star 
distributions~\cite{DeFazio:1999ptt}. We obtain, using 
(\ref{eq:Jmapprox}) and the numerical values of the group factors,
\begin{align}
J_{\rm int}(p^2,\mu) &= \delta(p^2) 
+  \frac{\hat{\alpha}_2(\mu)}{4 \pi} \,\Bigg\{
\,\delta (p^2)\,\left(\frac{70}{9} - 2 \pi^2\right) - \frac{19}{6} \left[\frac{1}{p^2}\right]^{[\mu^2]}_*  +  8 \left[\frac{\ln \frac{p^2}{\mu^2}}{p^2}\right]^{[\mu^2]}_*  \Bigg\}\, .
\end{align}
The definition of the star distributions is provided in 
(\ref{eq:stardistdef}) of Appendix \ref{app:recjetfnnarrow}.

The further treatment is very similar to the gluon jet function 
in QCD \cite{Becher:2010pd}. It is convenient to work with the 
Laplace-transformed jet function 
$j_{\rm int}$ since it renormalizes multiplicatively.
The Laplace transform of $J_{\rm int}(p^2, \mu)$ is defined by
\begin{align}
\label{eq:LapTrans}
j_{\rm int}\left(\ln \frac{\tau^2}{\mu^2}, \mu\right) =& 
\int_0^\infty dp^2 e^{-l p^2} J_{\rm int}(p^2, \mu), \, 
\end{align}
where $l = 1/(e^{\gamma_E} \tau^2)$ and the explicit result after renormalization reads
\begin{align}
j_{\rm int}\left(\ln \frac{\tau^2}{\mu^2}, \mu\right) = 1 + 
\frac{\hat{\alpha}_2(\mu)}{4\pi} \bigg(4 \ln^2 \frac{\tau^2}{\mu^2}- \frac{19}{6} \ln \frac{\tau^2}{\mu^2}  + \frac{70}{9}- \frac{4 \pi^2}{3}  \bigg).
\end{align}
The corresponding RG equation is the ordinary differential equation 
\begin{align}
\label{eq:RGE}
\frac{d}{d \ln \mu} j_{\rm int}\left(\ln \frac{\tau^2}{\mu^2}, 
\mu\right) = \gamma_{j}^\mu \, j_{\rm int}\left(\ln \frac{\tau^2}{\mu^2}, 
\mu\right) \,.
\end{align}
with Laplace-space anomalous dimension 
\begin{equation}
\gamma^\mu_{j} =-4 \gamma_{\text{cusp}} \ln \frac{\tau^2}{\mu^2} 
-2 \gamma_J \,.
\label{eq:anomDimJ}
\end{equation}
$\gamma_J$ is needed at the one-loop order for NLL' resummation,
\begin{equation}
\gamma_J = \frac{\hat{\alpha}_2}{4\pi} \gamma_J^{(0)} +\ldots 
\qquad\mbox{with}\qquad 
\gamma_J^{(0)} = - \beta_{0,\text{SU}(2)}\,.
\end{equation}
The RGE (\ref{eq:RGE}) is solved by
\begin{align}
j_{\rm int}\left(\ln \frac{\tau^2}{\mu^2}, \mu\right) =& 
\exp \bigg[ -\!\!\int\limits_{\ln \mu_j}^{\ln \mu} \hspace{-.1cm}d\ln\mu^{\prime} 
\left(\!4\gamma_{\text{cusp}}(\hat{\alpha}_2(\mu^{\prime})) 
\ln\frac{\tau^2}{\mu^{\prime \,2}} + 
2\gamma_J(\hat{\alpha}_2(\mu^{\prime})) \right) \!\bigg]
j_{\rm int}\left(\ln \frac{\tau^2}{\mu_j^2}, \mu_j \right) 
\nonumber \\
=& \exp \left[ 4\, S(\mu_j, \mu) + 2\, A_{\gamma_J}(\mu_j, \mu) \right] 
j_{\rm int}\left(\partial_{\eta}, \mu_j \right) 
\left( \frac{\tau^2}{\mu_j^2} \right)^{\eta} \,,
\label{eq:SolEvol}
\end{align}
where $\mu_j \sim \sqrt{m_\chi m_W}$ is the natural scale of the 
hard-collinear jet function and the integrals $S(\mu_j, \mu)$ 
and $A_{\gamma_J}(\mu_j, \mu)$ are defined as
\begin{align}
S(\mu_j, \mu) =& -\int_{\ln \mu_j}^{\ln \mu} d\ln\mu^{\prime} \, 
\gamma_{\text{cusp}}(\hat{\alpha}_2(\mu^{\prime}))\ln\frac{\mu_j^2}{\mu^{\prime \,2}} 
\,, \\
A_{\gamma}(\mu_j, \mu)=& -\int_{\ln \mu_j}^{\ln \mu} d\ln\mu^{\prime} 
\, \gamma(\hat{\alpha}_2(\mu^{\prime})) \, .
\label{eq:intAgamma}
\end{align}
The variable $\eta$ is defined by 
\begin{align}
    \eta = 4 A_{\gamma_{\text{cusp}}}(\mu_j,\mu)\, .
    \label{eq:etadef}
\end{align}

As mentioned before, at NLL' the integrals 
$S(\mu_j, \mu)$ and $A_{\gamma_J}(\mu_j, \mu)$ 
can only be solved numerically due to the appearence of several 
SM couplings in the $\beta$-function for $\hat{\alpha}_2$ 
beyond one loop. Note that in the second line of~(\ref{eq:SolEvol}), 
the logarithm in the argument of the Laplace-transformed
jet function has been traded for a derivative with respect to
$\eta$. The complete $\tau$-dependence of $j_{\rm int}$ is then 
contained in the factor $(\tau^2/\mu_j^2)^\eta$, and the inverse 
Laplace transform becomes simple. By exploiting the relation
\begin{align}
\int_{0}^{\infty} d p^2 e^{-p^2/(\tau^2 e^{\gamma_E})} \big(p^{2}\big)^{\eta-1} = \Gamma(\eta) e^{\gamma_E \eta} \left(\tau^2\right)^\eta\, ,
\label{eq:recoilJetInvLap}
\end{align}
one obtains the resummed jet function in momentum space 
in the form
\begin{align}
\label{eq:EvolvedJet}
J_{\rm int}(p^2, \mu) = \exp\left[ 4\, S(\mu_j, \mu) + 
2\,A_{\gamma_J}(\mu_j, \mu) \right]  
j_{\rm int}(\partial_{\eta}, \mu_j) 
\frac{e^{-\gamma_E \eta}}{\Gamma(\eta)} \frac{1}{p^2}
\left( \frac{p^2}{\mu_j^2} \right)^{\eta} \, .
\end{align}

The more complicated jet function for the narrow energy 
resolution, $J_{\rm nrw}$, was used in~\cite{Beneke:2018ssm}. In 
Appendix~\ref{app:recjetfnnarrow}, we provide details on its computation 
and give its full expression. 


\subsection{Soft function}
\label{sec:intressoftfn}

The soft function is defined by the vacuum amplitude 
(\ref{eq:softfnintdef}) of the soft operator (\ref{eq:softop}) 
with index contraction as specified in (\ref{eq:softtwoindexfn}). 
Let us recall that the soft operator is the product of soft 
Wilson lines arising from the decoupling of soft SU(2) gauge 
bosons from the four particles in the $2\to 2$ annihilation 
amplitude. The SU(2) indices are then contracted in a way 
that depends on the operator $\mathcal{O}_i$ and the external 
DM two-particle state $I$, resulting in the function 
$W_{IJ}^{ij}(\omega)$.  

The soft function is sensitive to physics at virtualities of 
order $m_W^2$, and therefore must be computed in the effective 
theory with broken SU(2)$_L\times$U(1)$_Y$ gauge symmetry and 
massive SM particles (unless the mass is much smaller than $ m_W$). 
For NLL' accuracy, the one-loop soft function 
and its NLL RG evolution is needed. Here we summarize these 
results. Technical details on the computation of virtual and 
real one-loop diagrams are given in Appendix~\ref{app:softFunction}, 
including the regularization, which involves rapidity regularization, 
together with some observations on partial virtual-real singularity 
cancellations. 

The virtual one-loop contributions to the soft 
function are the same as for the narrow resolution case and they were 
already computed in \cite{Beneke:2018ssm}. In the intermediate 
resolution range, the real emission of soft EW gauge bosons is 
kinematically allowed. The new contributions as well as the 
virtual diagram results are given explicitly in 
Appendix~\ref{app:softFunction}. 
To guide the discussion we present here the result for 
the $W^{22}_{(+-)(+-)}$ component of the soft function, which has the 
most complicated structure and allows us to explain the resummation 
procedure:
\begin{align}
W^{22}_{(+-)(+-)} (\omega,\mu,\nu) &= \delta(\omega) + 
\frac{\hat{\alpha}_2(\mu)}{4 \pi} 
\bigg[\delta(\omega) \left( - 8 \ln \frac{m_W}{\mu}
-16 \ln \frac{m_W}{\mu} \ln \frac{m_W}{\nu}
\right) 
\nonumber \\
&
- \,\frac{6}{\omega} \ln \left(\frac{m_W^2+\omega^2}{m_W^2}\right) -\frac{2\omega}{m_W^2 + \omega^2}
+ \left[\frac{1}{\omega}\right]^{[m_W]}_*
8 \ln \frac{\mu^2}{m_W^2} \bigg] \, ,
\label{eq:softW}
\end{align}
The complete set of soft function components for all operator and 
two-particle-state combinations is collected in Appendix~\ref{app:soft}. 

As for the unobserved-jet function, renormalization becomes 
multiplicative in Laplace space. The forward and inverse Laplace 
transforms are defined as
\begin{align}
\label{eq:softLapTrans}
w(s) &= \mathcal{L}\,\{W(\omega)\} = 
\int_0^{\infty} d \omega \, e^{-\omega s} \,W(\omega) \,, 
\\
W(\omega) &= \mathcal{L}^{-1}\left\{w(s)\right\} = 
\frac{1}{2 \pi i} \int^{c+i \infty}_{c-i \infty} 
d s\,  e^{s \omega} \,w(s) \, .
\end{align}
As can be seen from (\ref{eq:softW}), the  Laplace transforms 
required for the soft function are ($s=1/(e^{\gamma_E}\kappa)$)
\begin{align}
  \mathcal{L}\left\{\delta ( \omega )\right\}&=1 \, , \nonumber \\
  \mathcal{L}\left\{\left[\frac{1}{\omega}\right]_*^{\left[m_W\right]}\right\}&= \ln\frac{\kappa}{m_W} \, , \nonumber \\
  \mathcal{L}\left\{\frac{1}{\omega}\ln \left(\frac{m_W^2+\omega^2}{m_W^2}\right)\right\}&= 
  \text{si}^2\left(m_W s\right) + \text{ci}^2\left(m_W s\right)\equiv\tilde{G}(s) \, , \nonumber \\
  \mathcal{L}\left\{\frac{\omega}{m_W^2 + \omega^2}\right\} &= \cos(m_W s)\,\text{ci}(m_W s) - \sin(m_W s)\,\text{si}(m_W s) \equiv \tilde{Q}(s) \,,
  \label{eq:LaplaceTransforms}
\end{align}
where the functions si, ci are defined as
\begin{align}
\text{si}(x) \equiv -\int_{x}^{\infty}dt\, \frac{\sin(t)}{t}, \quad \text{and}\quad \text{ci}(x) \equiv -\int_{x}^{\infty} dt\, \frac{\cos(t)}{t}\, .
\end{align}
It is convenient to introduce the following vector notation
\begin{align}
\vec{w}_{IJ} = \left(w^{11}_{IJ},w^{12}_{IJ},w^{21}_{IJ},w^{22}_{IJ}
\right)^T 
\end{align}
for the Laplace transformed soft functions. The RRG equations for the 
soft functions take the form
\begin{align}
  \frac{d}{d \ln \nu} \vec{w}_{IJ}(s,\mu,\nu) &= \mathbf{\Gamma}^\nu_{W} 
\,\vec{w}_{IJ}(s,\mu,\nu) \, ,
\label{eq:softRRG}
\end{align}
where the fixed-order one-loop rapidity anomalous dimension is given by
\begin{align}
  \mathbf{\Gamma}^\nu_{W} &= \frac{\hat{\alpha}_2}{4\pi}\,4 \gamma_{\text{cusp}}^{(0)} \,\ln \frac{m_W}{\mu} \, \mathbf{1}_4\,.
\label{eq:softrapAD}
\end{align}
Note that the non-cusp piece of $\mathbf{\Gamma}^\nu_{W}$ is zero at one loop.
The discussion of the rapidity evolution factor from Section~\ref{sec:photonjet} equally applies to the soft function. We hence use~\eqref{eq:derivAnDim},~\eqref{eq:intPhotnAnDimRap} and~\eqref{eq:softRRG} to compute the 
rapidity-resummed soft function
\begin{align}
\vec{w}_{IJ}(s,\mu,\nu) &= \exp\left[\mathbf{\Gamma}^\nu_{W}(\mu) \ln \frac{\nu}{\nu_s} \right] \vec{w}_{IJ}(s,\mu,\nu_s) \,,
\label{eq:softRapidEvolved}
\end{align}
where $\mathbf{\Gamma}^\nu_{W}(\mu)$ is the integrated rapidity anomalous dimension for the soft function. As was discussed in the case of the photon jet function, the order of the anomalous dimensions included in~\eqref{eq:softRapidEvolved} depends on the resummation path in the $\mu - \nu$ plane.
The RRG is diagonal in both the operator index, encapsulated in the vector notation, and the two-particle state index pair $IJ$. Notice that only the soft functions and the photon jet function depend on the rapidity scale. We make the choice to evolve the photon jet function from the jet rapidity scale $\nu_h\sim 2 m_\chi$ down to $\nu_s\sim m_W$. This means that we can set $\nu = \nu_s$ for the soft function, which makes 
the rapidity evolution factor (\ref{eq:softRapidEvolved}) equal unity.

The virtuality RG equation for the Laplace-transformed soft function is
also diagonal in $IJ$, but its non-cusp piece is 
non-diagonal in operator space,
\begin{align}
  \frac{d}{d \ln \mu} \vec{w}_{IJ}(s,\mu,\nu) &=   \mathbf{\Gamma}_{W}^\mu  \, \vec{w}_{IJ}(s,\mu,\nu) \,,
\label{eq:softRGE}
\end{align}
with anomalous dimension
\begin{align}
\label{eq:anomDimSoft}
  \mathbf{\Gamma}_{W}^\mu &= 4\, \gamma_{\text{cusp}} \ln \frac{\kappa}{\nu} \, \mathbf{1}_4 + \begin{pmatrix}
0 & 0 & 0 & 0 \\
-2 \gamma_W & 3 \gamma_W & 0 & 0 \\
-2 \gamma^*_W & 0 & 3 \gamma^*_W & 0 \\
0 & -2\gamma_W^* & -2\gamma_W & 3 \gamma_W + 3\gamma_W^*
\end{pmatrix} \,.
\end{align}
As in the case of the photon jet function,~\eqref{eq:softrapAD} and~\eqref{eq:anomDimSoft} can be obtained from their definitions by taking the derivatives in $\mu$ and $\nu$, respectively, of $\vec{w}_{IJ}$.
At the one-loop order, which is enough for NLL' resummation, the 
anomalous dimension $\gamma_W$ evaluates to 
\begin{equation} 
\gamma^{(0)}_{W} = (2+2\pi i) c_2(j)\,.
\end{equation}
The solution to (\ref{eq:softRGE}) takes the form 
\begin{align}
\vec{w}_{IJ}(s,\mu,\nu) &= \bm{R}^{-1}\, \bm{U}_{W}(\mu,\mu_s)\, \bm{R} \, 
\vec{w}_{IJ}(s,\mu_s,\partial_\eta) \left(\frac{\kappa}{\nu}\right)^\eta \,.
\label{eq:softLaplaceResum}
\end{align}
The evolution matrix $\bm{U}_{W}$ is diagonal, 
\begin{align}
\bm{U}_{W} &= \begin{pmatrix}
1 & 0 & 0 & 0 \\
0 & \exp\left[3 A_{\gamma_W}\right] & 0 & 0 \\
0 & 0 & \exp\left[3 A_{\gamma_W^*}\right] & 0 \\
0 & 0 & 0 & \exp\left[3 (A_{\gamma_W} + A_{\gamma_W^*})\right]
\end{pmatrix}\,,
\end{align}
and the diagonalization matrix $\bm{R}$ and its inverse $\bm{R}^{-1}$ are 
given by
\begin{align}
\bm{R} = \begin{pmatrix}
\phantom{-}\frac{2}{3} & 0 & 0 & 0 \\
- \frac{2}{3} & 1 & 0 & 0 \\
-\frac{2}{3} & 0 & 1 & 0 \\
\phantom{-}\frac{2}{3} & -1 & -1 & \frac{3}{2} 
\end{pmatrix}
\,, \quad
\bm{R}^{-1} = \begin{pmatrix}
\frac{3}{2} & 0 & 0 & 0 \\
1 & 1 & 0 & 0 \\
1 & 0 & 1 & 0 \\
\frac{2}{3} & \frac{2}{3} & \frac{2}{3} & \frac{2}{3} 
\end{pmatrix} \,.
\end{align}
The integrals $A_{\gamma_W}$ and $\eta$ have been introduced in 
(\ref{eq:intAgamma}) and as already explained there, they can only be solved numerically at NLL'. As a last step we need to go back to momentum 
space and compute the inverse Laplace transform of 
(\ref{eq:softLaplaceResum}). The entire dependence on $\kappa$ is 
contained in $\vec{w}_{IJ}(s,\mu_s,\partial_\eta) \left(\frac{\kappa}{\nu}
\right)^\eta$. We therefore define $\hat{\vec{W}}_{IJ}(\omega,\mu_s,\nu)$ to 
be the inverse Laplace transform of $\vec{w}_{IJ}(s,\mu_s,\partial_\eta) \left(\frac{\kappa}{\nu}
\right)^\eta$: 
\begin{eqnarray}
\hat{\vec{W}}_{IJ}(\omega,\mu_s,\nu) = \mathcal{L}^{-1}\left[\vec{w}_{IJ}(s,\mu_s,\partial_\eta) \left(\frac{\kappa}{\nu}
\right)^\eta \, \right] \, .
\label{eq:softrescoeff}
\end{eqnarray}
The inverse transformation requires the
computation of 
\begin{eqnarray}
&& \mathcal{L}^{-1}\left[\left(\frac{\kappa}{\nu}\right)^\eta\right]  
= \frac{e^{-\gamma_E \eta} }{\Gamma(\eta)}\left(\frac{\omega}{\nu}\right)^\eta\frac{1}{\omega} \, , 
\label{eq:invTrans1} \\
&&F(\omega) \equiv \mathcal{L}^{-1}\left[ 
\left(\frac{\kappa}{\nu}\right)^\eta \tilde{G}\big(e^{-\gamma_E}/\kappa
\big)\right] 
\nonumber\\
&& \hspace*{1cm} = 
\left(\frac{e^{-\gamma_E}}{\nu}\right)^{\eta}\frac{\omega^{1+\eta}}{ \Gamma(2+\eta) m_W^2} \,
{}_4F_3\left(1,1,1,\frac{3}{2};1+\frac{\eta}{2},\frac{3}{2}+\frac{\eta}{2},2;-\frac{\omega^2}{m_W^2}\right) \label{eq:invTrans2} \,, \\
&&P(\omega) \equiv \mathcal{L}^{-1}\left[
\left(\frac{\kappa}{\nu}\right)^\eta \tilde{Q}\big(e^{-\gamma_E}/\kappa\big)\right]
\nonumber\\
&&\hspace*{1cm} =\left(\frac{e^{-\gamma_E}}{\nu}\right)^\eta \frac{\omega^{1+\eta}}{m_W^2 \Gamma(2+\eta)}\, {}_{3}F_2\left(1,1,\frac{3}{2};1+\frac{\eta}{2},\frac{3}{2}+\frac{\eta}{2};-\frac{\omega^2}{m_W^2}\right)
\label{eq:invTrans3}
\end{eqnarray}
For the above representative index and operator combination 
$IJ=(+-)(+-)$ and $ij=22$, the inverse Laplace transform gives 
\begin{align}
\hat{W}^{22}_{(+-)(+-)}(\omega,\mu_s,\nu)&= \left[1+\frac{\hat{\alpha}_2}{4 \pi} \left(\left(-16 \ln \frac{m_W}{\mu_s} \,\partial_\eta\right) - 8 \ln \frac{m_W}{\mu_s}\right)\right]\frac{e^{-\gamma_E \eta}}{\Gamma(\eta)} \frac{1}{\omega} \left(\frac{\omega }{\nu}\right)^\eta \nonumber \, \\
&\hspace{.5cm}+ \frac{\hat{\alpha}_2}{4 \pi} \left[-6 F(\omega) -2 P(\omega)\right]\, .
\end{align}
The results for $\hat{W}^{ij}_{IJ}$ in all possible index and operator combinations $IJ$ and $ij$ are collected in Appendix~\ref{app:softPrime}. Finally, using~\eqref{eq:softLaplaceResum} and~\eqref{eq:softrescoeff}, we find that the virtuality resummed soft function in momentum space takes the form
\begin{align}
    \vec{W}_{IJ}(\omega,\mu,\nu) = \bm{R}^{-1}\, \bm{U}_{W}(\mu,\mu_s)\, \bm{R} \, \hat{\vec{W}}_{IJ}(\omega,\mu_s,\nu) \,.
    \label{eq:evolvedSoft}
\end{align}
We emphasize that we did not include the rapidity evolution factor~\eqref{eq:softRapidEvolved} in~\eqref{eq:evolvedSoft}, since we evolve $Z_\gamma$ in $\nu$ from $\nu_h$ to $\nu_s$ which makes the soft function rapidity evolution factor unity. 


\subsection{RG and RRG invariance of the cross section}
\label{sec:RGI}

The factorization formula for the intermediate resolution case given 
in \eqref{eq:SIJGIJ} and \eqref{eq:factformulaint} puts constraints on 
the anomalous dimensions, since the physical photon energy spectrum 
has to be independent of the virtuality and rapidity factorization 
scales $\mu$ and $\nu$. This independence on the scales manifests itself in 
the two consistency equations
\begin{align}
\frac{d }{d \ln \mu} \,  \frac{d (\sigma v_{\text{rel}})}{d E_{\gamma}}&= 0 \, , 
\label{eq:virtconsistency} \\
    \frac{d}{d \ln \nu}\,  \frac{d (\sigma v_{\text{rel}})}{d E_{\gamma}}&= 0 \, . 
\label{eq:rapconsistency}
\end{align}
 Note that the Sommerfeld factor~\eqref{eq:SFdef} is computed at leading order, which makes it scale independent so it does not have to be taken into account when computing~\eqref{eq:virtconsistency} and~\eqref{eq:rapconsistency}. In previous subsections, we already made use of the fact that a Laplace transformation turns convolution into multiplication. It is thus easiest to derive the 
implications of~\eqref{eq:virtconsistency} and~\eqref{eq:rapconsistency} in Laplace space, by taking the Laplace transform of~\eqref{eq:SIJGIJ}, 
\eqref{eq:factformulaint} with respect to the variable 
$e_\gamma \equiv 2(m_\chi - E_\gamma)$. Calling the Laplace variable $t$, 
the Laplace transform of the convolution of the jet with the soft function is (for brevity, we omit $\mu$ and $\nu$ in the arguments, as well as operator and two-particle state indices)
\begin{align}
\mathcal{L}& \left[\int_{0}^{\infty} d\omega \,J_{\rm int} (2m_\chi(e_\gamma - \omega)) \,W(\omega)\right] \nonumber \\
&= \int_{0}^{\infty} d e_\gamma \,e^{-t e_\gamma} \int_{0}^{\infty} d\omega \,J_{\rm int} (2m_\chi(e_\gamma - \omega)) \,W(\omega) \nonumber \\
&= \int_{0}^{\infty} \frac{d p^2}{2m_\chi} \,e^{-t p^2/2m_\chi} J_{\rm int} (p^2) \int_{0}^{\infty} d\omega \,e^{-t\omega} W(\omega) \nonumber \\
&= \frac{1}{2m_\chi} \,j_{\rm int}\!\left(\ln\frac{2m_\chi}{t e^{\gamma_E} \mu^2}\right) \, w(t) \,.
\label{eq:convoLap}
\end{align}
When going from the second to the third line in~\eqref{eq:convoLap}, we made use of the substitution $p^2 = 2m_\chi(e_\gamma - \omega)$. Also, since $p^2$ is strictly positive, we can set the lower $p^2$-integration boundary to zero. Using the definitions~\eqref{eq:LapTrans}, (\ref{eq:softLapTrans}) of the Laplace-transformed jet function and soft function, respectively, we arrive at the fourth line of~\eqref{eq:convoLap}. We can therefore write~\eqref{eq:virtconsistency} as
\begin{align}
\frac{d }{d \ln \mu} \left[\vec{H}(\mu) \cdot \vec{w}(t,\mu,\nu) Z_{\gamma}(\mu,\nu) j_{\rm int}\left(\ln\frac{2m_\chi}{t e^{\gamma_E} \mu^2},\mu\right)
\right] = 0\, .
\label{eq:virtconsistencyp}
\end{align}
Taking the derivative and making use of the 
definitions~\eqref{eq:rgeH},~\eqref{eq:photnRRE},~\eqref{eq:RGE} 
and~\eqref{eq:softRGE} of the anomalous dimensions results in
\begin{align}
\mathbf{\Gamma}_H + \mathbf{\Gamma}^\mu_{W} + \gamma^\mu_{Z_\gamma} \mathbf{1}_4+ \gamma^\mu_{j} \mathbf{1}_4&=0 \, .
\label{eq:virtAnDimConsist}
\end{align}
The terms in~\eqref{eq:virtAnDimConsist} are matrices in operator space. Because the virtuality RG equation for the Laplace-transformed soft function is diagonal in $IJ$,~\eqref{eq:virtAnDimConsist} holds for every index pair $IJ$. We can now use the values of the anomalous dimensions, given in~\eqref{eq:anomDimH},~\eqref{eq:photnAnDimVirt},~\eqref{eq:anomDimJ} and~\eqref{eq:anomDimSoft}, to verify that~\eqref{eq:virtAnDimConsist} is indeed satisfied. For example, 
for the cusp 
terms, the consistency equation reads explicitly
\begin{align}
\left(4 \gamma_{\rm cusp} \ln\frac{4m_\chi^2}{\mu^2}  
+ 4 \gamma_{\rm cusp} \ln\frac{1}{t e^{\gamma_E} \nu}
+ 4 \gamma_{\rm cusp} \ln\frac{\nu}{2m_\chi} 
- 4 \gamma_{\rm cusp} \ln\frac{2m_\chi}{t e^{\gamma_E} \mu^2}\right) \mathbf{1}_4 = 0 \,.
\label{eq:virtAnDimConsistCusp}
\end{align}

The same steps can be applied for the evaluation of~\eqref{eq:rapconsistency}, except that in~\eqref{eq:virtconsistencyp} we differentiate with respect to $\ln \nu$. Since only the photon jet function and the soft function depend on the rapidity scale $\nu$, using the definitions~\eqref{eq:photnRRG} and~\eqref{eq:softRRG} results in the rapidity consistency equation
\begin{align}
\gamma^\nu_{Z_\gamma} \mathbf{1}_4 + \mathbf{\Gamma}^{\nu}_{W}&=0 \,.
\label{eq:rapAnDimConsist}
\end{align}
This can be shown to be satisfied by the values for the rapidity anomalous dimensions given in~\eqref{eq:photnAnDimRap} and~\eqref{eq:softrapAD}. 

Since~\eqref{eq:virtAnDimConsist} and~\eqref{eq:rapAnDimConsist} are fulfilled, we confirm that at the one-loop order the factorized cross section is independent of the scales $\mu$ and $\nu$. It should be noted that the cancellation of the off-diagonal non-cusp terms of $\mathbf{\Gamma}_H$ and $\mathbf{\Gamma}^\mu_{W}$ in~\eqref{eq:virtAnDimConsist} is non-trivial. In total, this provides a strong check of the consistency of the calculation. The corresponding consistency check for the factorization of the narrow resolution case is presented in Appendix~\ref{app:RGEconsistency}.

\subsection{Resummation schemes}
\label{sec:resscheme}

\begin{figure}[t]
	\centering
	\includegraphics[width=0.805\textwidth]{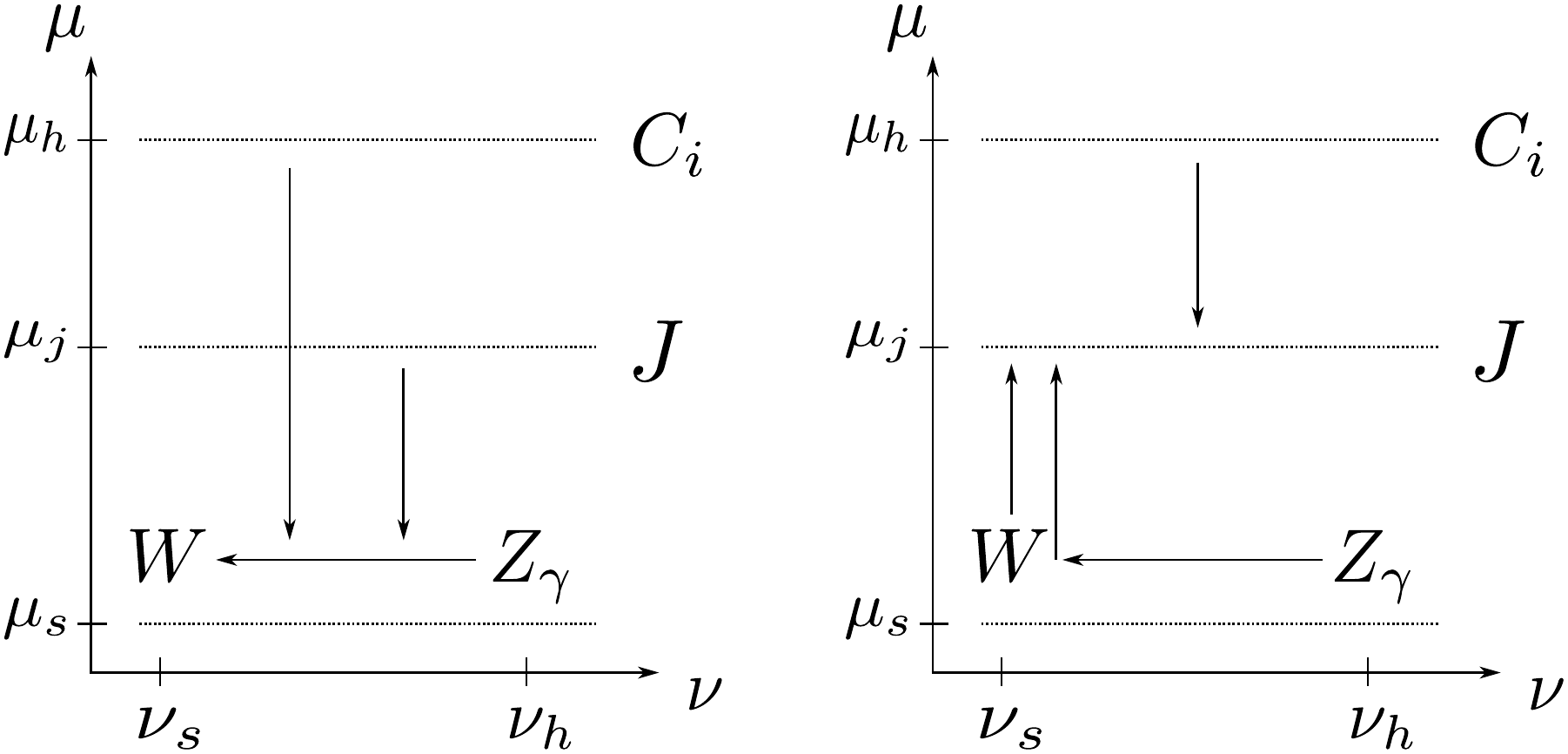} 
	\hspace*{0.1cm}
	\caption{Different possibilities for resumming the functions in the factorization theorem. Left: common reference scale is $\mu_s$. Right: common reference scale is $\mu_j$. In both cases, $Z_\gamma$ is evolved in $\nu$ from $\nu_h$ to $\nu_s$.
		\label{fig:resummation}}
\end{figure}

Having collected the RG equations for all the factors in the 
factorization formula, we show in Figure~\ref{fig:resummation} two 
different possibilities for the resummation of the functions appearing 
in the factorization theorem. For the first resummation scheme, shown 
in the left Figure~\ref{fig:resummation}, we choose $\mu_s$ as the 
common reference scale and evolve the Wilson coefficients $C_i$ and the 
unobserved-jet function $J$ down to the soft scale $\mu_s$, while the 
soft function $W$ and the photon jet function $Z_\gamma$ do not 
contain large logarithms when evaluated with $\mu=\mu_s$, 
and hence do not require 
resummation in $\mu$. Resummation in the rapidity scale is however 
necessary. We choose to evolve the photon jet function from $\nu_h$ 
to $\nu_s$. Equivalently one could also evolve the soft function from 
$\nu_s$ to $\nu_h$. This resummation scheme is close to the 
implementation of the narrow resolution case \cite{Beneke:2018ssm}, 
where there is no hard-collinear scale~$\mu_j$, and the hard 
functions are evolved all the way from the hard to the soft scale.

A more conventional implementation of resummation in the presence 
of an intermediate hard-collinear scale is the second resummation 
scheme illustrated in the right Figure~\ref{fig:resummation}. Here 
we choose $\mu_j$ as the common reference scale, and evolve $C_i$ 
down, and $Z_\gamma$ and $W$ up to $\mu_j$. $Z_\gamma$ is evolved in 
rapidity from $\nu_h$ to $\nu_s$ as before. Note that in this second 
case, as discussed in Section~\ref{sec:photonjet}, the specific form of 
the rapidity evolution factor $V$ depends on whether we first evolve 
in $\nu$ and then in $\mu$ or vice versa. Since we saw that $V$ takes 
a simpler form if we resum first in $\nu$ and then in $\mu$, we choose 
this ordering, as is also shown in Figure~\ref{fig:resummation} 
(right). 

Both schemes give the same results up to effects beyond 
the accuracy of the truncation of the RG equations.

\section{Results}
\label{sec:results}

In this section we present the results for the DM annihilation process 
$\chi^0\chi^0 \rightarrow \gamma + X$, assuming an intermediate energy 
resolution $E_{\text{res}}^\gamma$ of the instrument of order of the weak 
scale $m_W$. First we show $\langle \sigma v\rangle(E^\gamma_{\rm res})$, 
 as defined in (\ref{eq:defines}), as 
a function of the DM mass $m_\chi$ and then perform a numerical comparison 
of the present calculation with the narrow-resolution result 
of \cite{Beneke:2018ssm}. An analytic comparison of the two energy resolution 
cases is made in the next section, where we discuss the logarithms 
in the annihilation rates up to the two-loop order.

For the numerical results given in this section we use the couplings 
at the scale $m_Z = 91.1876 \,\text{GeV}$ in the $\overline{\text{MS}}$ 
scheme as input: $\hat{\alpha}_2(m_Z) = 0.0350009$, $\hat{\alpha}_3(m_Z) = 0.1181$, 
$\hat{s}_W^2(m_Z) = \hat{g}_1^2/(\hat{g}_1^2+\hat{g}_2^2)(m_Z) = 0.222958$, 
$\hat{\lambda}_t(m_Z) = 0.952957$, $\lambda (m_Z) = 0.132944$. The 
$\overline{\text{MS}}$ gauge couplings are in turn computed via one-loop relations 
from $m_Z, m_W = 80.385 \,\text{GeV}$, $\alpha_{\text{OS}}(m_Z) = 1/128.943$, and 
the top Yukawa and Higgs self-coupling, which enter our calculation only implicitly 
through the two-loop evolution of the gauge couplings, via tree-level relations
 to $\overline{m}_t = 163.35 \,\text{GeV}$ (corresponding to the top pole mass 
$173.2 \,\text{GeV}$ at four loops) and $m_H = 125.0 \,\text{GeV}$.


\subsection{Energy spectrum}

\begin{figure}[t]
\centering
\hspace*{-0.6cm}
\includegraphics[width=0.80\textwidth]{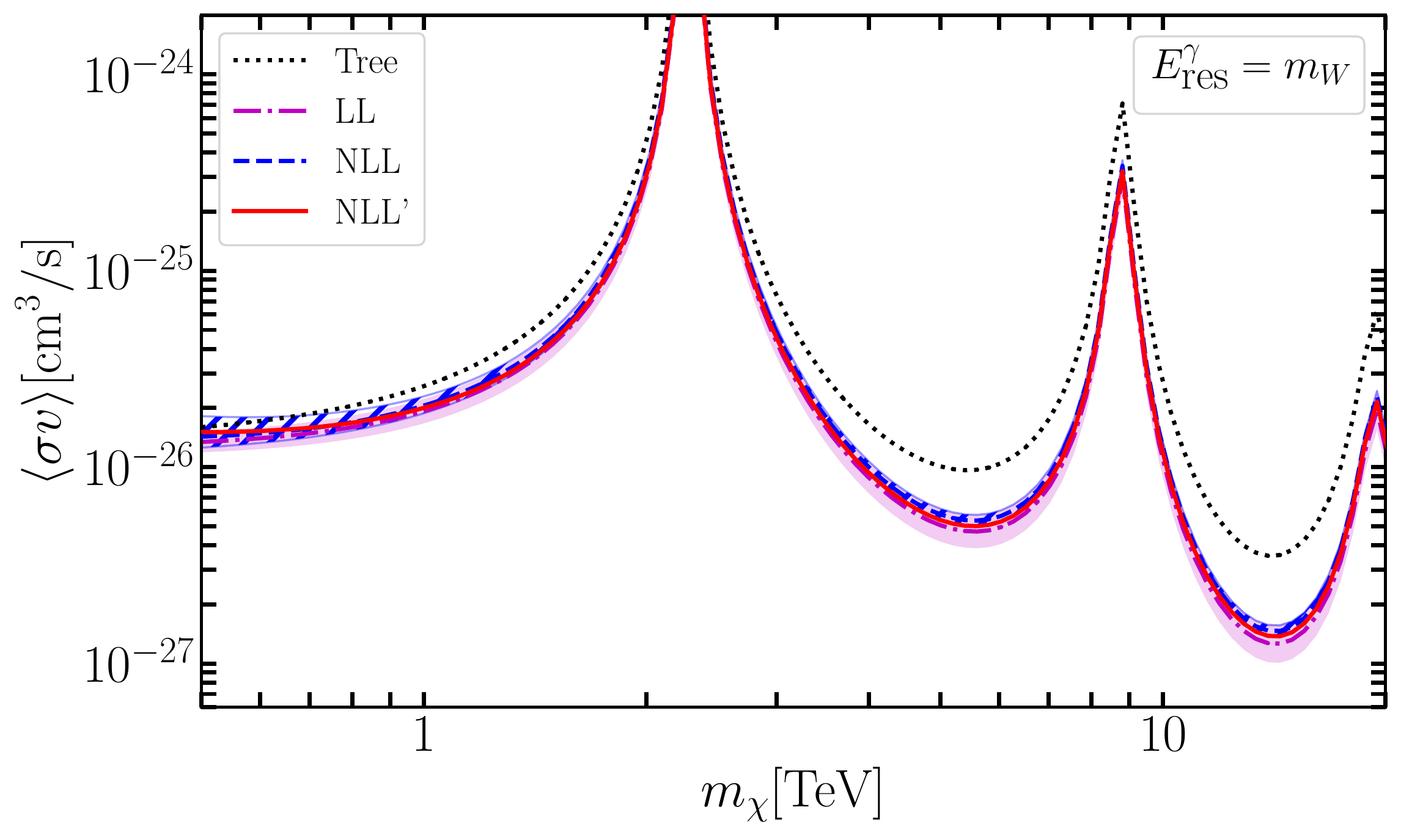} 
\vskip0.2cm
\includegraphics[width=0.77\textwidth]{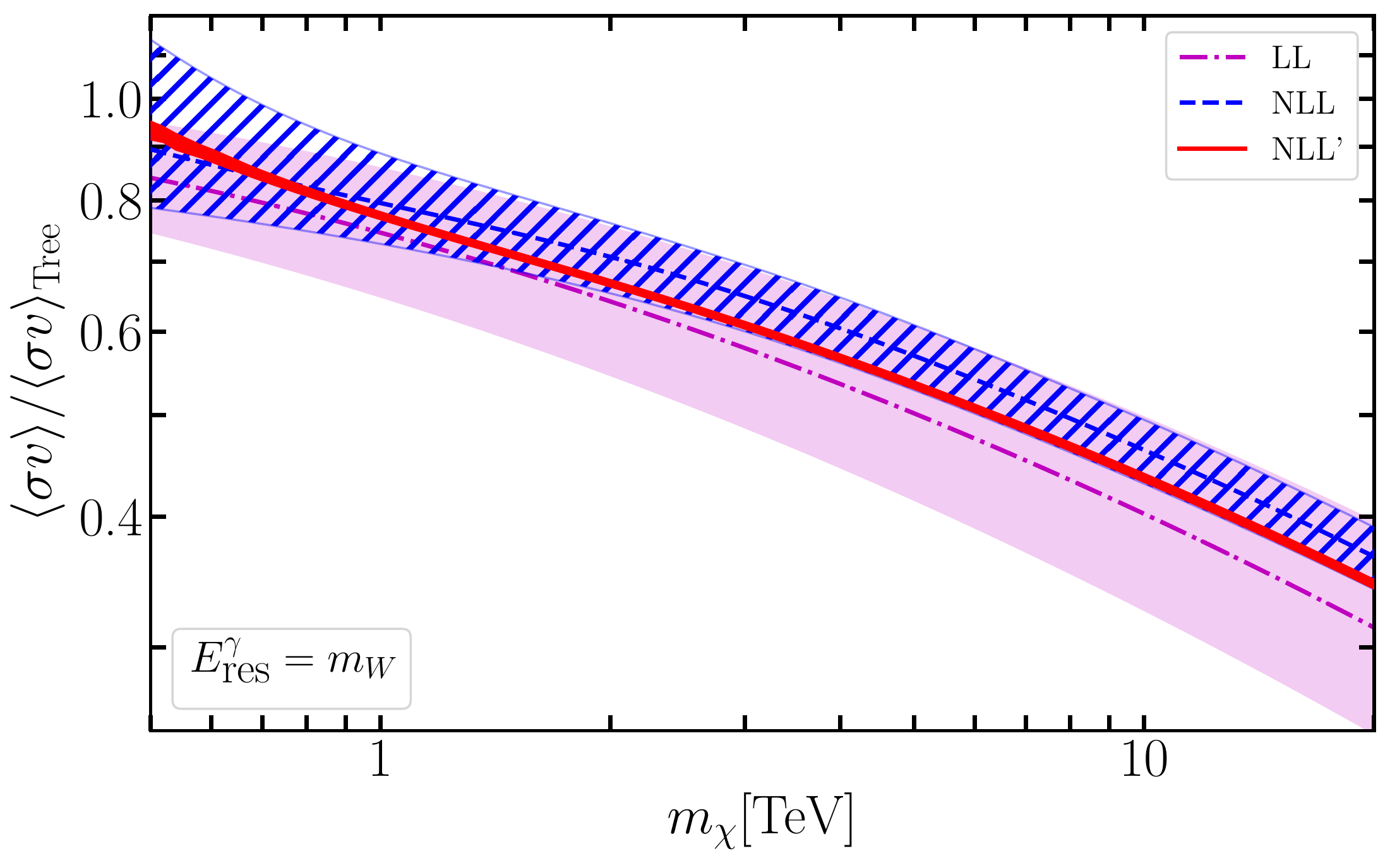}
\caption{Integrated photon energy spectrum within $E^\gamma_{\text{res}}$ 
from the endpoint $\mchi$ in the tree (Sommerfeld only) and LL, NLL, NLL' 
resummed approximation.  The energy resolution is set to $\Eres=m_W$. 
The shaded/hatched bands show the scale variation 
of the respective approximation as described in the text. For the NLL' 
result the theoretical uncertainty is given by the thickness of the red line.
\label{fig:result}}
\end{figure}

The upper panel of  Figure~\ref{fig:result} shows the cumulative endpoint 
annihilation rate $\langle \sigma v\rangle(E^\gamma_{\rm res})$, plotted as a 
function of the DM mass $m_\chi$.  The 
mass range includes the first two Sommerfeld resonances. The different lines 
refer to: the Sommerfeld-only calculation (black-dotted), also called ``tree'', 
since $\Gamma_{IJ}$ is evaluated in the tree approximation without any 
resummation, and multiplied with the Sommerfeld factor $S_{IJ}$ 
according to (\ref{eq:SIJGIJ}); the LL (magenta-dotted-dashed), 
the NLL (blue-dashed) and finally the NLL' (red-solid) resummed expression 
for $\Gamma_{IJ}$, the latter of which represents the calculation with the 
highest accuracy. The photon energy resolution is set to $\Eres=m_W$ in this 
figure.

The lower panel of the Figure shows the same LL, NLL and NLL' resummed 
annihilation rates, but normalized to the Sommerfeld-only result for 
better visibility of the resummation effect. We see that the resummation 
leads to a substantial reduction of the cross section, as is generally 
expected for Sudakov resummation. The size of the effect is consistent with the finding of
previous computations \cite{Beneke:2018ssm,Bauer:2014ula,Ovanesyan:2014fwa,Ovanesyan:2016vkk} of related 
observables or in different resolution regimes. In particular, in the 
interesting mass range around 3 TeV where wino DM accounts for the observed 
relic density, the rate is suppressed by about $30 - 40 \%$.

The resummed predictions are shown with theoretical uncertainty 
bands computed from a parameter scan with simultaneous variations of 
all scales. Specifically, the scales $\mu_h$, $\nu_h$ were varied in 
the interval $2m_\chi[1/2, 2]$, $\mu_j$ was varied in the interval 
$\sqrt{2m_\chi m_W}[1/2,2]$ and $\mu_s$, $\nu_s$ were varied in the 
interval $m_W[1/2, 2]$. The errors were then determined very 
conservatively by taking the maximum and minimum values in this 
five-dimensional parameter space. This scan was repeated for each mass point. 
For each parameter scan, we specified 21 values distributed 
logarithmically in the intervals given above, with ten values  
above and ten below the central values of the intervals. 

We find that the residual theoretical uncertainty at the NLL' order becomes 
negligible and is given by the width of the red-solid curve in 
Figure~\ref{fig:result}. It is also apparent that the different levels 
of resummation successively reduce the theoretical uncertainty 
considerably, from 
$17\%$ at LL, to $8\%$ at NLL and $1\%$ at NLL' at $m_\chi = 2 \,\text{TeV}$. 
Numerically, for the two mass values $m_\chi=2 \, \text{TeV} 
\left(10 \, \text{TeV}\right)$ the ratio to the 
Sommerfeld-only rate is $0.641^{+0.115}_{-0.097} \, (0.402^{+0.096}_{-0.077})$ at LL, $0.707^{+0.054}_{-0.054}\,(0.463^{+0.032}_{-0.033})$ at NLL and $0.667^{+0.007}_{-0.006} \, (0.435^{+0.005}_{-0.004})$ at NLL'. 
The central values correspond to central scales of the above intervals.

It is instructive to separate the integrated photon energy spectrum 
$\langle \sigma v \rangle(\Eres)$ into the contributions due to the different 
Sommerfeld factors in \eqref{eq:SIJGIJ}. Thus, we write 
\begin{equation}
\langle \sigma v \rangle = {\bf S_{(00)(00)}}[\sigma v]_{(00)(00)}
+2{\rm Re}[{\bf S_{\bm{(00)(+-)}}}[\sigma v]_{(00)(+-)}]
+{\bf S_{\bm{(+-)(+-)}}}[\sigma v]_{(+-)(+-)}\,,
\end{equation}
where 
\begin{equation}
{}[ \sigma v]_{IJ}(\Eres) =\int_{\mchi-\Eres}^{\mchi} d E_\gamma\,
\Gamma_{IJ}(E_\gamma)
\label{eq:gammaIJint}
\end{equation}
as in (\ref{eq:defines}). We find (Sommerfeld factors in bold), 
adopting  $\Eres=m_W$, 
\begin{align}
\langle \sigma v \rangle ={}&{}
2\times \bigg[\underbrace{{\bf 34.246}\times(1.2799)}_{\sim3\%}
+\underbrace{2\,{\rm Re}\left[{\bf42.100}\times(-0.9173 + 5.7918i)\right]}_{\sim-5\%} 
\nonumber\\
{}&{}+~\underbrace{{\bf 51.755}\times(29.907)}_{\sim102\%}\bigg]
\times10^{-28}~{\rm cm}^3/{\rm s}=3.0289\times10^{-25}~{\rm cm}^3/{\rm s}\ ,
\end{align}
for $m_\chi=$~2~TeV and
\begin{align}
\langle \sigma v \rangle ={}&{}
2\times \bigg[\underbrace{{\bf 1.1345}\times(1.2440)}_{\sim19\%}
+\underbrace{2\,{\rm Re}\left[{\bf0.35103}
	\times(-0.9553 + 7.7861i)\right]}_{\sim-9\%}\nonumber\\
{}&{}+~\underbrace{{\bf 0.10861}\times(62.788)}_{\sim90\%}\bigg]
\times10^{-27}~{\rm cm}^3/{\rm s}=1.5120\times10^{-26}~{\rm cm}^3/{\rm s}\ .
\end{align}
for the smaller DM mass value $m_\chi=$~500~GeV.
We observe that at $m_\chi=$~2~TeV (and similarly for larger masses), the 
Sommerfeld factors are large, as expected, and the annihilation rate is 
dominated by the $(+-)(+-)$ hard annihilation channel, which starts at 
tree level in the fixed-order expansion. The Sommerfeld factors are 
$\mathcal{O}(1)$ and even smaller than 1 for  $m_\chi=$~500~GeV for the 
off-diagonal annihilation contributions  $(00)(+-)$ and $(+-)(+-)$, 
for which the Sommerfeld enhancement does not compensate the loop 
suppression at small masses. 

The results shown in this section were computed with the more conventional 
second of the two resummation schemes discussed in 
Section~\ref{sec:resscheme}.  
We implemented both schemes and found full numerical 
agreement at NLL' at the 0.1\% level, as also follows from the 
analytic comparison, see (\ref{eq:diffresscheme}) below.

\subsection{Matching energy resolutions}
\label{sec:rescomparison}

In the introduction we identified three different regimes for the energy 
resolution $\Eres$, the narrow, the intermediate and the 
wide region. These cover the entire range of $\Eres$ for DM 
indirect detection experiments. In \cite{Beneke:2018ssm} we provided 
NLL' predictions for the photon-energy spectrum near the endpoint 
assuming a narrow energy resolution of $E_{\text{res}}^\gamma 
\sim m_W^2/m_\chi$, close to the line signal, while in this work 
we focus on $\Eres \sim m_W$, which is more realistic for 
present and future indirect DM searches in the TeV energy region. 
The two calculations differ in the structure of the 
unobserved jet function and the soft function, and exhibit different 
large logarithms.
The question arises whether the two computations can be 
matched to provide an accurate result for the entire range from 
$\Eres\sim 0$ to $\Eres \approx 4 m_W$, which we tentatively define 
as the upper limit of validity of the intermediate resolution case.

\begin{figure}[p]
	\centering
	\hspace*{-0.6cm}
	\includegraphics[width=0.66\textwidth]{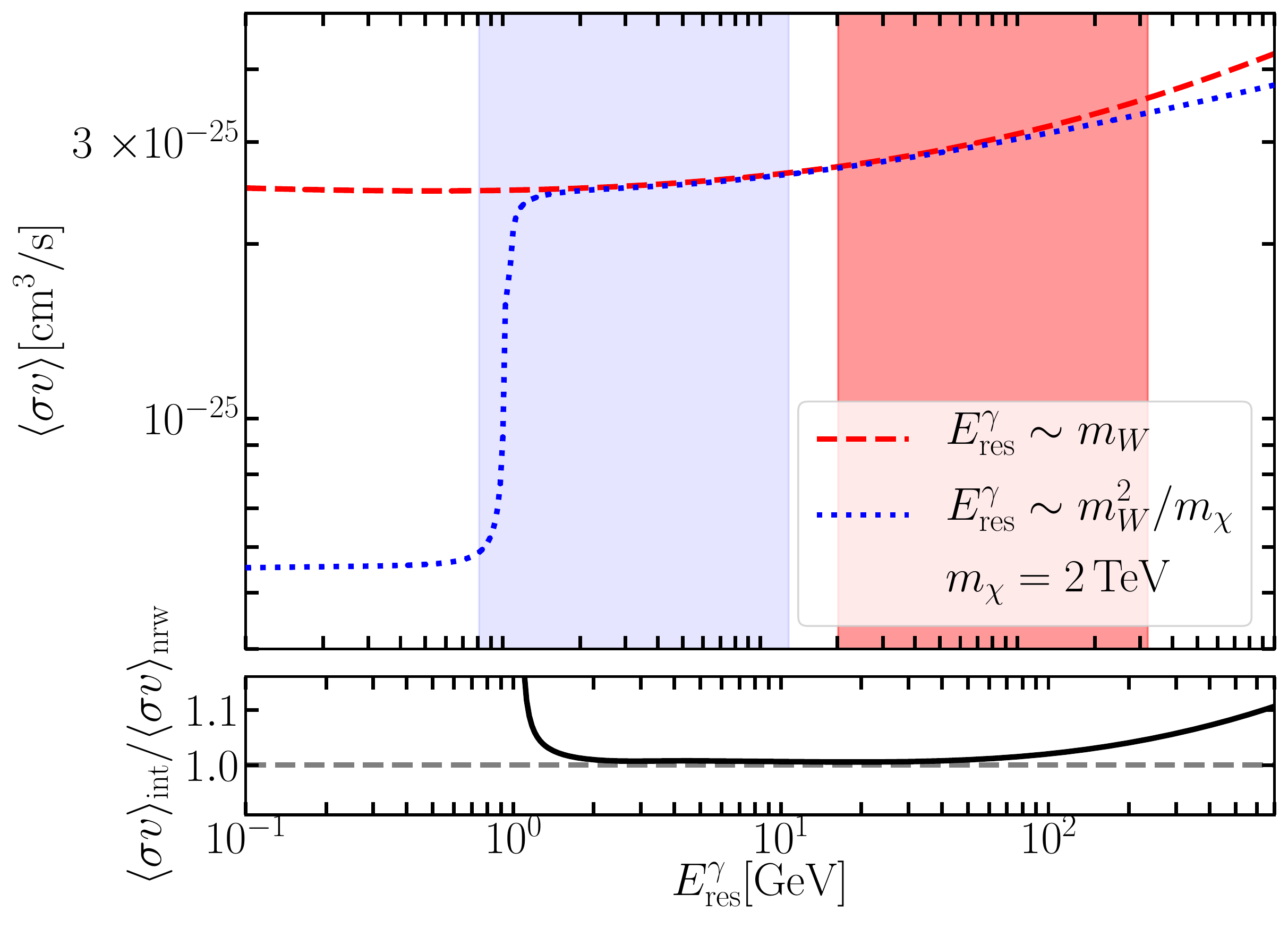} 
	\hspace*{-0.6cm}
	\includegraphics[width=0.66\textwidth]{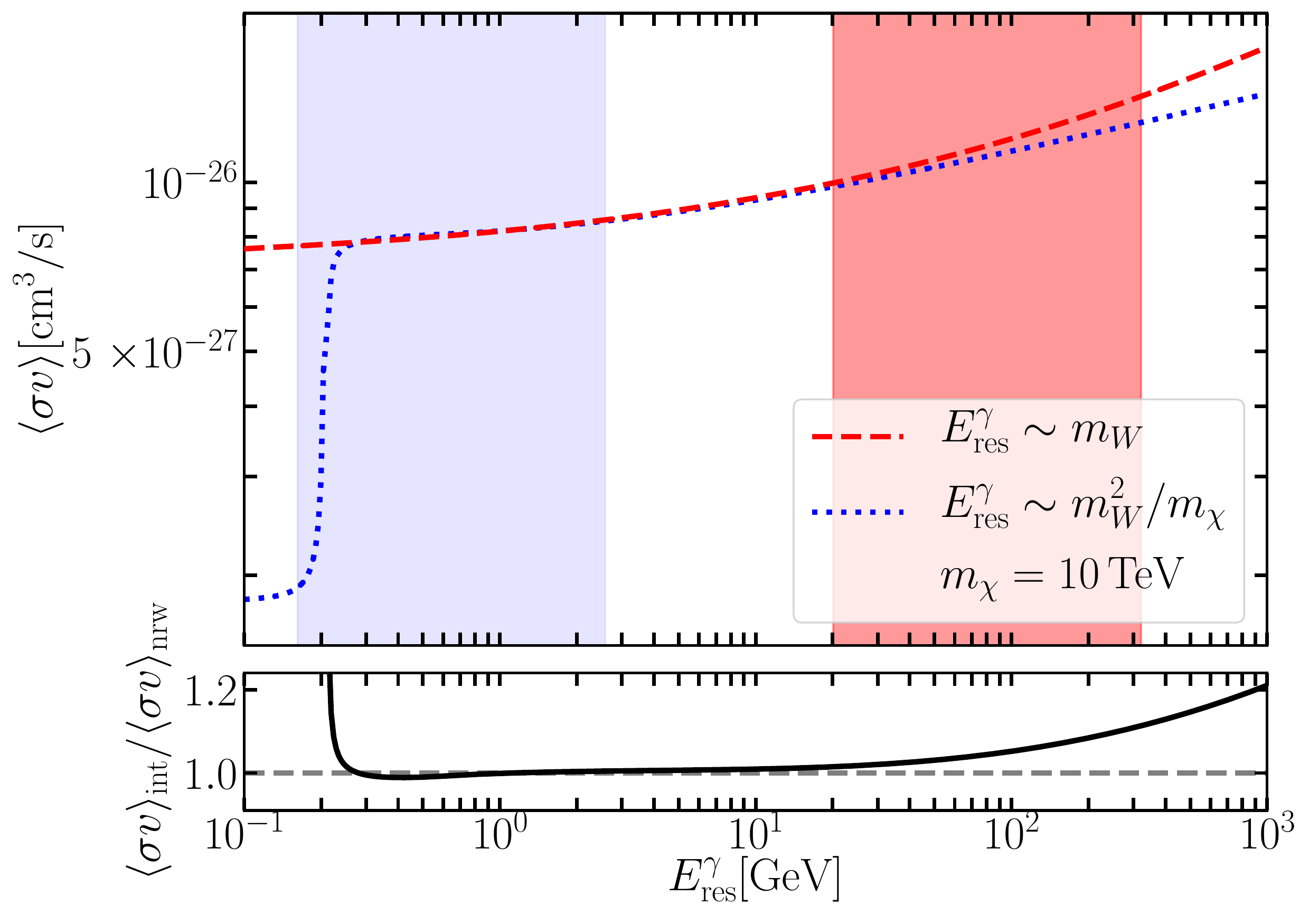} 
	\vskip0.1cm
\caption{Annihilation cross sections plotted as function of $E_{\text{res}}^\gamma$. The blue-dotted line shows the cross section for the narrow resolution computed in \cite{Beneke:2018ssm}. The red-dashed line shows the intermediate resolution cross section. The light-grey (blue) area represents the region of validity for the narrow resolution case and the dark-grey (red) area represents the region of validity for the intermediate resolution case. The ratio of the intermediate 
to narrow resolution annihilation cross section $\langle \sigma v \rangle_{\text{int}}/\langle \sigma v \rangle_{\text{nrw}}$ is added below each plot. The results are shown for DM masses of $m_\chi = 2 \,\text{TeV}$ (upper plot) and $m_\chi = 10 \, \text{TeV}$ (lower plot).
\label{fig:matching}}
\end{figure}

In Figure~\ref{fig:matching}, we show the annihilation cross sections for 
the narrow (blue-dotted) and the intermediate resolution 
(red-dashed) cases, plotted as functions of $\Eres$ for two representative 
DM mass values, $m_\chi = 2\, \text{TeV}$ (upper panel) and 
$m_\chi =10 \, \text{TeV}$ (lower panel). We also indicate the regions 
of validity of the narrow resolution (light-grey/blue) and the intermediate 
resolution (dark-grey/red) computations. The boundaries of these regions 
are defined by $m_W^2/m_\chi [1/4,4]$ (narrow resolution) and 
$m_W [1/4,4]$ (intermediate resolution).

We observe a wide interval in $E^\gamma_{\text{res}}$, covering the 
range of resolution in between the validity regions of the two 
calculations, for which the annihilation rates in both calculations 
agree with high precision. At low resolution there is a steep rise 
of the narrow resolution rate, which occurs at 
$\Eres \approx m_Z^2/(4 \mchi)$. Above this value the resolution is 
not enough to separate the $\gamma Z$ contribution, leading to a 
sharp increase of the semi-inclusive rate. Since the unobserved-jet 
function for the intermediate resolution cross section is computed under 
the assumption that the particles are massless, this feature is
absent in this curve (dashed/red), which is hence clearly not valid 
for very small resolution. In the narrow resolution regime the invariant 
mass of the unobserved-jet function also passes through the $W^+W^-$, 
$ZH$ and $t\bar{t}$ thresholds. However, these thresholds are not visible 
on the scale of the plot. The narrow resolution computation agrees 
very well with the intermediate one well into the regime of validity 
of the latter, and vice versa. As one moves to even higher 
$\Eres$, the intermediate resolution line starts to depart from the 
narrow resolution one. Here the narrow resolution computation clearly 
ceases to be accurate, because it fails to capture the effect of 
soft electroweak gauge boson radiation, which is now kinematically 
allowed. Nevertheless, even at the highest $\Eres =1\,$TeV shown 
in the plot, the difference stays below the 20\% level, as can be 
seen from the ratio plots at the bottom of the two panels in 
Figure~\ref{fig:matching}. We note that as the DM mass becomes larger, 
the separation between the two validity regimes (the shaded bands 
in the Figure) increases, but the matching continues to work well 
even for the 10 TeV DM mass example.

These observations show that the present work and 
\cite{Beneke:2018ssm} combined result in highly accurate theoretical 
predictions for the photon energy spectrum in dark matter annihilation, 
here for the wino model, in the entire energy resolution range  from 
$\Eres\sim 0$ to $\Eres \approx 4 m_W$. It would be 
interesting to perform a similar matching between the results of the  
present paper and the results of \cite{Baumgart:2017nsr}, which would 
extend the knowledge of the resummed energy spectrum to even wider 
resolution. As discussed in the introduction, with the 
anticipated energy resolution of the CTA experiment, we expect this 
to be necessary for DM searches only in the 10~TeV mass region and 
beyond.

\section{Fixed-order expansions}
\label{sec:logexpansions}

In this section we perform analytic expansions of the 
annihilation rate matrix $\Gamma_{IJ}$ up to the two-loop 
order. This provides some insight into the structure of 
large logarithms in the photon energy spectrum at large 
photon energy, depending on the energy resolution, and 
explains why the two computations agree remarkably well 
over a large interval of $\Eres$, as observed in the 
previous section. Readers interested only in the numerical 
result for the spectrum may skip this section.

\subsection{Double-logarithmic approximation}

Before moving to fixed-order expansions it is instructive to compare 
the NLL' result to the double-logarithmic approximation. This approximation 
is obtained by a) evaluating all functions in the tree-approximation, 
b) keeping only the $\hat{\alpha}_2\times\log^2$ terms in the exponents 
of the RG evolution factors. For the two resolution regimes discussed 
in this paper, the double-logarithmic approximations read
\begin{eqnarray}
\label{eq:dlognrw}
\langle\sigma v\rangle_{\rm nrw}(\Eres) &=& 
\frac{2\pi\hat\alpha_2^2\hat
	s_W^2}{m_\chi^2}\left[\hat s_W^2+\hat
c_W^2\Theta\!\left(\Eres-\frac{m_Z^2}{4m_\chi}\right)\right]
\,e^{-\frac{\hat{\alpha}_2}{\pi}\ln^2\frac{4m_\chi^2}{m_W^2}}\,
S_{(+-)(+-)}\,,\qquad
\\
\langle\sigma v\rangle_{\rm int}(\Eres) &=& 
\frac{2\pi\hat\alpha_2^2\hat s_W^2}{m_\chi^2}
\,e^{-\frac{3\hat{\alpha}_2}{4\pi}\ln^2\frac{4m_\chi^2}{m_W^2}}
\,S_{(+-)(+-)}\,.
\label{eq:dlogint}
\end{eqnarray}
The dependence of the coefficient of large logarithms on the energy 
resolution is already apparent from these equations. Since the `nrw' formula 
describes an observable that is more exclusive than the `int' one, the effect 
of the Sudakov double logarithm is, as expected, larger for the former. 
The exponents arise as follows in the first resummation scheme of 
Section~\ref{sec:resscheme}, where all functions are evolved to the soft scale.  
In both, the  `nrw' and `int' energy resolution formula the resummation of the 
hard function is responsible for the contribution  
$-\frac{\hat\alpha_2}{4\pi}\times 4\ln^2\frac{4m_\chi^2}{m_W^2}$ 
to the Sudakov exponent from the diagonal cusp logarithm in the anomalous dimension 
(\ref{eq:anDimWilson}). While in the `nrw' formula there are no further sources 
of double logarithms, the evolution of the unobserved-jet function from 
the hard-collinear to the soft scale for the `int' case adds the (positive) 
contribution $+\frac{\hat\alpha_2}{4\pi}\times\ln^2\frac{4m_\chi^2}{m_W^2}$,  
which partially compensates the Sudakov suppression associated with the 
hard-function resummation.
 
\begin{figure}[t]
\centering
\hspace*{-0.6cm}
\includegraphics[width=0.66\textwidth]{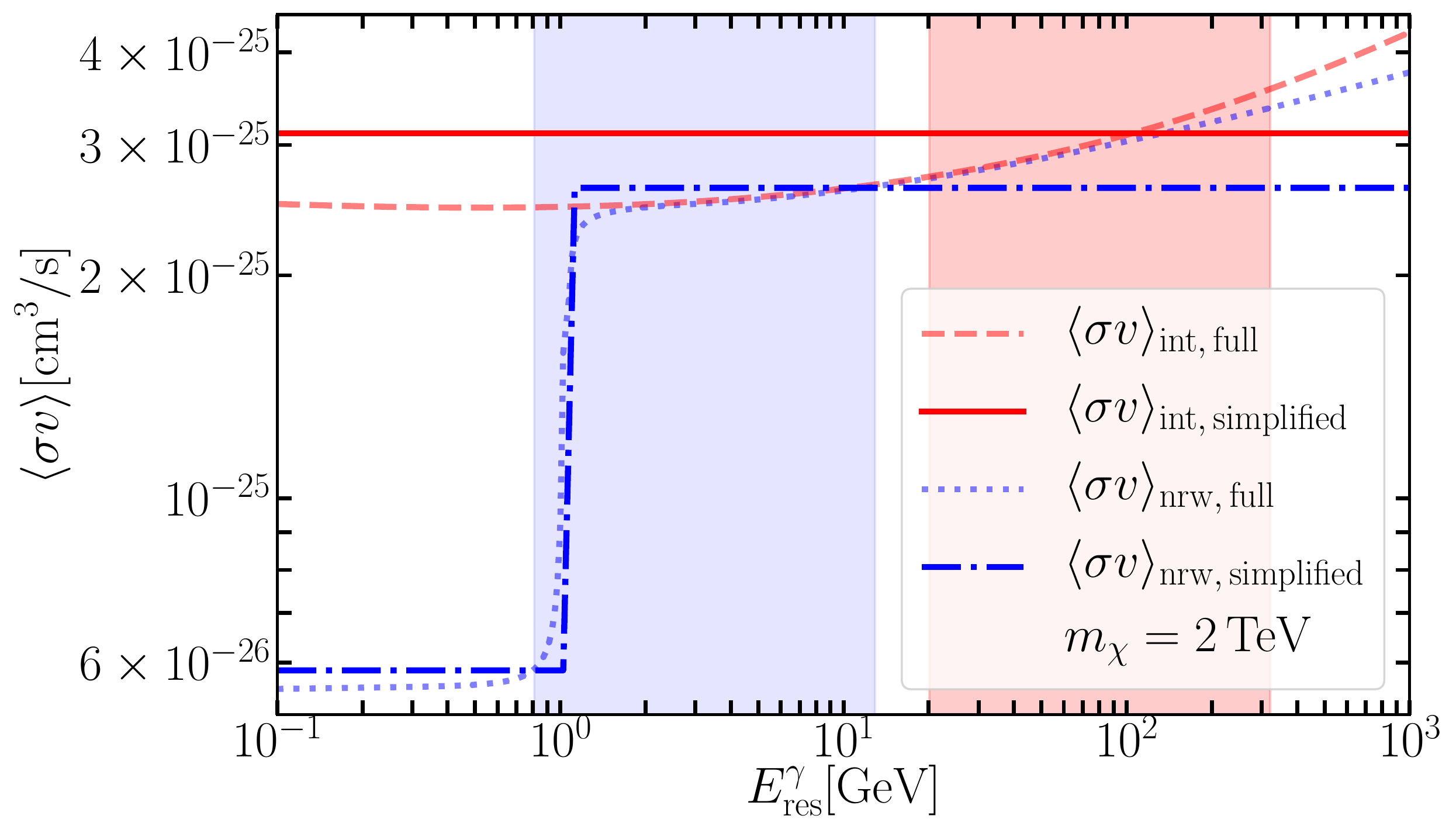} 
\hspace*{-0.6cm}
\includegraphics[width=0.66\textwidth]{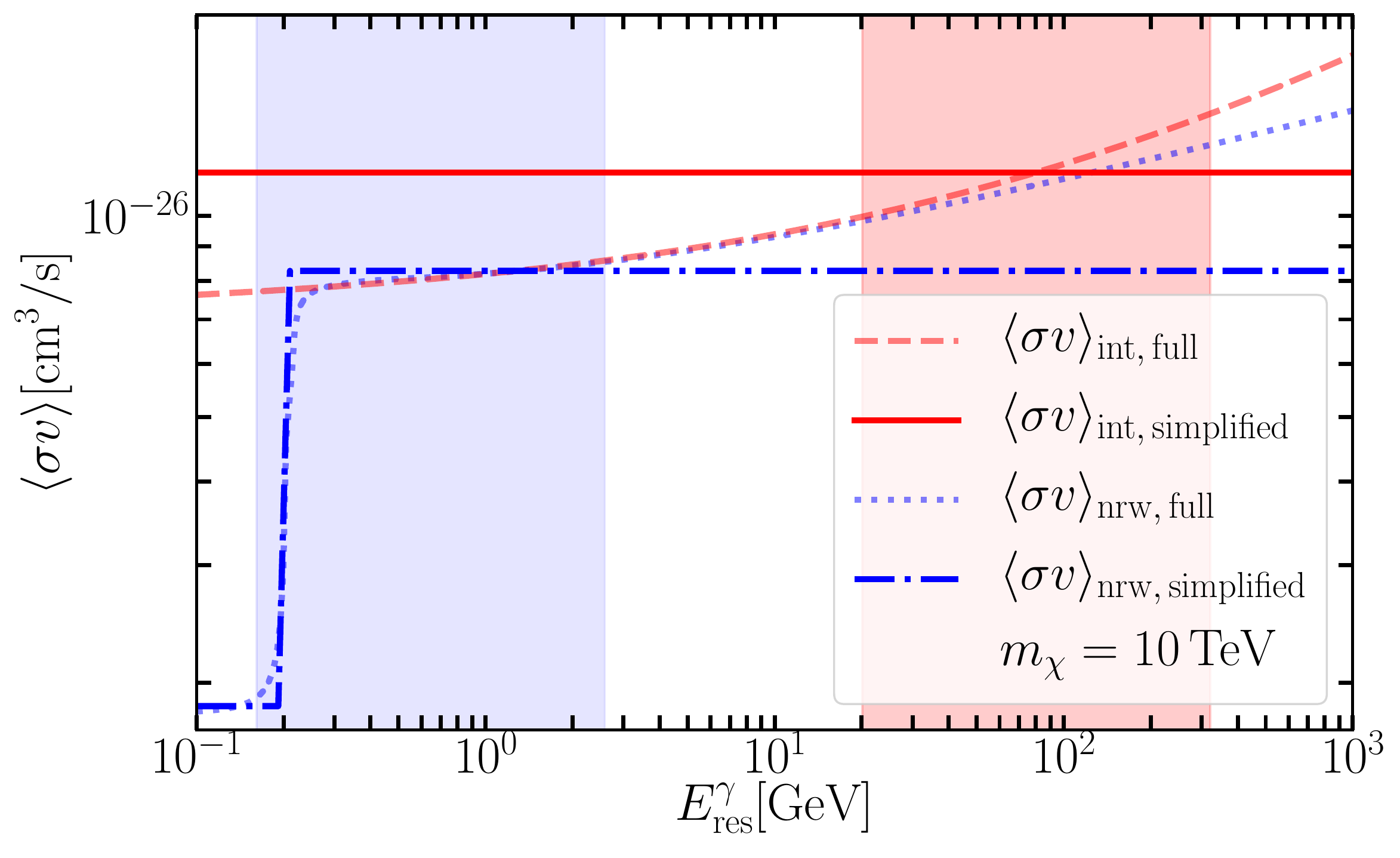} 
\vskip0.1cm
\caption{\label{fig:doublelog} Same as Figure~\ref{fig:matching} 
but in the double-logarithmic (``simplified") approximation. 
For comparison the complete (``full") NLL' 
results of Figure~\ref{fig:matching} are also 
included (dimmer dashed and dotted lines). Top: 
$m_\chi=2$~TeV. Bottom: $m_\chi=10$~TeV.}
\end{figure}

The double-logarithmic approximation is visualized in 
Figure~\ref{fig:doublelog}. It is seen that within their respective 
validity ranges (shaded areas in the plots) the double-log approximations 
of the intermediate and narrow resolution results are close to the full NLL' 
resummed results, shown for comparison 
(dimmer dashed/red and dotted/blue curves). In the narrow resolution case 
the step function in \eqref{eq:dlognrw} correctly describes the sharp 
rise of the annihilation cross section due to the opening of the 
$\gamma Z$ channel. 

However, Figure~\ref{fig:doublelog} also demonstrates that the precise 
shape of the cumulative annihilation rate in $\Eres$ and, in particular, 
the smooth matching of the two resolution regimes observed in the previous 
section cannot be explained in the double-logarithmic 
approximations~\eqref{eq:dlognrw}, \eqref{eq:dlogint}. We therefore 
analyze the subleading logarithms in the one- and two-loop order 
in the following subsection.

\subsection{Expansion of the resummed annihilation rate}

\begin{figure}[t]
\begin{center}
\includegraphics[width=.99\linewidth]{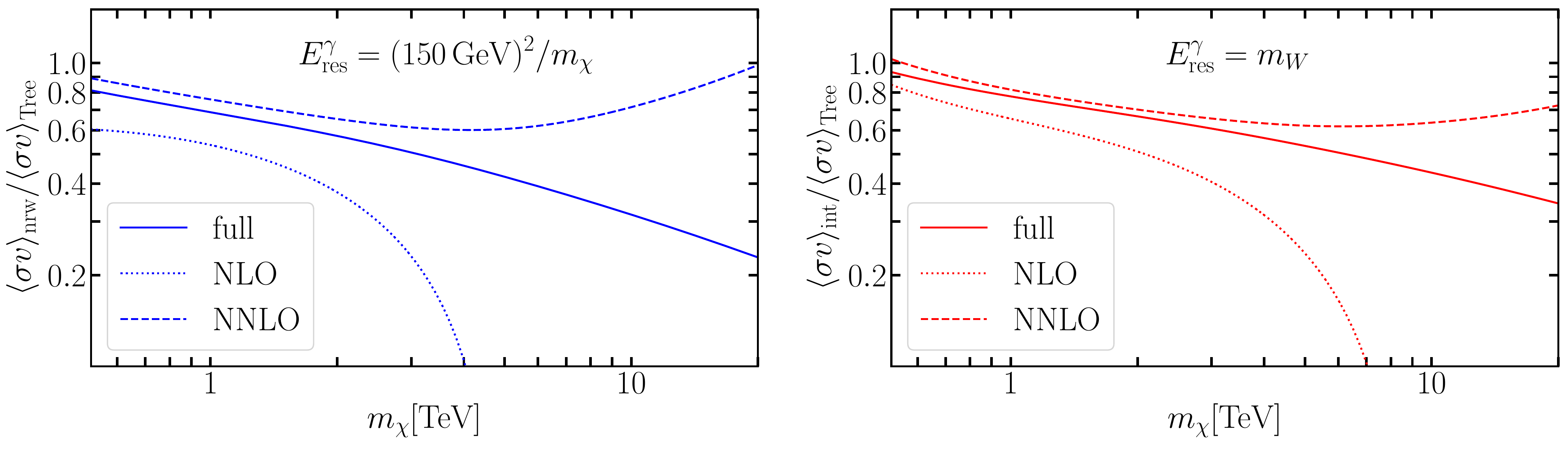}
\end{center}
\caption{\label{fig:logsratios} Left: Ratios to the tree-level cross section of the various fixed-order cross sections at NLO (dotted) and NNLO (dashed) and of the fully resummed NLL' cross section (solid) for the narrow resolution $\Eres = (150\, \text{GeV})^2/m_\chi$. Right: The same ratios as in the left panel, but in the intermediate resolution regime $\Eres = m_W$. As in Figure~\ref{fig:result}, in both cases the Sommerfeld factor $S_{IJ}$ is included and the formulas are evaluated at the central scales.}
\end{figure}

We re-expand the resummed annihilation rates (\ref{eq:factformulaint}), 
(\ref{eq:factformulanarrow}) for the intermediate and narrow resolution, 
respectively, in the number of loops. More precisely, we expand 
$[\sigma v]_{IJ}$ defined in (\ref{eq:gammaIJint}) in the form 
\begin{equation}
{}[\sigma v]_{IJ}(E_{\gamma}^{\rm res}) =\frac{2\pi\hat{\alpha}_2^2(\mu)\hat
s_W^2(\mu)}{\sqrt2^{n_{\rm id}}m_\chi^2}\sum_{n=0}^\infty
\sum_{m=0}^{2n}c_{IJ}^{(n,m)}(\Eres,\mu)\left(\frac{\hat{\alpha}_2(\mu)}{\pi}
\right)^{\!n}\,\ln^m\frac{4m_\chi^2}{m_W^2}\,
\label{eq:logexpand}
\end{equation}
where, by construction, the coefficients $c_{IJ}^{(n,m)}(\Eres,\mu)$ are $\mathcal
O(1)$ numbers, and the large logarithms $\ln (2\mchi/m_W)$ are made explicit. 
Note that the coefficients  $c_{IJ}^{(n,m)}(\Eres,\mu)$ are different for the two 
resolution regimes.\footnote{In the following we drop the arguments 
$\Eres,\mu$ for brevity.} The resummed rate depends on many scales, $\mu$ from 
the renormalization of the coupling, and the scales from the initial 
and final values of the RG evolution. To make the large logarithms 
explicit, we normalize scales by their natural values. For example, 
$\ln (\mu_j^2/m_W^2)$ is written as $\ln (\mu_j^2/(2 m_\chi m_W)) + 
\frac{1}{2}\ln (4\mchi^2/m_W^2)$, such that the first logarithm 
is $\mathcal{O}(1)$ and part of one of the $c_{IJ}^{(n,m)}$ coefficients.
For both factorization formulas we determine the
$c_{IJ}^{(n,m)}$ coefficients up to the two-loop level ($n=2$) for all
possible $m$ and $IJ$ combinations. These are listed in 
Appendix~\ref{app:logexpansions}, where some details about their determination
are also discussed. Figure~\ref{fig:logsratios} 
compares the numerical evaluation of the resulting fixed-order 
expressions with the full resummed result of 
Figure~\ref{fig:result}.\footnote{We use the following terminology: 
``$n$-loop'' refers to the $\mathcal{O}(\hat{\alpha}_2^n)$ correction only, 
while NLO refers to the sum of tree and one-loop, etc.} 
The Figure shows the breakdown of electroweak perturbation theory 
in the few TeV DM mass region, and makes the necessity of the
resummation evident.

Before discussing the behaviour of the $c_{IJ}^{(n,m)}$ coefficients as
functions of $\Eres$, let us clarify which logarithms in \eqref{eq:logexpand} 
are captured by NLL' resummation. After RG evolution, the resummed 
annihilation cross sections are obtained  in the form
\begin{eqnarray}
    \sigma v &\propto& (1+C_1\hat{\alpha}_2 +\ldots)
\,\exp\left[L f_0(\hat{\alpha}_2 L) +  f_1(\hat{\alpha}_2 L) +\ldots\right]
\end{eqnarray}
with functions $f_i(\hat{\alpha}_2 L)$ of the $\mathcal{O}(1)$ 
quantity $\hat{\alpha}_2 L \equiv 
\hat{\alpha}_2 \ln (4\mchi^2/m_W^2)$. The LL approximation 
amounts to keeping $f_0$, NLL adds $f_1$, while NLL' adds $C_1$. 
Other terms not written are beyond the NLL' accuracy. Expanding 
in $\hat{\alpha}_2$,  we observe that NLL' resummation determines 
the three highest powers of logarithms in any order of perturbation 
theory, specifically $c_{IJ}^{(n,2n)}$,
$c_{IJ}^{(n,2n-1)}$ and $c_{IJ}^{(n,2n-2)}$ in \eqref{eq:logexpand} for all $n$. 
In particular, for $n=1$ (one-loop) the NLL' resummation determines 
all the possible coefficients that exist at this order, 
including the non-logarithmic term $m=0$, while at two loops ($n=2$) 
all logarithms except the single logarithm are obtained. Since the 
dependence on the matching scales such as $\mu_j$ introduced by 
resummation must cancel at every fixed order, those fixed-order coefficients, 
which are obtained exactly from expanding the resummation formula, 
must be independent of these scales. On the other hand, at two loops, 
the single logarithmic and constant terms still depend on  
$\mathcal{O}(1)$ quantities such as $\ln (\mu_j^2/(2 m_\chi m_W))$ as can 
be seen from the explicit expressions in Appendix~\ref{app:logexpansions}.

In the following we discuss the logarithmic structure for the 
channel $IJ=(+-)(+-)$, which is the most interesting one, since the other 
channels do not have a tree-level coefficient.\footnote{
The other channels are listed in Appendix~\ref{app:logexpansions}.}
We then evaluate the 
coefficients outside of their validity range, for example we take a 
coefficient from the double Taylor expansion of the
narrow resolution formula and extrapolate it to $\Eres\sim m_W$ or
to the transition energy resolution scale $\Eres\sim
(m_W/m_\chi)^{1/2}m_W$ in order to study the numerical matching of 
the two resolution cases.  The extrapolation induces a reshuffling of
the logarithms in \eqref{eq:logexpand} because $\mathcal O(1)$ coefficients 
in one regime may develop large logarithms in the other.

\subsubsection{Tree level}

The tree-level coefficients in \eqref{eq:logexpand} are
$c_{(00)(00)}^{(0,0)}=c_{(00)(+-)}^{(0,0)}=0$ in both narrow and
intermediate resolution cases. The $\chi^+\chi^-\to\gamma + X$
tree-level cross sections, on the other hand, depend on which
factorization formula is being employed:
\begin{eqnarray}
\label{eq:nrwtree}
c_{(+-)(+-)}^{{\rm nrw}(0,0)} &=& \hat s_W^2+\hat
c_W^2\Theta\left(\Eres-\frac{m_Z^2}{4m_\chi}\right)\ ,\\
\label{eq:inttree}
c_{(+-)(+-)}^{{\rm int}(0,0)}&=& 1\ .
\end{eqnarray}
The narrow resolution formula distinguishes the contribution from the
$\gamma Z$ line from the $\gamma\gamma$ while the
intermediate resolution formula does not. When evaluated at
$\Eres>m_Z^2/(4m_\chi)$ both formulas yield the same result.

\subsubsection{One loop}

The one-loop term in \eqref{eq:logexpand} reads explicitly 
\begin{equation}
[\sigma v]_{(+-)(+-)}^{\rm1-loop}=\frac{2\pi\hat\alpha_2^2\hat
s_W^2}{m_\chi^2}\,\frac{\hat\alpha_2}{\pi}
\left[c_{(+-)(+-)}^{(1,2)}L^2+c_{(+-)(+-)}^{(1,1)}L+c_{(+-)(+-)}^{(1,0)}\right].
\label{eq:1loopexpand}
\end{equation}
For the presentation of the coefficients, some abbreviations will be 
helpful:
\begin{align*}
L\equiv\ln\frac{4m_\chi^2}{m_W^2},  \quad
x_\gamma\equiv&\frac{2\Eres}{m_W} \,\\ 
l_R\equiv\ln(x_\gamma)\,, \quad \kappa_R = \kappa_R{} (x_\gamma) = \frac{1}{2} \ln \left(1 + x_\gamma^2\right)& \,, \quad
\lambda_R=\lambda_R{} (x_\gamma)=-\frac12{\rm Li}_2(-x_\gamma^2)\ .
\end{align*}
The variable $L$ is the large logarithm in the expansion 
\eqref{eq:logexpand}. Note that $l_R$ is an $\mathcal{O}(1)$ quantity for 
intermediate resolution, but counts as a large logarithm in the 
narrow resolution case. The fixed-order expansion is performed in the 
running couplings $\hat\alpha_2(\mu)$, $\hat s_W^2(\mu)$ at the scale
$\mu$ of order $m_W$. We define the $\mathcal O(1)$ quantity 
$\lmu\equiv\ln(\mu^2/m_W^2)$. These explicit $\mu$-dependent logarithms 
cancel the implicit scale dependence of the couplings up to residual 
dependence of higher order than the NLL' accuracy of the approximation.
In addition to the variables introduced above, we define 
\begin{align}
z_\gamma \equiv {}&{} \left.\frac{4\pi}{\hat s_W^2(\mu)\hat\alpha_2(\mu)}Z_\gamma^{\rm 1-loop}(\mu,\nu)\right|_{\mu=m_W}\nonumber\\
{}={}&{}\left(-\frac{400}{27}+\frac23+\frac{16}9\ln\frac{m_t^2}{m_W^2}\right)\hat
s_W^2+\left(\frac{80}9\hat s_W^2
\ln\frac{m_Z^2}{m_W^2}-\frac{4\pi\Delta\alpha}{\hat\alpha_2}\right)\ ,
\end{align}
and the resolution-dependent function $j(\Eres)$ by means of the equation
\begin{equation}
j(E_\gamma^{\rm	res})\equiv \left.\frac{4\pi}{\hat\alpha_2(\mu)}\int_0^{4m_\chi E_\gamma^{\rm res}}\dif p^2J^{33,\,\rm 1-loop}_{\rm nrw}(p^2,\mu,\nu)\right|_{\mu=m_W}\ ,
\end{equation}
where the one-loop contributions to $Z_\gamma(\mu,\nu)$ and 
$J^{33}_{\rm nrw}(p^2,\mu,\nu)$ are given in \eqref{eq:photnJet} and 
\eqref{eq:J33nrw}, respectively. 
The function $j(\Eres)$ captures the complicated dependence of 
$J^{33}_{\rm nrw}(p^2)$ on the masses of the SM particles and $\Eres$ 
(see Appendix~\ref{app:recjetfnnarrow}),  
and is constructed such that it is independent of $\mu$ and $\nu$.

With these abbreviations at hand, we find 
\begin{eqnarray}
{}[\sigma v]_{(+-)(+-)}^{\rm nrw\ 1-loop} &=& \frac{2\pi\hat\alpha_2^2\hat
s_W^2}{m_\chi^2}\,\frac{\hat\alpha_2}{\pi}
\left[-L^2+L+c_{(+-)(+-)}^{{\rm nrw}(1,0)}\right],
\label{eq:nrw1loop}\\
{}[\sigma v]_{(+-)(+-)}^{\rm int\ 1-loop} &=&\frac{2\pi\hat\alpha_2^2\hat
s_W^2}{m_\chi^2}\,\frac{\hat\alpha_2}{\pi}
\left[-\frac34L^2+\left(l_R+\frac{29}{48}\right)L+c_{(+-)(+-)}^{{\rm
int}(1,0)}\right],
\label{eq:int1loop}
\end{eqnarray}
where
\begin{eqnarray}
c_{(+-)(+-)}^{{\rm nrw}(1,0)} &=&
\frac14\left(\frac{19}{6}-\frac{11}3s_W^2\right)\lmu-6+\frac{3\pi^2}4
+\frac14\left[j(E_\gamma^{\rm res})+z_\gamma\right]\ ,\\
c_{(+-)(+-)}^{{\rm int}(1,0)} &=&
\frac14\left(\frac{19}{6}-\frac{11}3s_W^2\right)\lmu-\frac{73}{18}+\frac{5\pi^2}{12}
+\frac14z_\gamma \nonumber \\
&&+l_R^2-\frac{19}{24}l_R-\frac32\lambda_R - \frac{1}{2} \kappa_R \ .
\end{eqnarray}
The $\lmu$ dependence in the $c_{(+-)(+-)}^{(1,0)}$ coefficients above
is compensated by the running couplings $\hat\alpha_2(\mu)$ and
$\hat s_W^2(\mu)$ in the corresponding LO terms.  

The coefficients depend 
on $\Eres$ through the
functions $l_R$, $\lambda_R$, $\kappa_R$ and $j$ defined above.
Therefore, in order to investigate the transition from the narrow to
the intermediate resolution formulas we need to
understand the asymptotic behaviour of these functions. For instance,
\begin{equation}
j(\Eres)\to
4\ln^2\frac{4m_\chi\Eres}{m_W^2}
-\frac{19}6\ln\frac{4m_\chi\Eres}{m_W^2}+\frac{70}9-\frac{4\pi^2}3
\quad \text{for }\ 4m_\chi\Eres\gg m_W^2
\end{equation}
up to corrections of order $m_W^2/(4m_\chi\Eres)$. 
This can be obtained from expanding the 
explicit expressions for the one-loop Wilson line and self-energy 
contributions given in Appendix~\ref{app:recjetfnnarrow}, or, more simply, 
by performing the expansion by regions \cite{Beneke:1997zp} 
before taking the integrals.

\begin{figure}[t]
\begin{center}
\includegraphics[width=.99\linewidth]{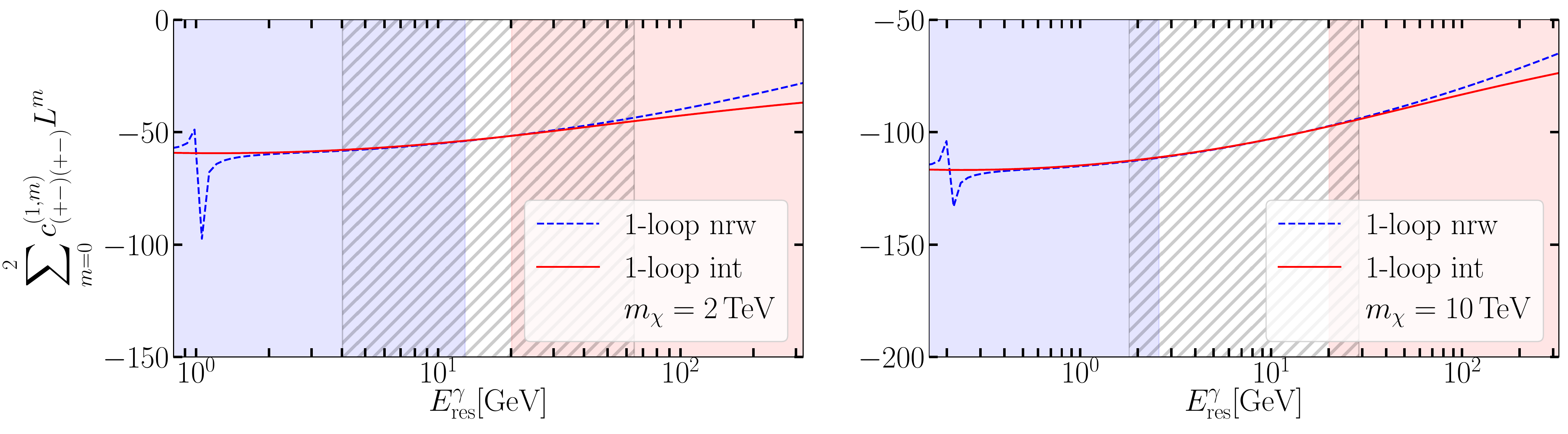}
\end{center}
\vskip-0.3cm
\caption{\label{fig:1loopnrwvsint} One-loop coefficient of the 
series \eqref{eq:logexpand} (including all logarithms) for the `nrw' 
(solid) and `int' (dashed)
factorization formulas. Left: $m_\chi=$2~TeV. Right: $m_\chi=$10~TeV.}
\end{figure}

When extrapolating \eqref{eq:nrw1loop} into the intermediate resolution 
regime, we can  write  $j(\Eres)$, using  $\ln(4m_\chi\Eres/m_W^2)=\frac12L+l_R$, 
as
\begin{equation}
\label{eq:jasymptote}
\frac14j(\Eres)\to
+\frac14L^2+\left(l_R-\frac{19}{48}\right)L+l_R^2-\frac{19}{24}l_R
+\frac{35}{18}-\frac{\pi^2}3\, .
\end{equation}
Then, for $\Eres\gg m_W^2/m_\chi$, 
\begin{equation}
\label{eq:sigmavdifference}
[\sigma v]_{(+-)(+-)}^{\rm nrw\ 1-loop} = [\sigma v]_{(+-)(+-)}^{\rm
int\ 1-loop}+ [\sigma v]^{\rm
tree}_{(+-)(+-)}\,\frac{\hat\alpha_2}\pi \left[\frac32\lambda_R\!\left(\frac{2\Eres}{m_W}\right) + \frac12\kappa_R\!\left(\frac{2\Eres}{m_W}\right) \right]\ .
\end{equation}
We note that due to the asymptotic behaviour of $j(\Eres)$, 
the large logarithms precisely match. The difference is a 
non-logarithmic term, which turns out to be quite small, and 
amounts to $\mathcal O$(1\%) of the tree-level cross sections
independent of the DM mass. 
This is visualized in Figure~\ref{fig:1loopnrwvsint}
where the one-loop coefficient (excluding the factor $\hat\alpha_2/\pi$) is
plotted for the two resolutions (narrow in dashed/blue, intermediate in 
solid/red). The absolute value of these dimensionless coefficients is, 
for both cases, large but the coefficients differ by no more than 
3\% in the hatched cross-over region. 
Similar results are found when $I$ or $J=(00)$ as can be 
verified from the coefficients $c_{IJ}^{(n,m)}$ listed
in Appendix~\ref{app:logexpansions}.

One may wonder why in (\ref{eq:sigmavdifference}) the narrow and 
intermediate resolution coefficients do not agree exactly, since 
by construction the NLL' approximation reproduces the full 
one-loop calculation. However, this is true only up to power 
corrections in $m_W/\mchi$. The difference in (\ref{eq:sigmavdifference}) 
arises from the $\lambda_R$, $\kappa_R$ terms in the intermediate resolution 
coefficient (\ref{eq:int1loop}). In the narrow resolution limit 
$\lambda_R$ and $\kappa_R$ are power-suppressed effects of order $m_W/\mchi$.

\subsubsection{Two loops}

The two-loop term in \eqref{eq:logexpand} reads  
\begin{eqnarray}
[\sigma v]_{(+-)(+-)}^{\rm2-loop} &=&\frac{2\pi\hat\alpha_2^2\hat
	s_W^2}{m_\chi^2}\,\frac{\hat\alpha_2^2}{\pi^2}
\,\bigg[{c_{(+-)(+-)}^{(2,4)}}L^4+{\
	c_{(+-)(+-)}^{(2,3)}}L^3+{ c_{(+-)(+-)}^{(2,2)}}L^2
\nonumber\\
&&+\,
c_{(+-)(+-)}^{(2,1)}L+c_{(+-)(+-)}^{(2,0)}\bigg]\,.
\label{eq:2loopexpand}
\end{eqnarray}
NLL' resummation determines all but the 
coefficients $c_{IJ}^{(2,1)}$ and $c_{IJ}^{(2,0)}$ of the series 
exactly. The expansion of the resummation formula also yields 
expressions for single logarithmic and constant terms, but these 
are incomplete. We find 
\begin{eqnarray}
c_{(+-)(+-)}^{{\rm nrw}(2,4)} &=&\frac1{2!}(-1)^2\,,\\
c_{(+-)(+-)}^{{\rm nrw}(2,3)} &=&-\frac{53}{72}\,,\\
c_{(+-)(+-)}^{{\rm nrw}(2,2)}
&=&\frac14(-1)\left[\frac{19}{3}-\frac{11}3s_W^2\right]\lmu
+\frac{671}{144}-\frac{13\pi^2}{12}-\frac{z_\gamma+j(\Eres)}4
\end{eqnarray}
in the narrow resolution case and
\begin{eqnarray}
c_{(+-)(+-)}^{{\rm int}(2,4)}
&=&\frac1{2!}\left(-\frac34\right)^2=\frac9{32}\,,\\
c_{(+-)(+-)}^{{\rm int}(2,3)} &=&-\frac29-\frac34l_R\,,\\
c_{(+-)(+-)}^{{\rm int}(2,2)}
&=&\frac14\left(-\frac34\right)\left[\frac{19}{3}-\frac{11}3s_W^2\right]\lmu+
\nonumber\\
&&+\,
\frac{4489}{2304}-\frac{37\pi^2}{48}-\frac3{16}z_\gamma+\frac98\lambda_R + \frac{3}{8}\kappa_R+l_R -\frac14l_R^2 \,,
\end{eqnarray}
for the intermediate resolution case. As before, the $\lmu$ dependence of the
$c_{(+-)(+-)}^{(2,2)}$ coefficients in both resolutions is compensated
by the scale dependence of the couplings. The $\Eres$
dependence of the coefficients is captured in the $j$,
$l_R$, $\lambda_R$ and $\kappa_R$ functions already encountered in the one-loop expansion. The
coefficients $c_{(+-)(+-)}^{(2,1)}$ and 
$c_{(+-)(+-)}^{(2,0)}$ are provided in the Appendix~\ref{app:logexpansions}.

\begin{figure}[t]
\begin{center}
\includegraphics[width=.99\linewidth]{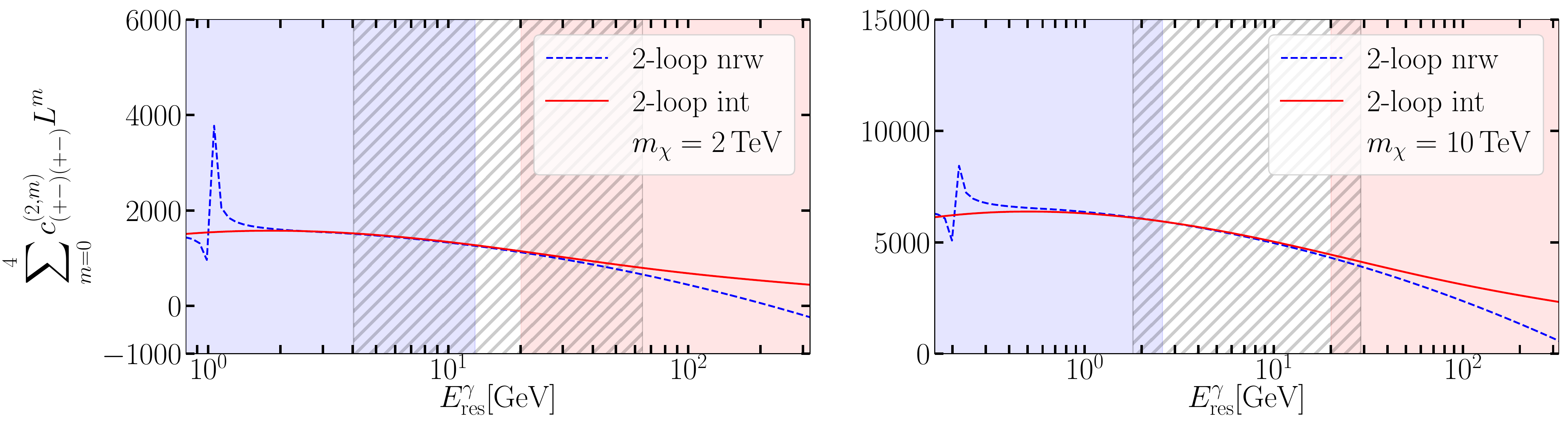}
\end{center}
\vskip-0.3cm
\caption{\label{fig:2loopnrwvsint} Two-loop coefficient of the series
\eqref{eq:logexpand} (including all logarithms and evaluated on the
central scales) for the `nrw' (solid) and `int' (dashed) factorization
formulas. Left: $m_\chi=$2~TeV. Right: $m_\chi=$10~TeV.}
\end{figure}

Figure~\ref{fig:2loopnrwvsint} compares for the two factorization formulas
the complete two-loop coefficient in \eqref{eq:2loopexpand} (including the 
$c_{(+-)(+-)}^{(2,m)}$ for all $m$).
These are evaluated at the central scales, i.~e. all $l_{\mu_i}$'s and 
$l_{\nu_i}$'s of the coefficients listed in the appendix are set to zero.

In order to understand the behaviour of the curves in the Figure
analytically, we use the asymptotic behaviour of the
$\Eres$-dependent functions within the coefficients.
The leading-logarithmic dependence of these for different $\Eres$ scaling
is shown in Table~\ref{tab:asymptotes}.
Besides the two energy resolution regimes associated with the `nrw' and
`int' factorization theorems, the transition scale constructed
from the geometric mean of the narrow and intermediate resolution scales
is also considered.

\begin{table}[t]
	\begin{center}
		\begin{tabular}{|c|c|c|c|}
			\hline
			{} & $\Eres\sim m_W^2/m_\chi$ & $\Eres\sim m_W \sqrt{m_W/m_\chi}$ &
			$\Eres\sim m_W$ \\
			\hline
			\hline
			$j(\Eres)$ & $j=\mathcal O(1)$ & $\frac14L^2+\mathcal O(L)$ &
			$L^2+\mathcal O(L)$\\
			\hline
			$l_R(\Eres)$ & $-\frac12L+\mathcal O(1)$ & $-\frac14L+\mathcal O(1)$ &
			$l_R=\mathcal O(1)$\\
			\hline
			$\lambda_R(\Eres),\kappa_R(\Eres)$ & 0 & 0 & $\lambda_R,\kappa_R=\mathcal O(1)$\\
			\hline		
		\end{tabular}
	\end{center}
	\caption{\label{tab:asymptotes} Leading-logarithmic dependence of the
		$\Eres$-dependent functions appearing in the fixed-order expansions when
		evaluated at the three $\Eres$-scales relevant to 
		Figure~\ref{fig:2loopnrwvsint}. Vanishing entries are to be understood as
		power-suppressed.}
\end{table}
\begin{table}[t]
	\begin{center}
		\begin{tabular}{|c|c|c|c|}
			\hline
			$\sum c_{(+-)(+-)}^{(2,m)}L^m$ & $\Eres\sim m_W^2/m_\chi$ & $\Eres\sim
			m_W \sqrt{m_W/m_\chi}$ & $\Eres\sim m_W$ \\
\hline
\hline
			`nrw' & $\frac{16}{32}L^4+\mathcal O(L^3)$ & $\frac7{16}L^4+\mathcal
			O(L^3)$ & $\frac8{32}L^4+\mathcal O(L^3)$\\[0.05cm]
			\hline
			`int' & $\frac{15}{32}L^4+\mathcal O(L^3)$ & $\frac7{16}L^4+\mathcal
			O(L^3)$ & $\frac9{32}L^4+\mathcal O(L^3)$\\[0.05cm]
\hline
`nrw'-`int' & $\frac{1}{32}L^4+\mathcal O(L^3)$ & $\mathcal O(L^3)$ &
$-\frac1{32}L^4+\mathcal O(L^3)$\\[0.05cm]
\hline
\end{tabular}
\end{center}
\caption{\label{tab:asympcoeff} Leading-logarithmic terms of the
two-loop coefficients in \eqref{eq:logexpand} for the 'nrw' and
`int' factorization formulas, and the difference of the two, 
at the scales relevant to Figure~\ref{fig:2loopnrwvsint}.}
\end{table}

In Table~\ref{tab:asympcoeff} we show the leading logarithm that results
from reevaluating the `nrw' and `int' two-loop coefficients at the
three scales in $\Eres$ relevant to Figure~\ref{fig:2loopnrwvsint}.
We verify the behaviour encountered in both panels of the
Figure. Namely, in Figure \ref{fig:2loopnrwvsint} the two-loop coefficient
associated with the `nrw' formula is larger than the corresponding one
from the `int' formula for the narrow resolution regime.
This property is supported by the positive difference of the leading
logarithmic term ($L^4/32$) between the two formulas
as evaluated at the `nrw' regime (last row and second column of Table
\ref{tab:asympcoeff}). Conversely, when $\Eres\sim m_W$ the opposite 
happens, and the `int' coefficient is larger than the `nrw' one, 
consistent with the last entry (last column from left to right and last row
from the top to the bottom) of Table \ref{tab:asympcoeff}.
The vanishing of the $\mathcal O(L^4)$ term for $\Eres\sim
m_W\sqrt{m_W/m_\chi}$ explains the almost perfect 
matching of the `nrw' and `int'
coefficients in the transition region as observed in 
Figure~\ref{fig:2loopnrwvsint}.

In summary, that the matching works so well over a wide range of energy resolution 
is a consequence of the smallness of the difference in the 
leading logarithms. For example, extrapolating the narrow resolution coefficient 
to intermediate resolution, we find at NNLO 
\begin{align}
\label{eq:nrwmintmW}
\frac{[\sigma v]_{(+-)(+-)}^{\rm nrw} - [\sigma v]_{(+-)(+-)}^{\rm
int}}{[\sigma v]_{(+-)(+-)}^{\rm tree}}=&
\frac{\hat \alpha_2}\pi \left[\frac32\lambda_R + \frac{1}{2}\kappa_R\right] \nonumber \\
&+\frac{\hat\alpha_2^2}{\pi^2}\left[-\frac{L^4}{32}
+\left(\frac{19}{144}-l_R\right)L^3+\mathcal{O}(L^2)\right].
\end{align}
At one loop, as discussed before, the difference lacks
large logarithms since $\lambda_R$, $\kappa_R$ are $\mathcal O(1)$ functions of
$\Eres$ provided $\Eres\sim m_W$. At the two-loop level we see a partial
cancellation of the $L^4$ coefficients (as $1/32\ll 1$).
In \eqref{eq:nrwmintmW} we therefore include the $L^3$ term, 
which constitutes the largest difference term at two loops when 
$L$ is not extremely large.

\subsection{Resummation schemes compared}
\label{sec:resschemecomp}
So far the discussion on the fixed-order expansions of the intermediate
resolution formula has been done using the first resummation 
scheme of Section~\ref{sec:resscheme}.
This is the most natural choice when comparing with the factorization
formula in the narrow resolution case.
We performed the same  fixed-order analysis for the second 
resummation scheme and found exact
agreement in most of the coefficients at two loops except for
$c_{IJ}^{(2,0)}$. Specifically,
\begin{align}
\frac{[\sigma v]_{(00)(00)}^{\rm Res. Sc. I}-[\sigma v]_{(00)(00)}^{\rm Res. Sc. II}}{[\sigma v]^{\rm tree}_{(00)(00)}}=&\frac{\hat \alpha_2^2}{\pi^2}\,4l_R\,\left(\varphi_{\lambda_R} - \varphi_{\kappa_R} \right)\,, \nonumber \\
\frac{[\sigma v]_{(00)(+-)}^{\rm Res. Sc. I}-[\sigma v]_{(00)(+-)}^{\rm Res. Sc. II}}{[\sigma v]^{\rm tree}_{(00)(+-)}}=&\frac{\hat \alpha_2^2}{\pi^2}\,2l_R\,\left(\varphi_{\lambda_R} + \varphi_{\kappa_R} \right)\,, \nonumber \\
\frac{[\sigma v]_{(+-)(+-)}^{\rm Res. Sc. I}-[\sigma v]_{(+-)(+-)}^{\rm Res. Sc. II}}{[\sigma v]^{\rm tree}_{(+-)(+-)}}=&-\frac{\hat \alpha_2^2}{\pi^2}\,l_R\,\left(3\varphi_{\lambda_R} + \varphi_{\kappa_R} \right)\,,
\label{eq:diffresscheme}
\end{align}
where $\varphi_{f_R}$ is defined in \eqref{eq:varphi}.
Numerically these differences are not larger than $\mathcal O(0.1\%)$
of the tree-level cross section. Note that since the single log coefficient 
is not obtained unambiguously by NLL' resummation, also the coefficients $c_{IJ}^{(2,1)}$  
could have depended on the resummation scheme, but this turns out not to 
be the case.

\section{Conclusion}
\label{sec:conclusion}

The search for high-energy photons plays an important role in detecting 
dark matter through its annihilation in the center of the Milky Way, or in 
dwarf galaxies. Connecting a possible signal to a DM model, or to place 
limits on the parameters of the model, including the DM mass itself, 
requires an accurate theoretical calculation of the annihilation rate. 
When the DM particle carries electroweak charges and its mass 
is much larger than the mass of the electroweak gauge bosons, standard 
perturbation theory in the small couplings of the SM breaks down. Large enhancements 
of loop diagrams due to non-relativistic scattering and due to soft and collinear 
gauge bosons must be summed to all orders in the coupling expansion.
In this paper we considered the photon energy spectrum of the 
semi-inclusive photon final state $\gamma+X$, integrated from the endpoint 
$E_\gamma = \mchi$ over an interval of size $\Eres$, which corresponds to 
the observable measured by $\gamma$-telescopes, when the flat integration 
in (\ref{eq:defines}) is replaced by the instrument-specific resolution  
function of characteristic width $\Eres$.

The main theoretical result is the factorization formula (\ref{eq:SIJGIJ}) 
for the annihilation rate for energy resolution $\Eres \sim m_W$ 
(\ref{eq:factformulaint}), and $\Eres\sim m_W^2/\mchi$ (\ref{eq:factformulanarrow}), 
respectively, and the calculation of the all-order resummed rate to 
NLL' accuracy in the electroweak Sudakov logarithms. The main results 
relevant to observations are summarized in Figures~\ref{fig:result}
and~\ref{fig:matching}. The corresponding result for narrow resolution 
has already been shown in \cite{Beneke:2018ssm}, but the derivation and 
the matching to the intermediate energy resolution  $\Eres \sim m_W$ 
has been presented here for the first time. While the theoretical formalism 
is more general, and so are some of the calculations, the complete 
NLL' calculation has been performed in the so-called pure wino model, 
where the SM is extended by a fermionic SU(2) triplet, of which the 
electrically neutral member is the DM particle. We highlight the following 
two observations:
\begin{itemize}
\item Electroweak Sudakov effects are large and reduce the annihilation 
rate to high-energy photons by about a factor of two in the multi-TeV 
region. As soon as the full one-loop effects are included, that is, the 
accuracy of the calculation elevated from NLL to NLL', the theoretical 
uncertainty, as measured by renormalization and factorization scale 
variation, becomes negligible (about or below 1\%), see Figure~\ref{fig:result}. 
\item The two separate calculations for narrow and intermediate energy 
resolution match very accurately, resulting in precise theoretical 
results from the line-like final state at $\Eres \sim 0$ to 
$\Eres \sim 4 m_W$ (perhaps, beyond), see Figure~\ref{fig:matching}.
While the calculations apply to any DM mass with $\mchi \gg m_W$, given 
the energy resolution of the H.E.S.S. and CTA experiments, they are most 
relevant for $\mchi$ in the range between 1 and 10~TeV. This is also 
the range where the wino model is most compelling.
\end{itemize}
In \cite{Baumgart:2017nsr,Baumgart:2018yed} a complementary approach has been 
pursued, which applies to what we called ``wide'' energy resolution 
$\Eres\gg m_W$. The available results are of NLL accuracy for the same wino 
model, and, given the observations above, it would be of interest to 
a) extend them to NLL' and b) match them to the intermediate resolution 
case discussed here.

The results shown here demonstrate the success of EFT techniques, 
non-relativistic and soft-collinear, to deal with the breakdown of 
electroweak perturbation theory in the high-energy regime. This 
opens the perspective to extend the calculations to models other than 
the wino model. Given the small uncertainty of $\leq 1\%$ from scale 
variation of the resummed perturbative expansion, it is 
probable that the largest theoretical uncertainty now arises from 
modifications of the Sommerfeld effect due to subleading effects in 
the non-relativistic effective theory, and, for smaller $\mchi$, 
from power-suppressed effects of order $m_W/\mchi$, which are 
systematically neglected in the present treatment. 

\subsubsection*{Note added}
The present arXiv version corrects a missing factor of two 
in the absolute annhilation cross section, adds an omitted 
(numerically negligible) soft function term, and fixes some typos. 
For an explicit list of errata see the JHEP 
erratum \cite{Beneke:2020erratum}. In addition, a typo 
affecting the sign in the exponent of (3.22) has been corrected.

\subsubsection*{Acknowledgement}

We thank C. Peset for careful reading of the text. 
This work has been supported in part by the 
Collaborative Research Center `Neutrinos and Dark Matter in Astro- and 
Particle Physics' (SFB 1258) and the Excellence Cluster `Universe' 
of the Deutsche Forschungsgemeinschaft. 
AB acknowledges the support by the ERC Starting Grant 
REINVENT-714788.


\appendix

\section{Hard matching coefficients}
\label{app:hardFunctions}

In this Appendix we provide more details on the calculation of the hard-matching 
coefficients given in (\ref{eq:hmatch1}) and (\ref{eq:hmatch2}).

\subsection{Amplitude in the full theory} 

The matching condition between the full theory (SM plus an isopsin-$j$ dark 
matter multiplet) and the effective theory requires that the on-shell amplitudes 
for $2\to 2$ annihilation of two dark matter fields to two SU(2) gauge bosons 
computed in the two theories must be equal:
\begin{align}
\label{eq:matching} 
\mathcal{M}^{AB}_{\text{full}}(\{p,s\}) = \frac{1}{2 m_\chi} 
\sum_{i=1,2} C^{\text{bare}}_i(\{p\}) \; (2m_\chi) 
\langle \mathcal{O}_i^{\text{bare}} \rangle^{AB} (\{p,s\})\, .
\end{align}
Here the left-hand side refers to the UV-renormalized amplitude in 
the full theory. The symbol $\{p,s\}$ indicates the dependence of the 
amplitudes on the momenta and the spin/polarization orientations of 
the four external particles.
The operators $\mathcal{O}_i$ are S-wave operators. To extract their 
coefficient we can set the relative momenta of the annihilating 
particles to zero. We choose $p_1=p_2=\mchi (1,\bm{0})$ for the 
initial state, and $p_3=\mchi n_-$, $p_4=\mchi n_+$ for the 
final state.
We define projectors applied to the full theory amplitude such that
\begin{align}
\label{eq:subamp}
\sum_{s}\mathcal{P}^{AB}_{i}(\{p,s\})\, \mathcal{M}^{AB}_{\text{full}}(\{p,s\})   
= \mathcal{M}_{i, \, \text{full}}(4 m^2_\chi)\,, \quad i=1,2\, ,
\end{align}
where $\mathcal{M}_{i, \, \text{full}}$ are the full-theory projected amplitudes
corresponding to the gauge and spin structures of the two operators 
$\mathcal{O}_{1,2}$ defined in (\ref{eq:opbasis1}) and (\ref{eq:opbasis2}). The expressions in (\ref{eq:subamp}) directly correspond to the bare matching coefficients since the loop diagrams in the effective theory are all scaleless and vanish in dimensional regularization.
The two projectors have the explicit expressions
\begin{align}
\mathcal{P}^{AB}_{1}(\{p,s\}) = \,&  \frac{1}{(3-4 c_2(j))(2 j+1)}\Bigg(\frac{1-2 c_2(j)}{2} T^{AB}_1+ T^{AB}_2\Bigg)\;\nonumber\, \\
& \times \frac{\bar{u}(p_1,s_1) (\slashed{n}_+-\slashed{n}_-) [\gamma^\sigma,\gamma^{\rho}]v(p_2,s_2) \varepsilon_{\rho}(p_3,s_3)\varepsilon_{\sigma}(p_4,s_4)}{32 \,m_\chi\, (1-3 \epsilon +2 \epsilon^2)}\, ,\nonumber \\
\mathcal{P}^{AB}_{2}(\{p,s\}) = \, &  \frac{1}{(3-4 c_2(j))(2 j+1)} \Bigg( T^{AB}_1 +\frac{-3}{c_2(j)}\, T^{AB}_2\Bigg)\, \nonumber\\
& \times \frac{\bar{u}(p_1,s_1) (\slashed{n}_+-\slashed{n}_-) [\gamma^\sigma,\gamma^{\rho}]v(p_2,s_2) \varepsilon_{\rho}(p_3,s_3)\varepsilon_{\sigma}(p_4,s_4)}{32 \, m_\chi\, (1-3 \epsilon +2 \epsilon^2)}\, ,
\label{eq:proj}
\end{align}
where $\epsilon = (4-d)/2$, $d$ is the space-time dimension and $c_2(j)= j (j+1)$ 
for an isospin-$j$ representation. The projectors differ only in the SU(2) part, 
since both operators have the same Dirac and Lorentz index structure which projects only on the spin-singlet contribution of the amplitude.
The projectors can be considered as operators in spin space and the same is 
true for the amplitude.

\begin{figure}[t]
\begin{center}
\includegraphics[width=.75\linewidth]{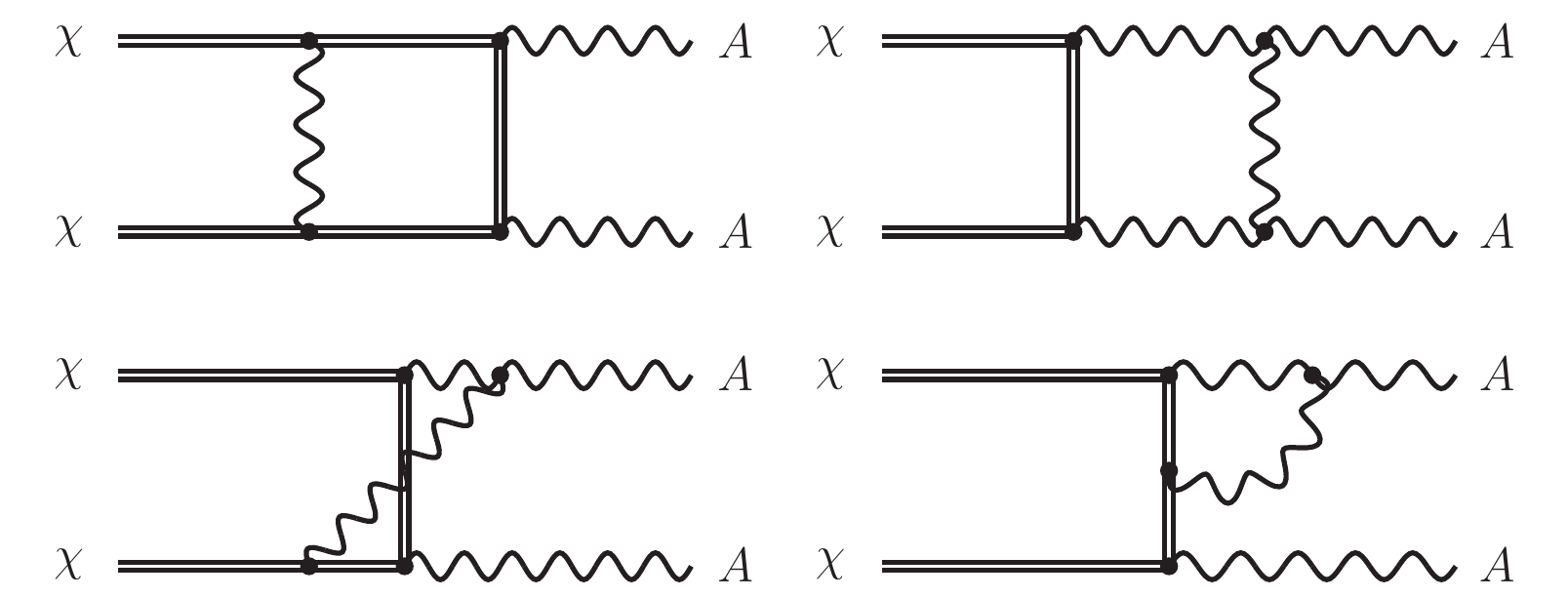}
\end{center}
\caption{\label{fig:hardMatchingDiags} Representative sample of one-loop diagrams contributing to the computation of the Wilson coefficients.}
\end{figure}

We compute the matching coefficients at the one-loop order (see 
Figure~\ref{fig:hardMatchingDiags} for a sample of diagrams). We use 
dimensional regularization for both ultraviolet and infrared singularities. 
The calculation of the bare full theory amplitudes has been carried out by using a 
set of computer-algebra tools. {\tt{FeynRules}} \cite{Alloul:2013bka}, 
{\tt{FeynArts}} \cite{Hahn:2000kx} and {\tt{FormCalc}} \cite{Hahn:1998yk} were 
used in combination for the model implementation and the amplitude generation. 
The algebraic manipulations and simplifications have been carried out with a 
private code written in {\tt{FORM}} \cite{Kuipers:2012rf}. The reduction to master integrals at threshold was performed with {\tt{Reduze}} 
\cite{vonManteuffel:2012np}. We calculate the Feynman diagrams in the 
unbroken SU(2) gauge theory and find for the 
bare projected full-theory amplitudes
\begin{align}
\mathcal{M}^{\text{bare}}_{1\, \text{full}}(4 m^2_\chi) = &\, \frac{g_{2\, \text{bare}}^4}{16 \pi^2} (4 m^2_\chi)^{-\epsilon} (e^{-\gamma_E} 4 \pi)^\epsilon\, \nonumber \\
&\times\, \Bigg[\frac{1}{\epsilon} c_2(j) (2 - 2 i\pi ) - c_2(j) \Big(4 - \frac{\pi^2}{2}\Big) + \mathcal{O}(\epsilon)\Bigg] \, ,\nonumber \\
\mathcal{M}^{\text{bare}}_{2\, \text{full}}(4 m^2_\chi) = &\, g^2_{2\, \text{bare}} + \frac{g_{2\, \text{bare}}^4}{16 \pi^2}  (4 m^2_\chi)^{-\epsilon}(e^{-\gamma_E} 4 \pi)^\epsilon\, \nonumber \\
\times &\,  \Bigg[- \frac{4}{\epsilon^2} + \frac{1}{\epsilon} (-6 + 2i \pi) + 16 - \frac{\pi^2}{6} - c_2(j) \Big(10 -\frac{\pi^2}{2}\Big) + \mathcal{O}(\epsilon)\Bigg] \, ,
\quad
\label{eq:barecoeffs}
\end{align}
where
\begin{align}
g_{2\, \text{bare}} = Z_{g_2} \tilde{\mu}^\epsilon \hat{g}_2(\mu),\quad \tilde{\mu}^2=\frac{\mu^2 e^{\gamma_E}}{4 \pi} \, .
\end{align}
We find the same expressions, both for the case of Dirac and Majorana fermions. 
This is no surprise since a possible difference could arise only  
from $s$-channel diagrams with a fermion-fermion-gauge boson vertex. At threshold 
these diagrams do not contribute to the amplitude.

In the following, we find it convenient to suppress the $\mu$ dependence of the 
renormalized SU(2) coupling in intermediate results. We remove the UV divergences 
by coupling, field and DM mass renormalization. The coupling constant is 
renormalized in the $\overline{\text{MS}}$ scheme while the mass and field 
renormalization is done in the on-shell scheme so that no further residue factor 
is required to obtain the on-shell amplitude. The SU(2) coupling, DM mass and 
field renormalization, and the SU(2) gauge boson field renormalization constants 
are given, respectively, by
\begin{align}
Z_{g_2} & = 1 + \frac{\hat{g}^2_2}{16 \pi^2} \, \frac{1}{\epsilon}\, \Big[\frac{2}{3}c(j) r-\frac{43}{12}+\frac{2}{3} n_G\Big]\, ,\label{eq:zg}\\
Z_{m_\chi} & = 1 - \frac{\hat{g}^2_2}{16 \pi^2} e^{\gamma_E \epsilon}\Gamma(1+\epsilon) \bigg(\frac{\mu^2}{m^2_\chi}\bigg)^{\epsilon} c_2(j) \frac{3-2 \epsilon}{\epsilon (1- 2 \epsilon)} \, , \\
Z_\chi & = 1 - \frac{\hat{g}^2_2}{16 \pi^2} e^{\gamma_E \epsilon}\Gamma(1+\epsilon) \bigg(\frac{ \mu^2}{m^2_\chi}\bigg)^{\epsilon} c_2(j) \frac{3-2 \epsilon}{\epsilon (1- 2 \epsilon)}\, , \\
Z_A & = 1 + \frac{\hat{g}^2_2}{16 \pi^2} \bigg(\frac{ \mu^2}{m^2_\chi}\bigg)^{\epsilon} \Big[-\frac{4}{3 \epsilon} c(j) r+\mathcal{O}(\epsilon) \Big]\, ,
\end{align}
where $c(j) = c_2(j)(2j+1)/3$ and $n_G=3$ is the number of fermion generations. 
In (\ref{eq:zg}) the term $2 c(j) r/3$ 
corresponds to the heavy DM fermion contribution, 
the term $-43/12$ to the gauge boson and Higgs contributions, while 
the $2 n_G /3$ piece arises from the SM fermion loops. The parameter $r$ 
assumes the values $r=\{1,1/2\}$ for Dirac and Majorana fermions, respectively.
In the effective theory the heavy fermion is integrated out and does not 
contribute to the running of the gauge coupling anymore.
Hence, similarly to switching between schemes with different massless quark 
flavours in QCD, we decouple the DM contribution from the running of the gauge coupling $\hat{g}_2$ through the substitution
\begin{align}
\label{eq:decoupling}
\hat{g}^2_2 \rightarrow  \hat{g}^2_2 + \frac{\hat{g}^4_2}{16 \pi^2} \Big[\frac{4}{3}c(j) r \ln \frac{\mu^2}{m^2_\chi} \Big]\, .
\end{align}
in (\ref{eq:barecoeffs}).
After this replacement the $r$ dependence drops out, and the 
final result will be independent of the Dirac or Majorana nature of the fermion. 
The UV-renormalized projected full-theory amplitudes, which 
equal the bare Wilson coefficients, read
\begin{align}
C_1^{\text{bare}} = &\, \frac{\hat{g}_{2}^4}{16 \pi^2} \Bigg\{\frac{c_2(j)}{\epsilon} (2 - 2 i\pi ) - c_2(j) \Big(4 - \frac{\pi^2}{2}\Big) 
+ c_2(j) (2 - 2 i\pi ) \ln \frac{\mu^2}{4 m^2_\chi}
+ \mathcal{O}(\epsilon)\Bigg\} \, ,\nonumber \\
C_2^{\text{bare}}= &\, \hat{g}^2_{2} + \frac{\hat{g}_{2}^4}{16 \pi^2} \,\Bigg\{- \frac{4}{\epsilon^2} + \frac{1}{\epsilon} \Big[-\frac{79}{6} + \frac{4 n_G}{3}+ 2i\pi -4 \ln \frac{\mu^2}{4 m^2_\chi} \Big] \nonumber\\
&\hspace*{0cm} 
+ 16 - \frac{\pi^2}{6}  - c_2(j) \Big(10 -\frac{\pi^2}{2}\Big) 
- (6 -2 i \pi) \ln \frac{\mu^2}{4 m^2_\chi}  -2 \ln^2 \frac{\mu^2}{4 m^2_\chi}  +  \mathcal{O}(\epsilon)\Bigg\}  \, .
\end{align}
The remaining IR divergences must be cancelled by matching. This will be done in 
the next subsection by operator renormalization in the effective theory.

\subsection{Operator renormalization in the effective theory}
\label{sec:OpRen}

In the effective theory the loop diagrams are all scaleless and therefore 
vanish in dimensional regularization. The EFT matrix element is therefore given by 
the tree matrix element and the tree diagrams with counterterm insertions. 
To compute the UV counterterms in the effective theory, we need to regulate 
the IR-divergences with a different regulator than dimensional regularization. To this purpose we take slightly off-shell momenta for the incoming \
DM fermions and the final-state gauge bosons.
By direct calculation of the effective theory diagrams we obtain for the UV poles 
in the $\overline{\text{MS}}$ scheme
\begin{eqnarray}
\langle \mathcal{O}^{\text{bare}}_1\rangle &=&
\,\langle\mathcal{O}_1\rangle^{\text{tree}}\, \,\Bigg\{1+ \frac{\hat{g}^2_{2}}{16 \pi^2} \Bigg[\frac{4}{\epsilon^2} + \frac{1}{\epsilon}\Big(4-2 c_2(j)+ 4 i \pi - 4 \ln\frac{4 m^2_\chi}{\mu^2}\Big)\Bigg] \Bigg\}  +\mathcal{O}(\epsilon^0)\, ,\qquad\;\; \\
\langle \mathcal{O}^{\text{bare}}_2\rangle &=&
\, \langle\mathcal{O}_1\rangle^{\text{tree}}\,\,\frac{\hat{g}^2_{2}}{16 \pi^2} \frac{1}{\epsilon}\Big[-c_2(j) (2- 2 i\pi) \Big] +\mathcal{O}(\epsilon^0)  \, 
\nonumber \\
&& \hspace*{-1cm}
+ \,  \langle\mathcal{O}_2\rangle^{\text{tree}}\, \Bigg\{1 + \frac{\hat{g}^2_{2}}{16 \pi^2}\Bigg[ \frac{4}{\epsilon^2} + \frac{1}{\epsilon}\Big(10 -2 c_2(j) -2 i \pi - 4 \ln\frac{4 m^2_\chi}{\mu^2}\Big)\Bigg]\Bigg\} +\mathcal{O}(\epsilon^0)\, ,\qquad
\end{eqnarray}
where $\langle\mathcal{O}_i\rangle^{\text{tree}}$ correspond to the 
tree-level matrix elements of the first or second operator.
Notice that the divergent parts shown do not depend on the infrared regulator 
and that they only depend on the hard scale $2 m_{\chi}$.
We still need to add the external field $\overline{\rm MS}$ renormalization 
factors for the effective theory fields, which read
\begin{align}
Z_{\chi_v} =&\, 1+ 
\frac{\hat{g}^2_2}{16 \pi^2} \frac{1}{\epsilon}\Big[2 c_2(j)\Big]\,,
\\
Z_{A} =&\, 1+ 
\frac{\hat{g}^2_2}{16 \pi^2} \frac{1}{\epsilon}\Big[\frac{19}{6} 
- \frac{4}{3} n_G\Big]\, .
\end{align}
By combining everything we arrive at
\begin{eqnarray}
Z_{\chi_v} Z_{A} \langle \mathcal{O}^{\text{bare}}_1\rangle &=&
\langle\mathcal{O}_1\rangle^{\text{tree}}\,\Bigg\{1+ \frac{\hat{g}^2_{2}}{16 \pi^2} \Bigg[\frac{4}{\epsilon^2} + \frac{1}{\epsilon}\bigg(\frac{43}{6} -\frac{4}{3} n_G  + 4 i \pi + 4 \ln\frac{\mu^2}{4 m^2_\chi}\bigg)\Bigg] \Bigg\} \! +\mathcal{O}(\epsilon^0)\, , 
\nonumber\\[-0.2cm]
&&\\[-0.2cm]
Z_{\chi_v} Z_{A} \langle \mathcal{O}^{\text{bare}}_2\rangle &=&
\langle\mathcal{O}_1\rangle^{\text{tree}}\,\frac{\hat{g}^2_{2}}{16 \pi^2} \frac{1}{\epsilon}\Big[- c_2(j)(2- 2 i \pi) \Big]  +\mathcal{O}(\epsilon^0) \, 
\nonumber \\
&& \hspace*{-1.5cm}+ \,  \langle\mathcal{O}_2\rangle^{\text{tree}}\, \Bigg\{1 + \frac{\hat{g}^2_{2}}{16 \pi^2} \Bigg[ \frac{4}{\epsilon^2} + \frac{1}{\epsilon}\bigg(\frac{79}{6} -\frac{4}{3} n_G -2 i \pi +4 \ln\frac{\mu^2}{4 m^2_\chi} \bigg)\Bigg]\Bigg\}\! +\mathcal{O}(\epsilon^0)\, .
\qquad
\end{eqnarray}
Coupling renormalization contributes only at 
higher orders in $\hat{g}_2$.

The $\overline{\rm MS}$ operator renormalization constants $Z_{ij}$ are a 
matrix in operator space such that $\hat{\mathcal{O}}^{\text{bare}}_i
= Z_{ij} \hat{\mathcal{O}}^{\text{ren}}_j(\mu)$, $i,j=1,2$. We obtain 
from the above
\begin{align}
Z_{11} =&\, 1 + \frac{\hat{g}^2_2}{16 \pi^2} \bigg[\frac{4}{\epsilon^2} + \frac{1}{\epsilon}\,\bigg(\frac{43}{6} -\frac{4}{3} n_G  + 4 i \pi + 4 \ln \frac{\mu^2}{4 m^2_\chi}\bigg)\bigg] \, , \nonumber \\
Z_{12} =&\, 0 \, , \nonumber \\
Z_{21} =&\, \frac{\hat{g}^2_2}{16 \pi^2}  \frac{1}{\epsilon}
\Big[- c_2(j)(2- 2 i \pi) \Big] \, , \nonumber \\
Z_{22} =&\, 1 + \frac{\hat{g}^2_2}{16 \pi^2} \bigg[ \frac{4}{\epsilon^2} + \frac{1}{\epsilon}\,\bigg(\frac{79}{6} -\frac{4}{3} n_G -2 i \pi +4 \ln \frac{\mu^2}{4 m^2_\chi}\bigg)\bigg]\label{eq:zdirect} \, .
\end{align}
By making use of the matching condition in (\ref{eq:matching}), the decoupling relation in (\ref{eq:decoupling}) and
\begin{align}
C_i(\mu) = Z_{ji} \, C^{\text{bare}}_j(\mu)\, ,\quad i=1,2\, ,
\end{align}
we find that all $1/\epsilon$ poles cancel and we obtain the 
explicit results for the hard matching coefficients given in 
(\ref{eq:hmatch1}) and (\ref{eq:hmatch2}).

\subsection{\boldmath Operator $Z$-factors from the anomalous 
dimension}

A second way to obtain the operator renormalization $Z_{ij}$ factor is to 
adapt the anomalous dimension known for QCD processes 
\cite{Ferroglia:2009ii,Beneke:2009rj} to the SU(2) gauge group. We switch to the operator basis where the DM bilinear is in a definite 
isospin representation (the DM bilinear can be either in a singlet or in a 
quintuplet representation)
\begin{align}
\label{eq:newopbasis}
\mathcal{O}^\prime = \hat{V}^T \mathcal{O},\qquad
\hat{V}=\left(\begin{array}{cc} {\displaystyle 1} \quad&\ {\displaystyle -\frac{c_2(j)}{3}} \\  
{\displaystyle 0 } \quad & \; \displaystyle{1}\end{array}\right).
\end{align}
The advantage of this basis is that the anomalous dimension at threshold 
is diagonal \cite{Beneke:2009rj},
\begin{align}
\label{eq:andim}
\Gamma = \frac{1}{2} \gamma_{\text{cusp}}\Bigg[2 c_2(\text{ad})\bigg(\ln\frac{4 m^2_\chi}{\mu^2}-i \pi\bigg)+i \pi c_2(J)\Bigg] + 2 \gamma_{\text{ad}} + \gamma^{J}_{H,s} \, ,
\end{align}
where $c_2(\text{ad})$ is the Casimir value of the gauge boson in the 
adjoint representation, and $c_2(J)$ the one for the DM fermion pair in 
the representation $J=0$ (singlet) or $J=2$ (quintuplet).
The quantity $\gamma_{\text{ad}}$ is the gauge boson anomalous dimension 
and $ \gamma^{J}_{H,s}$ is the anomalous dimension of the heavy fermion pair.
The anomalous dimensions have perturbative expansions in terms of 
$\hat{\alpha}_2$ (and, possibly, other couplings in higher orders 
than given)
\begin{align}
\gamma_{\text{cusp}}(\hat{\alpha}_2) &= \gamma_{\text{cusp}}^{(0)} 
\,\frac{\hat{\alpha}_2}{4\pi} 
+ \gamma_{\text{cusp}}^{(1)} \left( \frac{\hat{\alpha}_2}{4\pi} \right)^2 
+ \mathcal{O} (\hat{\alpha}_2^3) \,, 
\\ 
\gamma^{(0)}_{\text{cusp}} &= 4, \qquad
\gamma^{(1)}_{\text{cusp}} = \left(\frac{268}{9} - 
\frac{4 \pi^2}{3}\right)c_2(\text{ad}) - \frac{80}{9}n_G - \frac{16}{9}  \,,
\\
\gamma_{\text{ad}}(\hat{\alpha}_2) &= \gamma^{(0)}_{\text{ad}} 
\,\frac{\hat{\alpha}_2}{4\pi} + \mathcal{O} (\hat{\alpha}_2^2) \, , \\
\gamma^{(0)}_{\text{ad}} &= - \beta_{0, \text{SU(2)}} = 
- \left( \frac{43}{6} - \frac{4}{3}n_G \right) ,
\\
\gamma^J_{H,s}(\hat{\alpha}_2) &= \gamma^{(0)}_{H,s}\, c_2(J) 
\,\frac{\hat{\alpha}_2}{4\pi} + \mathcal{O} (\hat{\alpha}_2^2) \, , \\
\gamma^{(0)}_{H,s} &= -2 \,.
\end{align}
The Higgs contribution $-16/9$ to the two-loop cusp anomalous dimension has 
been extracted from the $\epsilon$-scalar contribution computed 
in \cite{Broggio:2015dga}.

The operator $Z$-factor in the $\overline{\rm MS}$ scheme 
can be obtained from the anomalous dimension. 
Up to order $\hat{g}^2_2$ it reads
\begin{align}
Z = 1 - \frac{\hat{g}^2_2}{(16 \pi^2)}
\bigg(\frac{\Gamma^{\prime\,(0)}}{4 \epsilon^2}
+ \frac{\Gamma^{(0)}}{2 \epsilon}\bigg) + 
\mathcal{O}(\hat g^4_2)\, , 
\end{align}
where $\Gamma^\prime = -2 c_2(\text{ad}) \gamma_{\text{cusp}}$.
In the diagonal basis defined in (\ref{eq:newopbasis}) we find 
\begin{align}
Z_{11} =&\, 1 + \frac{\hat{g}^2_2}{16 \pi^2} \bigg[\frac{4}{\epsilon^2}+ \frac{1}{\epsilon}\, \bigg(\frac{43}{6}-\frac{4}{3} n_G+ 4 \ln \frac{\mu^2}{4 m^2_\chi} +4 i \pi\bigg)\bigg] \, , \nonumber \\
Z_{12} =&\, 0 \, , \nonumber \\
Z_{21} =&\, 0 \, , \nonumber \\
Z_{22} =&\, 1 + \frac{\hat{g}^2_2}{16 \pi^2} \bigg[\frac{4}{\epsilon^2}+ \frac{1}{\epsilon}\, \bigg(\frac{79}{6}-\frac{4}{3} n_G+ 4 \ln\frac{\mu^2}{4 m^2_\chi}-2 i \pi\bigg)\bigg] \, .
\end{align}
Transforming back to the unprimed basis (\ref{eq:opbasis1}), 
(\ref{eq:opbasis2}) we find agreement with (\ref{eq:zdirect}).


\section{Collinear functions and rapidity regularization}

In this Appendix provide some details on the rapidity regularization, 
which is required for collinear and soft functions, on collinear 
integrals, and we supply the lengthy expressions for the narrow 
resolution jet function that were not given in \cite{Beneke:2018ssm}.

\subsection{Rapidity regularization}
\label{app:rapreg}
We employ the rapidity regulator introduced in \cite{Chiu:2012ir}, which amounts 
to the following replacements in the eikonal Feynman rules that originate from 
soft and (anti-) collinear Wilson lines
\begin{align}
\text{collinear emission}: \, \frac{n_+^\mu}{n_+ k}\,
&\to\,\frac{n_+^\mu}{n_+k}\,\frac{\nu^{\eta}}{|n_+ k|^{\eta}}\, ,
\qquad\\
\text{anti-collinear emission}: \, \frac{n_-^\mu}{n_- k}\,
&\to\,\frac{n_-^\mu}{n_-k}\, \frac{\nu^{\eta}}{ |n_- k|^{\eta}} \, ,\\
\text{soft emission from (anti-) collinear direction}: \, \frac{n_\pm^\mu}{n_\pm k}\,
&\to\,\frac{n_\pm^\mu}{n_\pm k}\, \frac{\nu^{\eta/2}}{ |2 k^3|^{\eta/2}}\, , \\
\text{soft emission from the heavy line}: \, \frac{v^\mu}{v\cdot k}\,
&\to\,\frac{v^\mu}{v\cdot k}\,\frac{\nu^{\eta/2}}{ |2 k^3|^{\eta/2}}\, .
\end{align}
$\eta$ is the rapidity regulator and $\nu$ is a newly introduced rapidity 
scale, the equivalent of $\mu$ in dimensional regularization.
Notice that the rapidity regulator in the above expressions is consistent 
between soft and collinear integrals since $2k^3\to n_+k$ ($2k^3\to n_-k$) 
in the (anti-) collinear limit. In our case all the soft and 
photon jet functions always require rapidity regularization, but the 
unobserved-jet function only in the narrow resolution case.
In the following we focus on the collinear and anti-collinear scalar 
integrals, which appear, respectively, in the unobserved jet and photon jet 
function calculation. In Appendix \ref{app:softFunction} we present 
the computation of the relevant soft virtual and real integrals.

As an example we compute the off-shell collinear scalar integral ($p^2\neq 0$) 
\begin{align}
\label{eq:collintp2}
I_c(p^2) = \int [dk] \frac{\nu^\eta}{[k^2-m^2_W+i \varepsilon] 
[(p+k)^2-m^2_W+i \varepsilon] [n_+ k+i\varepsilon] |n_+ k|^{\eta}} \, ,
\end{align}
relevant to the unobserved-jet function in the narrow resolution 
regime. From contour integration we find that $n_+ k < 0$, such that we can 
replace the absolute value by $- n_+ k$.
We proceed with the loop integration by introducing the Feynman 
parametrization 
\begin{align}
\frac{1}{abc^{1+\eta}} = \int^{\infty}_0 dx_1 \, \int^{\infty}_0 d x_2\, \frac{(2+\eta) (1+\eta)}{(c+a x_1 + b x_2)^{3+\eta}}\, .  
\end{align}
After performing the loop integration we arrive at
\begin{eqnarray}
I_c(p^2) &=&\big(\mu^2 e^{\gamma_E}\big)^\epsilon \frac{(-i)}{16 \pi^2} \frac{\Gamma (1+\eta+\epsilon)}{\Gamma(1+\eta)} \, \nonumber \\
&& 
\hspace*{-1cm}\times \int^{\infty}_0 dx_1 \, \int^{\infty}_0 d x_2 \,
\frac{\nu^\eta}{ (x_1+x_2)^{1-\eta-2 \epsilon}\, [(x_1+x_2)^2 m^2_W-x_2 x_1 p^2+n_+ p x_2]^{1+\epsilon+\eta}}\, .\qquad
\label{eq:colint}
\end{eqnarray}
We make the substitution $x_1\to x^\prime_1 x_2$ and integrate first over $x_2$.
For convenience we drop the $+i \varepsilon$ in the intermediate expressions, and we identify $p^2 \to p^2+i \varepsilon$ as follows from the definition of the integral.
After integrating over $x_2$ we obtain
\begin{align}
    I_c(p^2) =\left(\mu^2 e^{\gamma_E}\right)^{\epsilon} \frac{(-i)}{16 \pi^2} \left(\frac{\nu}{n_+ p}\right)^{\!\eta} \, \frac{\Gamma(\epsilon)}{n_+ p}\,\int^{\infty}_0  dx^\prime_1\,  (1+x^\prime_1)^{-1+2\epsilon+\eta}\, \big[m^2_W (1+x^\prime_1)^2 -p^2 x^\prime_1\big]^{-\epsilon} .
\end{align}
We rewrite part of the integrand as
\begin{equation}
\big[m^2_W (1+x^\prime_1)^2 -p^2 x^\prime_1\big]^{-\epsilon} 
= (-p^2)^{-\epsilon} \bigg[r \,(1+x^\prime_1)^2+x^\prime_1\bigg]^{-\epsilon}\, ,
\end{equation}
where $r\equiv m^2_W/(-p^2)$, and perform the variable change 
$x^\prime_1=(1-y)/y$. The $y$ integration amounts to 
\begin{align}
\int^1_0 dy\, y^{-1-\eta} (r+y-y^2)^{-\epsilon} = -\frac{r^{-\epsilon}}{\eta} F_1\bigg(-\eta,\epsilon,\epsilon,1-\eta;\frac{2}{1+\sqrt{1+4 r}},\frac{2}{1-\sqrt{1+4 r}}\bigg)\, ,
\label{eq:appellf1}
\end{align}
where $F_1$ is the $F_1$-Appell hypergeometric function.
This gives $I_c(p^2)$ to all orders in $\eta$ and $\epsilon$. Numerical checks were 
performed to ensure the consistency of (\ref{eq:colint}) and (\ref{eq:appellf1}). 
We need to expand the result first in $\eta\to 0$, using the formula
\begin{align}
y^{-1-\eta} &= -\frac{\delta(y)}{\eta} + \sum_{m=0}^{\infty} \frac{(-\eta)^m}{m!}\bigg[\frac{\ln^m(y)}{y}\bigg]_+ = -\frac{\delta(y)}{\eta} + \bigg[\frac{1}{y}\bigg]_+ + \ldots\, 
\label{eq:eta_exp}
\end{align}
and then in $\epsilon$. The $+$-distributions acting on a test function are defined as
\begin{align}
    \int_0^1 d x \left[\frac{\ln^n(x)}{x}\right]_+ f(x) = \int_0^1 d x \, \frac{\ln^n(x)}{x} \left(f(x) - f(0)\right) \, .
\end{align}
For the $y$-integral \eqref{eq:appellf1} we find 
\begin{align}
\int^1_0 dy\, y^{-1-\eta} (r+y-y^2)^{-\epsilon} = -\frac{r^{-\epsilon}}{\eta} 
+ \epsilon \,\bigg[-\frac{\ln^2(-x)}{2}\bigg] 
+\mathcal{O}(\eta, \epsilon^2) \, .
\label{eq:y_integr}
\end{align}
We find this compact result after introducing the variables
\begin{align}
x \equiv \frac{1-\beta}{1+\beta}\,, \quad \quad 
\beta = \sqrt{1-\frac{4 m^2_W}{p^2}} \, , \label{eq:betax}
\end{align}
and with the help of relations between polylogarithms of different arguments in intermediate 
steps.  We used the package {\tt{NumExp}} \cite{Huang:2012qz} for the numerical 
expansion of the Appell $F_1$
function in (\ref{eq:appellf1}) to check (\ref{eq:y_integr}). In total we find
\begin{align}
\label{eq:collp2}
I_c(p^2) &= \frac{i\, (n_+ p)^{-1}}{16 \pi^2}\bigg[\frac{1}{\epsilon \eta} 
-\frac{1}{\eta} \ln\frac{m^2_W}{\mu^2} -\frac{1}{\epsilon} \ln\frac{n_+ p}{\nu} 
+ \ln\frac{m^2_W}{\mu^2}\, \ln\frac{n_+ p}{\nu}\nonumber \\
& + \frac{\ln^2(-x)}{2}\bigg] 
+ \mathcal{O}(\eta,\epsilon) \, .
\end{align}
The integral above is real in the region $p^2<0$, but we need to extract potential imaginary parts in the regions $p^2>4m^2_W$ and $0<p^2<4 m^2_W$.
To obtain the result in the region $p^2>4m^2_W$ from (\ref{eq:collp2}) one needs to perform the substitution
\begin{align}
\ln(-x)\to \ln(x) +i \pi \, .
\end{align}
In the region $0<p^2<4 m^2_W$ the result does not develop an imaginary part and 
it can be obtained from (\ref{eq:collp2}) by making the substitution
\begin{align}
\ln(-x)\to i \big(- 2 \arctan(\bar{\beta}) + \pi \big) \, ,
\end{align}
where we define
\begin{align}
    \bar{\beta} = \sqrt{\frac{4 m^2_W}{p^2}-1}\, . \label{eq:betabar}
\end{align}

In the intermediate resolution case the external momentum $p$ 
has hard-collinear scaling $p^\mu\sim m_\chi (1,\lambda,\sqrt{\lambda})$ such 
that $p^2\sim \lambda m^2_\chi$, while the square of the gauge boson mass 
scales as $m^2_W\sim \lambda^2 m^2_\chi$. Hence the expansion for 
$p^2 \gg m^2_W$ becomes relevant.  Directly expanding the result in 
(\ref{eq:collp2}) yields 
\begin{align}
\label{eq:collexp}
I_c(p^2) = \frac{i\, (n_+ p)^{-1}}{16 \pi^2}\bigg[&\frac{1}{\epsilon \eta} -\frac{1}{\eta} \ln\frac{m^2_W}{\mu^2} -\frac{1}{\epsilon} \ln\frac{n_+ p}{\nu} + \ln\frac{m^2_W}{\mu^2}\, \ln\frac{n_+ p}{\nu}\nonumber \\
& + \frac{1}{2} \ln^2\bigg(-\frac{m^2_W}{p^2}\bigg)\bigg] + 
\mathcal{O}(\eta, \epsilon) + \mathcal{O}\bigg(\frac{m^2_W}{p^2}\bigg)
\end{align}
up to power corrections, which seems to be at variance with the gauge-boson 
mass independence of the result for the hard-collinear jet 
function in the main text. However, the integral (\ref{eq:collintp2}) is 
now a two-scale object. We find that there are two regions contributing to this 
integral, namely the hard-collinear and the soft region, 
$k^\mu\sim (\lambda,\lambda,\lambda)$.  To extract the soft contribution, we 
need to expand the propagator 
\begin{align}
\big[(p+k)^2-m^2_W\big] = p^2+ n_+ p \, n_- k + \mathcal{O}(\lambda^2)\, .
\end{align}
in this region. The soft integral then reads
\begin{align}
\label{eq:collsoft}
I_{c\text{-}s}(p^2) = \int [dk] \frac{\nu^\eta}{[k^2-m^2_W+i \varepsilon] [p^2+ n_+ p \, n_- k+i \varepsilon] [n_+ k+i\varepsilon] |n_+ k|^{\eta}} \, .
\end{align}
Calculating the integral in a similar way as above, we obtain
\begin{align}
I_{c\text{-}s}(p^2) &=   \frac{i\, (n_+ p)^{-1}}{16 \pi^2}\bigg[ -\frac{1}{\epsilon^2}+\frac{1}{\epsilon \eta} +\frac{\pi^2}{12} + \frac{1}{\epsilon}\ln \frac{m^2_W}{\mu^2} + \frac{1}{\epsilon} \ln\bigg(\frac{-p^2 \nu}{m^2_W n_+ p}\bigg)- \frac{1}{\eta}\ln \frac{m^2_W}{\mu^2} \nonumber \\
                    &-\frac{1}{2} \ln^2 \frac{m^2_W}{\mu^2} -\ln\frac{m^2_W}{\mu^2}\, \ln \bigg(\frac{-p^2 \nu}{m^2_W n_+ p}\bigg)\bigg] + \mathcal{O}(\eta, \epsilon)\, .
\label{eq:collsoft2}
\end{align}
The rapidity regulator is only needed in the soft contribution and not in the 
hard-collinear one. In the hard-collinear region we can drop the gauge boson 
mass at leading power,  and the integral evaluates to
\begin{align}
\label{eq:hardcoll}
I_{c\text{-}hc}(p^2) = \frac{i\, (n_+ p)^{-1}}{16 \pi^2}\bigg[\frac{1}{\epsilon^2} +\frac{1}{\epsilon} \ln\bigg(-\frac{\mu^2}{p^2}\bigg)+ \frac{1}{2} \ln^2\bigg(-\frac{\mu^2}{p^2}\bigg)-\frac{\pi^2}{12} \bigg] + \mathcal{O}(\epsilon)\, .
\end{align}
By adding up the two contributions (\ref{eq:collsoft2}) and  (\ref{eq:hardcoll}) 
we reproduce (\ref{eq:collexp}). After dressing the collinear-soft scalar integral 
(\ref{eq:collsoft}) with the proper tree-level factors to obtain the soft 
contribution to the jet function Wilson line diagram, and after taking its 
imaginary part, we find that the virtual (single-particle cut) piece evaluates to a 
scaleless integral while the real emission (two-particle cut) piece is  
non-vanishing. It can be shown by direct comparison that this last term equals the soft emission diagram in (\ref{eq:softnpnn}) after the convolution with the 
tree-level jet function has been done.
This shows that in the small mass limit, $m^2_W \ll p^2$, the soft region 
of the jet function integral is correctly reproduced by the soft function in the 
factorization formula for intermediate resolution and should not be assigned 
to a mass-dependent collinear function. The hard-collinear region only
contributes the mass-independent unobserved-jet function. 

The photon jet function also requires rapidity regularization. In this case only 
virtual diagrams contribute. As an example, we compute the 
rapidity and dimensionally regulated on-shell anti-collinear 
scalar integral 
\begin{align}
I_{\bar{c}}(0)  = \int [dk] \frac{\nu^\eta}{[k^2-m^2_W+i \varepsilon] [k^2 + 2 p\cdot k-m^2_W+i \varepsilon] [n_- k+i\varepsilon] |n_- k|^{\eta}} \, ,
\end{align}
which can be obtained from (\ref{eq:collintp2}) by setting $p^2=0$ and 
replacing  $n_+k\to n_-k$. We parametrize the integration measure by
\begin{align}
d^d k = \frac{1}{2}\,  d n_-k\, dn_+k \, d^{d-2}k_\perp\, ,
\end{align}
and rewrite the integrand as
\begin{align}
\frac{\nu^\eta}{n_-k[n_+k - \frac{k^2_T+m^2_W-i\varepsilon}{n_-k}] (n_-k+2 m_\chi) [n_+k-\frac{k^2_T+m^2_W-i \varepsilon}{n_-k+2 m_\chi}][n_-k+i\varepsilon] |n_-k|^\eta}.
\end{align}
We perform the $n_+ k$ integral first by closing the contour in the upper half 
plane and pick the pole $(k^2_T+m^2_W-i \varepsilon)/n_-k$ for $-2 m_\chi < n_-k< 0$. 
The integral vanishes for $n_- k$ outside this range. After performing the $n_- k$ 
and $k_\perp$ integrals, we obtain the final result
\begin{align}
I_{\bar{c}}(0) = \frac{i}{32 \pi^2 m_\chi} \, \left(\frac{\mu}{m_W}\right)^{\!\!2\epsilon} 
\left(\frac{\nu}{2 m_\chi}\right)^{\!\!\eta}  
\frac{e^{\gamma_E \epsilon}\,  \Gamma(\epsilon)}{\eta}\, .
\end{align}

\subsection{Unobserved-jet function for the narrow resolution case}
\label{app:recjetfnnarrow} 

Here we supply the lengthy expressions for the narrow 
resolution jet function that were not given in \cite{Beneke:2018ssm}.
The jet function of the unobserved final state in the narrow resolution regime 
$E^\gamma_{\text{res}}\sim m^2_W/m_\chi$ is defined as
\begin{align}
(-g_{\perp\, \mu \nu}) J^{BC}(p^2)\equiv \frac{1}{2\pi} \int d^4x \,e^{ip\cdot x}\langle 0|\mathcal{A}^B_{\perp \,c,\mu}(x) \,\mathcal{A}_{\perp\,c,\nu}^C(0) |0\rangle \, .
\end{align}
This is equivalent to computing the total discontinuity
\begin{align}
J^{BC}(p^2) = \frac{1}{\pi}\text{Im}\big[i\mathcal{J}^{BC}(p^2) \big]
\end{align}
of the gauge boson two-point function
\begin{align}
(-g_{\perp\, \mu \nu}) \mathcal{J}^{BC}(p^2) \equiv \int d^4x \,e^{ip\cdot x}\langle 0|\mathbf{T}\big\{\mathcal{A}^B_{\perp\,c, \mu}(x) \,\mathcal{A}_{\perp\,,c \nu}^C(0)
\big\} |0\rangle \, ,
\end{align}
where $\mathcal{A}^B_{\perp\, \mu}$ is the collinear gauge-invariant collinear 
building block of SCET. While formally the definition appears the same 
as (\ref{eq:jetfndef}) for the intermediate resolution, in the present case 
$p^2\sim m_W^2$ rather than $p^2\gg m_W^2$. The implication of this difference 
for the computation of the collinear integrals have been discussed in the 
previous subsection.

\begin{figure}[t]
\centering
\includegraphics[width=0.8\textwidth]{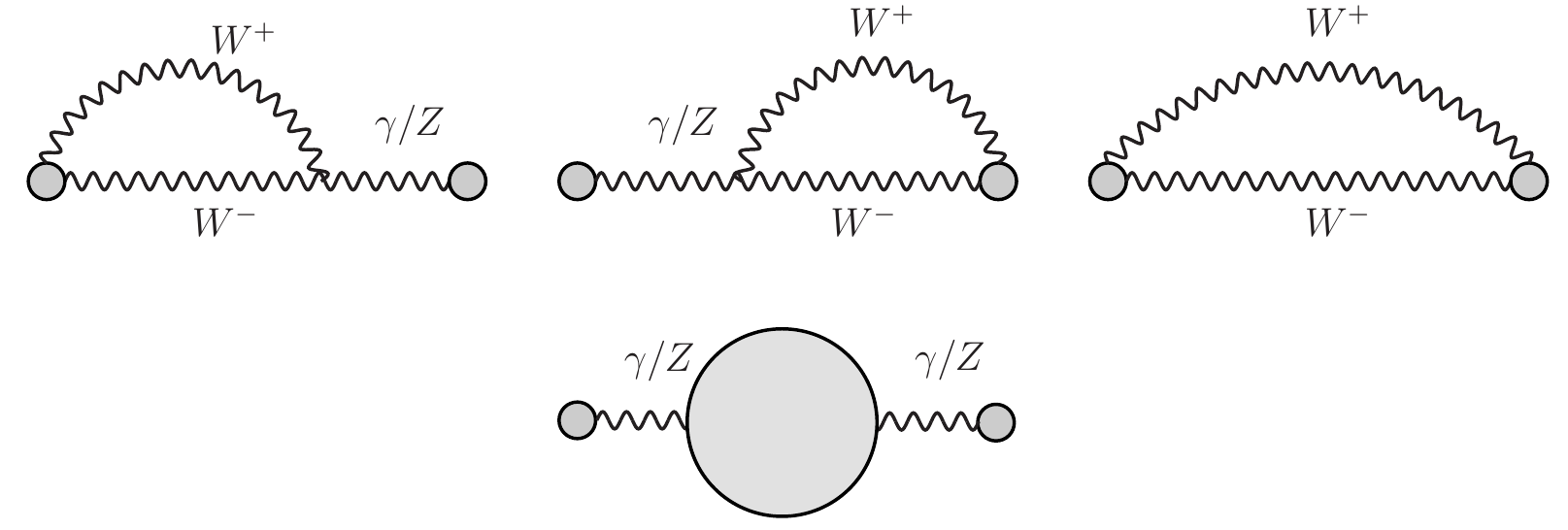} 
\caption{Wilson line and self-energy type Feynman diagrams contributing to the 
narrow resolution jet function. 
\label{fig:jetfun}}
\end{figure}

Since we are considering the $\chi^0\chi^0$ , $\chi^+\chi^-$ initial states and 
since we require a single photon in the anti-collinear final state, electric 
charge conservation implies that we only need to calculate the 33 component of 
$J^{BC}$. To the one loop-order, we can write $J^{33}(p^2)$ as
\begin{align}
\label{eq:J33nrw} 
J^{33}(p^2,\mu,\nu) = \hat{s}^2_W(\mu) \delta(p^2)+\hat{c}^2_W(\mu) 
\delta(p^2-m^2_Z)+J^{33}_{\text{Wilson}} (p^2,\mu,\nu)
+J^{33}_{\text{se}} (p^2,\mu) \, .
\end{align}
where we split the result into Wilson line and a self-energy type contributions 
as shown in the first and second line of Figure~\ref{fig:jetfun}, respectively .
Only the Wilson line diagrams require rapidity regularization.
After subtracting both dimensional and rapidity regularization poles their 
sum is given by
\begin{align}
\label{eq:J33nrwWilson}
J^{33}_{\text{Wilson}}&(p^2,\mu,\nu) =  - \frac{\hat{s}^{2}_W(\mu) \hat{g}^2_2(\mu)}{16 \pi^2} \bigg\{ \delta(p^2)\bigg[ - 16\ln\frac{m_W}{\mu} \ln\frac{2 m_\chi}{\nu}+ 8\ln\frac{m_W}{\mu}\bigg] \, \nonumber \\
&+ \frac{1}{p^2} \theta(p^2-4 m^2_W)\bigg[ 4 \beta + 8\ln\frac{1-\beta}{1+\beta} \bigg]\bigg\} \, \nonumber \\
&  - \frac{\hat{c}^{2}_W(\mu) \hat{g}^2_2(\mu)}{16 \pi^2} \bigg\{ \delta(p^2-m^2_Z) \bigg[- 16\ln\frac{m_W}{\mu} \ln\frac{2 m_\chi}{\nu} + 8\ln\frac{m_W}{\mu}- 8 + 4\pi^2   \, \nonumber\\
&+ 4 \pi \bar{\beta}_Z  - (16 \pi+8 \bar{\beta}_Z)\arctan(\bar{\beta}_Z)  + 16 \arctan^2(\bar{\beta}_Z)\bigg] \, 
\nonumber \\
&+ \frac{1}{p^2-m^2_Z} \,\theta(p^2-4 m^2_W)
\bigg[ 4 \beta + 8\ln\frac{1-\beta}{1+\beta} \bigg]\bigg\} \, ,
\end{align}
where
\begin{align}
\beta=\sqrt{1-\frac{4 m^2_W}{p^2}}, \quad\quad \bar{\beta}_Z = \sqrt{\frac{4\, m^2_W}{m^2_Z}-1} \, .
\end{align}
The self-energy contribution $J^{33}_{\text{se}} (p^2,\mu)$ is expressed in terms 
of standard one-loop gauge-boson self-energies which can be found 
in \cite{Denner:1991kt} in the Feynman gauge. We take the fermions to be massless 
except for the top quark. Hence we further separate the massless 
fermion contribution from the massive contributions,
\begin{align}
J^{33}_{\text{se}} (p^2,\mu) = J^{33}_{\text{se},\, f\neq t\, \text{only}}(p^2,\mu) + J^{33}_{\text{se},\, f\neq t\, \text{excluded}}(p^2,\mu)\, ,
\end{align}
where the second term includes the $W^+W^-$, $ZH$ and $t \bar{t}$ loops.
For the massless fermion contribution we obtain 
\begin{align}
\label{eq:J33nrwsefneqtonly}
J^{33}_{\text{se},\, f\neq t\, \text{only}}(p^2,\mu) &= \frac{\hat{s}^{2}_W(\mu) \hat{g}^2_2(\mu)}{16 \pi^2} \Bigg\{ \hat{s}^2_W(\mu )  \frac{80}{9} \bigg[-\delta(p^2) \frac{5}{3}  + \bigg[\frac{1}{p^2}\bigg]^{[\mu^2]}_*\bigg]\nonumber \\
& + 2  \,\bigg(\frac{10}{3} -\frac{80}{9} \hat{s}^2_W(\mu) \bigg) \nonumber \times \bigg[\bigg[\frac{1}{p^2-m^2_Z}\bigg]_* - \delta(p^2-m^2_Z)\nonumber \bigg( \frac{5}{3}-\ln\frac{m^2_Z}{\mu^2}\bigg)\bigg] \nonumber \\
&+  \bigg(-\frac{20}{3} +\frac{7}{2}\frac{1}{\hat{s}^2_W(\mu)}+\frac{80}{9}\hat{s}^2_W(\mu)\bigg) \nonumber \\
&\times \bigg[\bigg[\frac{1}{(p^2-m^2_Z)^2}\bigg]_{**} p^2 - \bigg(\frac{2}{3}-\ln\frac{m^2_Z}{\mu^2}\bigg) \delta(p^2-m^2_Z)\bigg]\Bigg\} \, .
\end{align}
The star distributions are defined as
\begin{eqnarray}
\label{eq:stardistdef}
&&
\int_0^{p_{\text{max}}^2} d p^2 \left[\frac{\ln^n \frac{p^2}{\mu^2}}{p^2}\right]_*^{[\mu^2]} f(p^2)= \nonumber \, \\
&&\hspace*{3cm}\,\int_0^{p_{\text{max}}^2} d p^2 \, \frac{f(p^2)-f(0)}{p^2} \ln^n \frac{p^2}{\mu^2} + \frac{f(0)}{n+1}\, \ln^{n+1} \frac{p^2_{\text{max}}}{\mu^2} \, ,
\\
\label{eq:Zstar}
&& \int^{p^2_{\text{max}}}_0 dp^2\,  
\bigg[\frac{1}{p^2-m^2_Z}\bigg]_*  f(p^2) =
\nonumber \,\\ 
&&\hspace*{3cm}
\,\int^{p^2_{\text{max}}}_0 dp^2 \, \frac{f(p^2)-f(m^2_Z)}{p^2-m^2_Z} + f(m^2_Z)\ln \bigg(\frac{p^2_{\text{max}}-m^2_Z}{m^2_Z}\bigg)\, ,
\\
&&\int^{p^2_{\text{max}}}_0 dp^2 \, 
\bigg[\frac{1}{(p^2-m^2_Z)^2}\bigg]_{**} f(p^2)  = 
\int^{p^2_{\text{max}}}_0 dp^2 \, 
\frac{\big[f(p^2)-f(m^2_Z)-(p^2-m^2_Z)f^\prime(m^2_Z)\big]}{(p^2-m^2_Z)^2} 
\nonumber \\
&& 
\label{eq:Zdoublestar}
\hspace*{3cm} 
- \,f(m^2_Z)\bigg(\frac{1}{m^2_Z}+\frac{1}{p^2_{\text{max}}-m^2_Z}\bigg) +f^\prime(m^2_Z) \ln\bigg(\frac{p^2_{\text{max}}-m^2_Z}{m^2_Z}\bigg)\, , 
\end{eqnarray}
where $f(p^2)$ is a test function and $p^2_{\text{max}}>m^2_Z$ 
in the last two equations. For 
$p^2_{\text{max}}<m^2_Z$ the introduction of star distributions for the 
$Z$-boson propagators is not necessary. The above expressions diverge as 
$p^2_{\text{max}} \to m_Z^2$, see Appendix~\ref{app:zpole} for the 
treatment of the $Z$ resonance. 
The massive piece instead reads
\begin{align}
\label{eq:J33nrwsefneqtexc}
&\hspace*{-1cm} J^{33}_{\text{se},\, f\neq t\, \text{excluded}}(p^2,\mu) = \nonumber \, \\
&2 \hat{s}_W(\mu) \hat{c}_W(\mu) \bigg[\frac{\text{Re}\big[\Sigma^{\gamma Z}_T(0)\big]_{t,W}}{m^2_Z}\delta(p^2) - \frac{\text{Re}\big[\Sigma^{\gamma Z}_T(m^2_Z)\big]_{t,W}}{m^2_Z} \delta(p^2-m^2_Z)\bigg] \nonumber \\
-& \hat{s}^2_W(\mu) \text{Re}\frac{\partial \Sigma^{\gamma \gamma}_T(p^2)_{t,W}}{\partial p^2}\bigg|_{p^2=0} \delta(p^2) - \hat{c}^2_W(\mu) \text{Re}\frac{\partial \Sigma^{ZZ}_T(p^2)_{t,W,Z,H}}{\partial p^2}\bigg|_{p^2=m^2_Z}\delta(p^2-m^2_Z) \nonumber \\
+& 2 \hat{s}_W(\mu) \hat{c}_W(\mu) \bigg[-\frac{1}{m^2_Z}\frac{1}{p^2} \frac{\text{Im}\big[\Sigma^{\gamma Z}_T(p^2)\big]_{t,W}}{\pi} +\frac{1}{m^2_Z}\frac{1}{p^2-m^2_Z} \frac{\text{Im}\big[\Sigma^{\gamma Z}_T(p^2)\big]_{t,W}}{\pi}\bigg] \nonumber \\
& + \hat{s}^2_W(\mu) \frac{1}{\big(p^2\big)^2} \frac{\text{Im} \big[\Sigma^{\gamma \gamma}_T(p^2)\big]_{t,W}}{\pi} + \hat{c}^2_W(\mu) \frac{1}{\big(p^2-m^2_Z\big)^2} \frac{\text{Im} \big[\Sigma^{Z Z}_T(p^2)\big]_{t,W,Z,H}}{\pi} \, .
\end{align}
In this case we do not need to introduce star distributions, 
because the imaginary parts vanish below the massive thresholds 
indicated by the subscripts, and hence there are no singularities 
at $0$ and $m^2_Z$. 

For convenience we collect below the explicit expressions for the 
gauge-boson self energies and their derivatives in the Feynman 
gauge. Their transverse parts are taken from \cite{Denner:1991kt}. 
The derivatives have been computed in a straightforward way.
\begin{align}
\Sigma^{\gamma \gamma}_{T}(p^2)  
&=   - \frac{\hat{g}^2_2 \hat{s}^2_W}{16\pi^2} \Bigg\{
\frac{2}{3} \sum_{f,i} N_{C}^{f}2Q_{f}^{2}
\Big[-
(p^{2}+2m_{f,i}^{2})B_{0}(p^{2},m_{f,i},m_{f,i}) \nonumber \\
&+2m_{f,i}^{2}B_{0}(0,m_{f,i},m_{f,i}) +\frac{1}{3}p^{2} \Big] \nonumber \\
&+ \bigg\{
\Big[3 p^{2} + 4m_{W}^{2}\Big] B_{0}(p^{2},m_{W},m_{W})
- 4m_{W}^{2}B_{0}(0,m_{W},m_{W}) \bigg\}
\Bigg\},\label{eq:SigmaGG}
\end{align}
\begin{align}
\frac{\partial\, \Sigma^{\gamma \gamma}_{T}(p^2)}{\partial p^2}\bigg|_{p^2\to 0}  &=   - \frac{\hat{g}^2_2 \hat{s}^2_W}{16\pi^2} \Bigg\{
\frac{2}{3} \sum_{f,i} N_{C}^{f}2Q_{f}^{2}
\Big[- B_{0}(p^2,m_{f,i},m_{f,i})\nonumber \\ &
-(p^2+2m_{f,i}^{2})\frac{\partial \, B_{0}(p^2,m_{f,i},m_{f,i})}{\partial p^2} +\frac{1}{3} \Big] \nonumber \\
&+ \bigg\{
3 B_{0}(p^2,m_{W},m_{W}) + (3p^2+4m_{W}^{2}) \frac{\partial \, B_{0}(p^2,m_{W},m_{W})}{\partial p^2} \bigg\}
\Bigg\}\Bigg|_{p^2\to 0} \,,
\end{align}
\begin{align}
\label{eq:sigmagammaZ}
\Sigma^{\gamma Z}_{T}(p^2)  &=   - \frac{\hat{g}^2_2 \hat{s}^2_W}{16\pi^2} \Bigg\{
\frac{2}{3} \sum_{f,i} N_{C}^{f}(-Q_{f})\Big(\hat{g}_{f}^{+}+\hat{g}_{f}^{-}\Big)
\Big[-
(p^{2}+2m_{f,i}^{2})B_{0}(p^{2},m_{f,i},m_{f,i}) \nonumber\\
&+2m_{f,i}^{2}B_{0}(0,m_{f,i},m_{f,i}) +\frac{1}{3}p^{2} \Big] \nonumber \\
&+\frac{1}{3\hat{s}_{W}\hat{c}_{W}} \bigg\{
\Big[\left(9\hat{c}_{W}^{2} + \frac{1}{2}\right) p^{2} + (12\hat{c}_{W}^{2} + 4) m_{W}^{2}\Big]
B_{0}(p^{2},m_{W},m_{W}) \nonumber \\
& -(12\hat{c}_{W}^{2} - 2) m_{W}^{2}B_{0}(0,m_{W},m_{W})
+ \frac{1}{3}p^{2} \bigg\}
\Bigg\},
\end{align}

\begin{align}
\label{eq:sigmaZZ}
\Sigma^{ZZ}_{T}(p^2)  &=   - \frac{\hat{g}^2_2 \hat{s}^2_W}{16\pi^2} \Bigg\{
\frac{2}{3} \sum_{f,i} N_{C}^{f}
\bigg\{\Big((\hat{g}_{f}^{+})^{2}+(\hat{g}_{f}^{-})^{2}\Big)
\Big[-(p^{2}+2m_{f,i}^{2})B_{0}(p^{2},m_{f,i},m_{f,i})
\nonumber \\
& +2m_{f,i}^{2}B_{0}(0,m_{f,i},m_{f,i}) +\frac{1}{3}p^{2} \Big]
+\frac{3}{4\hat{s}_{W}^{2}\hat{c}_{W}^{2}} m_{f,i}^{2}
B_{0}(p^{2},m_{f,i},m_{f,i}) \bigg\} \nonumber \\
& +\frac{1}{6\hat{s}_{W}^{2}\hat{c}_{W}^{2}} \bigg\{
\Big[\bigg(18\hat{c}_{W}^{4} + 2\hat{c}_{W}^{2} -\frac{1}{2}\bigg) p^{2}
+ (24\hat{c}_{W}^{4} + 16\hat{c}_{W}^{2} -10) m_{W}^{2}\Big]
B_{0}(p^{2},m_{W},m_{W}) \nonumber \\
&  -(24\hat{c}_{W}^{4} - 8\hat{c}_{W}^{2} + 2)
m_{W}^{2}B_{0}(0,m_{W},m_{W})
+ (4\hat{c}_{W}^{2}-1) \frac{1}{3}p^{2} \bigg\} \nonumber \\
& +\frac{1}{12\hat{s}_{W}^{2}\hat{c}_{W}^{2}} \bigg\{
\Big(2m_{H}^{2} - 10m_{Z}^{2} -p^{2}\Big) B_{0}(p^{2},m_{Z},m_{H}) \nonumber \\
& -2m_{Z}^{2}B_{0}(0,m_{Z},m_{Z}) -
2m_{H}^{2}B_{0}(0,m_{H},m_{H}) \nonumber \\
& -\frac{(m_{Z}^{2}-m_{H}^{2})^{2}}{p^{2}}
\Big( B_{0}(p^{2},m_{Z},m_{H}) - B_{0}(0,m_{Z},m_{H})\Big)
-\frac{2}{3}p^{2} \bigg\}
\Bigg\} ,
\end{align}
\begin{eqnarray}
\label{eq:sigmazz}
\frac{\partial\, \Sigma^{ZZ}_{T}(p^2)}{\partial p^2}\bigg|_{p^2=m^2_Z}  &=&
-\frac{\hat{g}^2_2 \hat{s}^2_W}{16\pi^2} \Bigg\{
\frac{2}{3} \sum_{f,i} N_{C}^{f}
\bigg\{\Big((\hat{g}_{f}^{+})^{2}+(\hat{g}_{f}^{-})^{2}\Big)
\Big[-B_{0}(p^2,m_{f,i},m_{f,i})
\nonumber \\
&& \hspace*{-2.5cm}
-\,(m^{2}_Z+2m_{f,i}^{2})\frac{\partial B_{0}(p^2,m_{f,i},m_{f,i})}{\partial p^2} +\frac{1}{3} \Big]
+\frac{3}{4\hat{s}_{W}^{2}\hat{c}_{W}^{2}} m_{f,i}^{2}
\frac{\partial B_{0}(p^2,m_{f,i},m_{f,i})}{\partial p^2} \bigg\} \nonumber \\
&&  \hspace*{-2.5cm}
+\,\frac{1}{6\hat{s}_{W}^{2}\hat{c}_{W}^{2}} \bigg\{
\bigg(18\hat{c}_{W}^{4} + 2\hat{c}_{W}^{2} -\frac{1}{2}\bigg)
B_{0}(p^{2},m_{W},m_{W}) \nonumber \\
&&  \hspace*{-2.5cm}
+ \,\Big[\bigg(18\hat{c}_{W}^{4} + 2\hat{c}_{W}^{2} -\frac{1}{2}\bigg) m^{2}_Z
+ (24\hat{c}_{W}^{4} + 16\hat{c}_{W}^{2} -10) m_{W}^{2}\Big]
\frac{\partial B_{0}(p^{2},m_{W},m_{W})}{\partial p^2}\nonumber \\
&& \hspace*{-2.5cm}
+\, (4\hat{c}_{W}^{2}-1) \frac{1}{3} \bigg\} \nonumber \\
&& \hspace*{-2.5cm}
 +\,\frac{1}{12\hat{s}_{W}^{2}\hat{c}_{W}^{2}} \bigg\{
- B_{0}(p^{2},m_{Z},m_{H}) + \Big(2m_{H}^{2} - 11m_{Z}^{2} \Big) \frac{\partial B_{0}(p^{2},m_{Z},m_{H})}{\partial p^2}  \nonumber \\
&& \hspace*{-2.5cm}
 +\, \frac{(p^{2}-m_{H}^{2})^{2}}{m^{4}_Z}
\Big( B_{0}(p^{2},m_{Z},m_{H}) - B_{0}(0,m_{Z},m_{H})\Big) \nonumber \\
&& \hspace*{-2.5cm}
-\,\frac{(m_{Z}^{2}-m_{H}^{2})^{2}}{m^{2}_Z}
\Big( \frac{\partial B_{0}(p^{2},m_{Z},m_{H})}{\partial p^2}\Big)
-\frac{2}{3} \bigg\}
\Bigg\}\Bigg|_{p^2=m^2_Z} \,  . \label{eq:DSigmaZZ}
\end{eqnarray}
Here $m_{f,i}$ is the mass of a fermion, where $i$ indicates the generation index and $f$ refers to the fermions within 
a generation. $N^f_C$ is the number of fermion colors, $N^f_C=1$ in the case of 
leptons and $N^f_C=3$ in the case of quarks. The electroweak couplings are written 
in terms of the charge and the third SU(2) generator
\begin{align}
\hat{g}^+_f = \frac{\hat{s}_W}{\hat{c}_W} Q_f, \quad \quad \quad \hat{g}^-_f = 
\frac{\hat{s}^2_W Q_f-I^3_{W,f}}{\hat{s}_W \hat{c}_W}\, .
\end{align}
We also provide the explicit expressions for the $B_0$ and 
$\partial B_0/\partial p^2$ functions that are required for the jet function 
computation. In the expressions \eqref{eq:SigmaGG} to \eqref{eq:DSigmaZZ} the poles in the expressions below are subtracted. In the following $p^2>0$, since the imaginary parts are made explicit:
\begin{align}
B_0(0,m,m) &=  \frac{1}{\epsilon} - 2 \ln\frac{m}{\mu}\, , \\
B_0(0,0,m) &= \frac{1}{\epsilon} + 1 - 2 \ln\frac{m}{\mu} \, , \\
B_0(0,m_1,m_2) &=  \frac{1}{\epsilon} + 1 + \frac{m^2_1+m_2^2}{m^2_1-m^2_2}\ln\frac{m_2}{m_1}+\ln\frac{\mu^2}{m_1m_2}\, ,  \\
B_0(p^2,0,0) &=  \frac{1}{\epsilon} + 2  + \ln\frac{\mu^2}{p^2}+i \pi \, , \\
\frac{\partial B_0(p^2,0,0)}{\partial p^2} &= -\frac{1}{p^2}\, ,\\
B_0(p^2,m,m) &= \bigg\{ \theta(4 m^2-p^2)\bigg[\frac{1}{\epsilon} + 2 -2 \ln\frac{m}{\mu}- 2 \bar{\beta} \arctan\frac1{\bar{\beta}} \bigg] \nonumber \\
&+ \theta(p^2-4 m^2)\bigg[\frac{1}{\epsilon} + 2 -2 \ln\frac{m}{\mu}+ \beta \ln(x) + i \beta \pi\bigg]\bigg\}\, ,\\
\frac{\partial B_0(p^2,m,m)}{\partial p^2} & = \bigg\{\theta(4 m^2-p^2)\frac1{p^2}\left[\frac{1+\bar\beta^2}{\bar\beta}\arctan\frac1{\bar\beta}-1\right] \nonumber \\
                                           &+\theta(p^2-4 m^2)\left[-\frac{1}{p^2}+ \frac{2 m_W^2}{p^4 \beta} \left( \ln(x) + i\pi\right)\right]\bigg\}\, , \\
\left.\frac{\partial B_0(p^2,m,m)}{\partial p^2}\right|_{p^2=0} &= \frac{1}{6m^2}\, , \\
B_0(p^2,0,m) &=  \bigg[\frac{1}{\epsilon} + 2 - 2\ln\frac{m}{\mu} - \bigg(1-\frac{m^2}{p^2}\bigg) \bigg[\theta(m^2-p^2) \ln\bigg(1-\frac{p^2}{m^2}\bigg) \nonumber \\
&+ \theta(p^2-m^2) \left(\ln\bigg(\frac{p^2}{m^2}-1\bigg) - i\pi\right) \bigg]\bigg]\, ,\\
B_0(p^2,M,m) & = \bigg[\frac{1}{\epsilon} +2 -\frac{M^2-m^2}{p^2} \ln\frac{M}{m} +\ln\frac{\mu^2}{mM} \, \nonumber \\
& +\frac{\sqrt{|\kappa(p^2,m^2,M^2)|}}{p^2}F(p^2,M,m)\bigg]\, , \\
    \left.\frac{\partial B_0(p^2,m_H,m_Z)}{\partial p^2}\right|_{p^2=m_Z^2} &=   \bigg[- \frac{1}{m^2_Z} -\frac{m^2_H-m^2_Z}{m^4_Z}\ln\bigg(\frac{m_Z}{m_H}\bigg) \, \nonumber \\
& - \frac{(m^2_H - 3 m^2_Z ) \arctan{\left[\sqrt{\frac{4 m^2_Z}{m^2_H}-1}\right]}}{m^4_Z \sqrt{\frac{4 m^2_Z}{m^2_H}-1}}\bigg]
\, ,
\end{align}
where for $M>m$
\begin{align*}
F(p^2,M,m) =\left\{
\begin{array}{rl}
\ln\frac{\sqrt{(M+m)^2-p^2}+\sqrt{(M-m)^2-p^2}}
{\sqrt{(M+m)^2-p^2}-\sqrt{(M-m)^2-p^2}} &\qquad p^2<(m-M)^2\\[0.6cm]
-2\arctan\sqrt{\frac{p^2-(M-m)^2}{(M+m)^2-p^2}} &\qquad (M-m)^2<p^2<(m+M)^2\\[0.6cm]
\ln\frac{\sqrt{p^2-(M-m)^2}-\sqrt{p^2-(M+m)^2}}
{\sqrt{p^2-(M-m)^2}+\sqrt{p^2-(M+m)^2}}+i\pi &\qquad p^2>(m+M)^2 
\end{array}	\right  . \, 
\end{align*}
and 
\begin{align}
\kappa(x,y,z) = x^2+y^2+z^2-2 x y-2 x z - 2 y z
\end{align}
is the K\'allen function. $\beta,x$ were defined in \eqref{eq:betax} and $\bar{\beta}$ in \eqref{eq:betabar}.


\subsection{\boldmath 
Treatment of the $Z$ resonance}
\label{app:zpole}

The `nrw' jet function \eqref{eq:J33nrw} requires the introduction 
of the star and ``double-star" distributions to deal with the 
singular $Z$-boson propagators when $p_{\rm max}^2>m_Z^2$. 
The distributions can be integrated against smooth test functions. 
However, as is evident from the definitions~\eqref{eq:Zstar}, 
\eqref{eq:Zdoublestar}, as the integration limit $p_{\rm max}^2$ 
approaches the singular value $m_Z^2$, the integrals diverge, 
which was already pointed out in~\cite{Beneke:2018ssm} (see for 
instance Figure~3 there). 

The singularity arises from the $Z$-boson resonance in the 
narrow width jet function,\footnote{The issue is absent 
for the hard-collinear intermediate resolution jet function. In 
this case, the gauge boson masses can be neglected and the 
$Z$-boson resonance does not appear in the regime of validity, 
$p^2=\mathcal{O}(\mchi m_W)$.} and can be cured by the standard 
Dyson resummation. Inspection of the expressions in the 
previous subsection shows that the divergent terms in the 
integral $\int_0^{p_{\rm max}^2}dp^2\,J^{33}(p^2)$ arise from 
the light-fermion self-energy diagrams as these give rise to the 
star distributions in the one-loop result  \eqref{eq:J33nrw}. 
Reordering this expression as 
\begin{eqnarray}
J^{33}(p^2,\mu,\nu) &=& 
\hat s_W^2(\mu)\delta(p^2) + J^{33}_{\rm Wilson}(p^2,\mu,\nu)
\nonumber\\
&& +\,\hat c_W^2(\mu)\delta(p^2-m_Z^2)
+ J^{33}_{{\rm se}
}(p^2,\mu)\ ,
\label{eq:Dysonresjet}
\end{eqnarray}
we therefore focus on the last two terms.

It is not necessary to perform the full Dyson resummation, resumming
the $Z$-boson propagator insertions is enough. We obtain (dropping 
the argument $\mu$ from the coupling and the jet function)
\begin{eqnarray}
\hat c_W^2\delta(p^2-m_Z^2)
+ J^{33,\, \rm Dyson}_{\text{se}}(p^2) &=& 
\frac{\hat{s}^{2}_W \hat{g}^2_2}{16 \pi^2} \hat{s}^2_W  \frac{80}{9} \bigg[-\delta(p^2) \frac{5}{3}  + \bigg[\frac{1}{p^2}\bigg]^{[\mu^2]}_*\bigg]\nonumber \\
&& \hspace*{-5cm} +\, 
 \hat{s}_W^2\,\frac{1}{\pi}\,\frac{\mbox{Im}\left[\Sigma_T^{\gamma\gamma}(p^2)_{t,W}\right]}{(p^2)^2}-\hat{s}_W^2{\rm Re}\left[\frac{\partial\Sigma_T^{\gamma\gamma}(p^2)_{t,W}}{\partial p^2}\right]_{p^2=0}\delta(p^2)\nonumber \\
&& \hspace*{-5cm} +\, \frac{1}{\pi}\,\mbox{Im}\left[
 2 \hat{s}_W\hat{c}_W\,\frac{\Sigma_T^{\gamma Z}(p^2)}{-p^2}
\,\frac{1}{-p^2+m_Z^2+{\rm Re}\left[\Sigma_T^{ZZ}(m_Z^2)\right]-\Sigma_T^{ZZ}(p^2)}\,\right]
\nonumber \\
&& \hspace*{-5cm} +\, \frac{1}{\pi}\,\mbox{Im}\left[
\hat{c}^{2}_W \,\frac{1}{-p^2+m_Z^2+{\rm Re}\left[\Sigma_T^{ZZ}(m_Z^2)\right]-\Sigma_T^{ZZ}(p^2)}
\,\right]\, .
\label{eq:JDyson33}
\end{eqnarray}
Note that \eqref{eq:JDyson33} also includes the tree-level $Z$-boson 
contribution to the jet function through the last line. 
The terms ${\rm Re}\left[\Sigma_T^{ZZ}(m_Z^2)\right]$ in the 
denominator ensure that the real part of the renormalized 
$Z$-boson self-energy vanishes at $p^2=m_Z^2$ as is required 
in the adopted on-shell scheme for the $Z$ mass. 
The imaginary parts of the square brackets in 
\eqref{eq:JDyson33} can be further simplified by 
noting that the Dyson resummation is necessary only when 
$p^2\approx m_Z^2$. We then obtain
\begin{eqnarray}
\hat c_W^2\delta(p^2-m_Z^2)
+ J^{33,\, \rm Dyson}_{\text{se}}(p^2) &=& 
\frac{\hat{s}^{2}_W \hat{g}^2_2}{16 \pi^2} \hat{s}^2_W  \frac{80}{9} \bigg[-\delta(p^2) \frac{5}{3}  + \bigg[\frac{1}{p^2}\bigg]^{[\mu^2]}_*\bigg]\nonumber \\
&& \hspace*{-5cm} +\, 
 \hat{s}_W^2\,\frac{1}{\pi}\,\frac{\mbox{Im}\left[\Sigma_T^{\gamma\gamma}(p^2)_{t,W}\right]}{(p^2)^2}-\hat{s}_W^2{\rm Re}\left[\frac{\partial\Sigma_T^{\gamma\gamma}(p^2)_{t,W}}{\partial p^2}\right]_{p^2=0}\delta(p^2)\nonumber \\
&& \hspace*{-5cm} +\, 2\hat{s}_W\hat{c}_W
\bigg[\,\frac{\mbox{Re}\left[\Sigma_T^{\gamma Z}(0)_{t,W}\right]}{m_Z^2}\,\delta(p^2)-\,\frac{\mbox{Re}\left[\Sigma_T^{\gamma Z}(m_Z^2)_{t,W}\right]}{m_Z^2}\,\delta(p^2-m_Z^2)\nonumber \\
&& \hspace*{-5cm} -\,\frac1{\pi}\frac{\mbox{Im}\left[\Sigma_T^{\gamma Z}(p^2)_{t,W}\right]}{m_Z^2}\frac1{p^2}+\,\frac1{\pi}\frac{\mbox{Im}\left[\Sigma_T^{\gamma Z}(p^2)_{t,W}\right]}{m_Z^2}\frac1{p^2-m_Z^2}\bigg]\nonumber \\
&& \hspace{-5cm}+\,
\frac{\hat{s}^{2}_W \hat{g}^2_2}{16 \pi^2}\,2 
\left(\frac{10}3-\frac{80}9\hat s_W^2\right)\left[\frac{p^2-m_Z^2}{(p^2-m_Z^2)^2+m_Z^2\Gamma_Z^2}-\delta(p^2-m_Z^2)\left(\frac53-\ln\frac{m_Z^2}{\mu^2}\right)\right]\nonumber\\
&& \hspace*{-5cm} +\, 
 \hat{c}_W^2\,\frac{1}{\pi}\,\frac{\mbox{Im}\left[\Sigma_T^{ZZ}(p^2)_{t,W,Z,H}\right]}{(p^2-m_Z^2)^2}-\hat{c}_W^2{\rm Re}\left[\frac{\partial\Sigma_T^{ZZ}(p^2)_{t,W,Z,H}}{\partial p^2}\right]_{p^2=m_Z^2}\delta(p^2-m_Z^2)\nonumber \\
&& \hspace*{-5cm} +\, \frac{\hat{c}_W^2\Gamma_Z}{\pi m_Z}\left[-\left(\frac23-\ln\frac{m_Z^2}{\mu^2}\right)\delta(p^2-m_Z^2)+\frac{p^2}{(p^2-m_Z^2)^2+(p^2)^2\Gamma_Z^2/m_Z^2}\right]\, ,
\label{eq:JDyson33tilde}
\end{eqnarray}
where
\begin{equation}
    \Gamma_Z =\frac{\text{Im}\left[ \Sigma_T^{ZZ}(m_Z^2)|_{f\neq t \text{only}}\right]}{m_Z}
 = \frac{\hat{g}_2^2 m_Z}{16 \pi\hat c_W^2}
\left[-\frac{20}{3}\hat s^2_W+\frac{7}{2}+\frac{80}{9}\hat s^4_W
\right]
\end{equation}
denotes the tree-level decay width of the $Z$ boson into the 
light fermions of the SM (all, except the top quark, masses set 
to zero). In deriving (\ref{eq:JDyson33tilde}), we used 
the identities
\begin{align}
     {\rm Re}\!\left[\Sigma_T^{ZZ}(p^2)-\Sigma_T^{ZZ}(m_Z^2)\right]\!\delta'(p^2-m_Z^2) &=\text{Re}\!\left[\frac{\partial\Sigma_T^{ZZ}(m_Z^2)}{\partial p^2}\right](p^2-m_Z^2)\,\delta'(p^2-m_Z^2)\, , \\
    \text{Re}\left[ \frac{\partial \Sigma_{T,f\neq t}^{ZZ}(p^2)}{\partial p^2}\right]_{p^2=m_Z^2}&\!\!=  \frac{\Gamma_Z}{\pi m_Z} \left(\frac{2}{3} - \ln \frac{m_Z^2}{\mu^2}\right)
\end{align}
and $x\delta'(x)=-\delta(x)$, valid for non-singular test functions 
at $p^2=m_Z^2$. 

The Dyson-resummed expression \eqref{eq:JDyson33tilde} can be 
obtained from the fixed-order expressions \eqref{eq:J33nrw}, 
\eqref{eq:J33nrwsefneqtonly} and \eqref{eq:J33nrwsefneqtexc} by 
employing the substitution rules
\begin{eqnarray}
\left[\frac1{p^2-m_Z^2}\right]_*{} & \to &{}\frac{p^2-m_Z^2}{(p^2-m_Z^2)^2+m_Z^2\Gamma_Z^2}\, ,\\
\delta(p^2-m_Z^2)+\frac{\Gamma_Z}{\pi m_Z}\left[\frac1{(p^2-m_Z^2)^2}\right]_{**}p^2{} & \to &{}\frac1{\pi}\frac{p^2\,\Gamma_Z/m_Z}{(p^2-m_Z^2)^2+(p^2)^2\Gamma_Z^2/m_Z^2}\ .\quad
\end{eqnarray}


\section{Soft function}
\label{app:softFunction}
In this Appendix we discuss the one-loop computation of the soft function. We start by discussing the scalar integrals for the virtual and real parts of the soft function. The final result is given by linear combinations of these integrals. We also illustrate how the rapidity divergences change between the two factorization theorems presented in the main text. Furthermore we give the inverse Laplace transforms of the 
resummed soft functions $\hat{W}$.

\begin{figure}[t]
\centering
\includegraphics[width=0.5\textwidth]{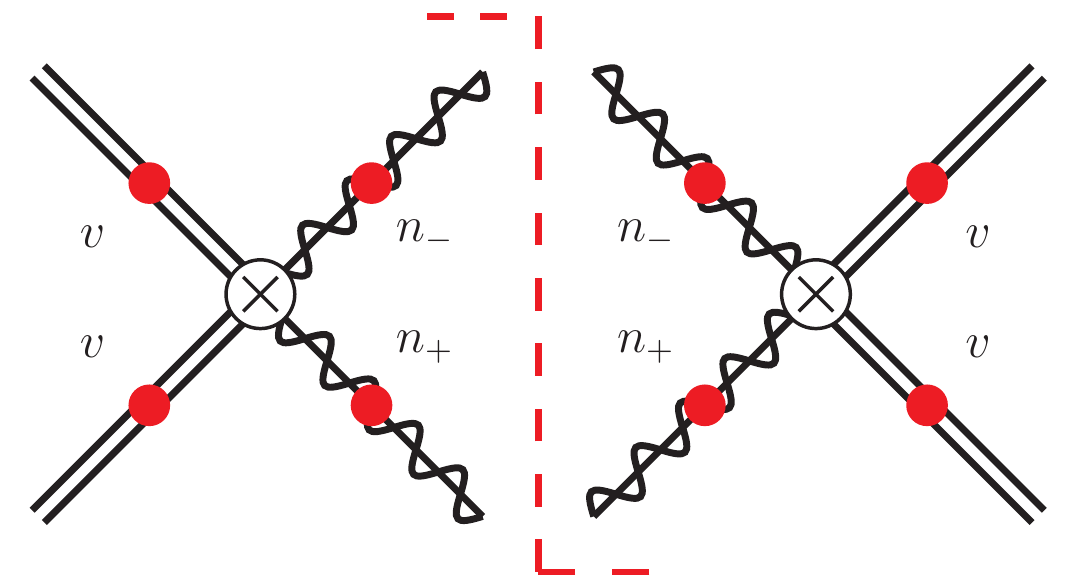}
\caption{Diagrammatic representation of the one-loop soft function.}
\label{fig:soft_function}
\end{figure}

The integrated soft function was defined in \eqref{eq:softfnintdef} and the index-contracted version in \eqref{eq:softtwoindexfn}. For the calculation of the integrals and the soft coefficients we find it convenient to shift the position of the Wilson line to 0 and to perform the integration in $n_+ y$. This leads to
\begin{align}
    W^{ij}_{IJ}(\omega)&= 
    \sumint_{X_s} \delta(\omega-n_- p_{X_s})\left\langle 0\middle|\,\mathbf{\bar{T}}[[\mathcal{S}^{\dagger}]^{j}_{J,V3}(0)] \middle| X_s \right\rangle\left\langle X_s \middle|\mathbf{T}[\mathcal{S}^i_{I,V3}(0)]\,|0\right\rangle \, . \label{eq:softfndelta}
\end{align}
Diagrammatically, the one-loop soft function is shown in 
Figure~\ref{fig:soft_function}, where a single soft gauge boson attaches to any two distinct (red) dots on the external legs. In the following, we categorize the integrals according to which external legs the soft radiation attaches to. If, for example, the soft gauge boson connects the collinear ($n_{-}^\mu$) and the anti-collinear ($n_{+}^\mu$) external leg, we call this the $n_{+} n_{-}$ \textit{virtual} or \textit{real integral}, depending on whether the soft gauge boson passes through the cut.

\subsection{Virtual soft integrals}
In this section we report the calculation of the relevant scalar integrals for the virtual soft correction. We define the integration 
measure (and implicitly $\tilde{\mu}$) as
\begin{align}
[dk] = \tilde{\mu}^{2 \epsilon}\frac{d^dk}{(2 \pi)^d} =\bigg(\frac{\mu^2 e^{\gamma_E}}{4 \pi}\bigg)^{\!\epsilon}\frac{d^dk}{(2 \pi)^d}\, ,
\label{eq:loopmeasure}
\end{align}
where $d=4-2\epsilon$ and $\gamma_E$ is the Euler-Mascheroni 
constant. For the real integrals discussed later, we also use 
the phase-space measure in terms of light-cone coordinates
\begin{eqnarray}
 \int d^d k \, \theta(k^0) \delta(k^2 - m_W^2) &=& 
\frac{1}{2}\int_0^\infty d n_+ k  \int_0^\infty d n_- k  
\int d^{d-2}k_\perp \delta(n_+ k \,n_- k + k_\perp^2 - m_W^2) \nonumber \\ 
&& \hspace*{-3cm} = \,\frac{\Omega_{d-2}}{2}\int_0^\infty\! dn_+ k 
\int_0^\infty\! dn_-k \int_0^\infty \! d k_T \, k_T^{d-3} 
\delta(n_+ k \,n_- k-k_T^2-m_W^2) \, ,
\end{eqnarray}
where $k_T^2=-k_\perp^2>0$, and the delta- and theta-functions enforce 
$n_+ k, n_- k\geq0$.

\paragraph{The $n_+ n_-$ virtual integral}\mbox{}\\
We start by analyzing the virtual $n_+ n_-$ integral
\begin{align}
    I_{n_+ n_-}^{\text{virt.}} =-i \hat{g}_2^2\delta(\omega) (n_+ \cdot n_-)\int [dk] \frac{ \nu^{\eta}}{[k^2-m^2_W+i\varepsilon] [n_- k+i\varepsilon] [n_+ k-i\varepsilon] |2 k^3|^{\eta}}\, . 
\end{align}
It is convenient to proceed by first doing the contour integration in the variable $k^0$. To this purpose we rewrite the integral as
\begin{align}
    I_{n_+ n_-}^\text{virt.} &=- 2 i \hat{g}_2^2 \delta(\omega) \tilde{\mu}^{2 \epsilon} \int \frac{dk^0 dk^3 d^{d-2}k_\perp}{(2 \pi)^d}\frac{\nu^\eta}{|2k^3|^\eta}  \nonumber \\
&\hspace{3cm}\times \frac{1}{[(k^0)^2-E^2_k+i \varepsilon] [k^0-k^3+i \varepsilon] [k^0+k^3-i \varepsilon]} \, ,
\end{align}
where $E^2_k = (k^3)^2+k^2_T+m^2_W$.
If $k^3>0$ one finds four poles in the $k^0$ complex plane situated at $\pm (E_k-i \varepsilon)$, $k^3-i \varepsilon$ and $-k^3+i \varepsilon$. Two of these poles are in the upper half plane and the other two are in the lower half plane. We close the integration contour in the lower half plane (notice that by doing this we pick up a factor $-2\pi i$). For $k^3<0$ we find again two poles in the upper half plane and two poles in the lower half plane. The poles in $k^3-i \varepsilon$ and $-k^3 +i \varepsilon$ moved from the positive to the negative $k^0$ domain and vice versa, respectively. We obtain
\begin{align}
    I_{n_+ n_-}^\text{virt.} &=2i \hat{g}_2^2 \delta(\omega) \tilde{\mu}^{2 \epsilon} 
\int \frac{d^{d-2}k_\perp }{(2 \pi)^d}\left\{ \,
\int^\infty_0  dk^3 \frac{\nu^\eta}{(2 k^3)^\eta}\frac{2 \pi i}{k_T^2 +m_W^2} \left[\frac{1}{2 E_k } -\frac{1}{ 2 k^3-i \varepsilon  }\right] \right. 
\nonumber \\
                             &\hspace{4.2cm} \left.+\int^0_{-\infty} dk^3\frac{\nu^\eta}{(-2k^3)^\eta}\frac{2 \pi i}{k_T^2 + m_W^2} \left[\frac{1}{2 E_k} -\frac{1}{ 2 k^3-i \varepsilon }\right] \,\right\}\, . 
\label{eq:soft1}
\end{align}
By summing the first terms in the two square brackets of (\ref{eq:soft1}) (which give the same contribution as can be easily seen by making the variable transformation $k^3\to -k^3$ in the second line) and after performing the $k^3$ and $k_\perp$ integrations one obtains for this part
\begin{align}
    -\frac{ \hat{g}_2^2}{4 \pi^{2}} \delta(\omega) \left(\frac{\mu}{m_W}\right)^{\!2\epsilon}  e^{\gamma_E \epsilon}\left(\frac{\nu}{m_W}\right)^{\!\eta} \,
\frac{\Gamma\big(\frac{1}{2}-\frac{\eta}{2}\big) \Gamma\big(\epsilon+\frac{\eta}{2}\big)}{2^\eta \, \pi^{\frac{1}{2}} \, \eta}\,.
\end{align}
A pure imaginary part comes from the sum of the remaining terms in the upper and lower line of (\ref{eq:soft1}). The $k^3$ integral of these two terms is 
\begin{eqnarray}
&& \int dk^3  \frac{(-2 \pi i) \nu^{\eta}}{[2 k^3-i \varepsilon] |2 k^3|^{\eta}  } 
= \int^{\infty}_0 dk^3 \frac{(-2 \pi i) \nu^\eta}{(2 k^3)^\eta}\bigg[\frac{1}{2k^3-i \varepsilon} + \frac{1}{-2k^3-i \varepsilon}\bigg] \nonumber \\
&& \hspace*{1cm}\, 
= (- i \pi) \nu^\eta \pi  \csc (\pi \eta) \big((-i \varepsilon)^{-\eta} - (i \varepsilon)^{-\eta}\big)
=(2 \pi^2) \nu^\eta  \csc (\pi \eta) \, \varepsilon^{-\eta} \sin(\eta\, \pi/2)
\qquad
\nonumber\\
&& \hspace*{1cm}\, = \pi^2 + \mathcal{O}(\eta)\, ,
\end{eqnarray}
where the result is independent of the small imaginary part $i \varepsilon$ at $\mathcal{O}(\eta^0)$.
After performing the $k_\perp$ integration and summing the two contributions we obtain the final result
\begin{align}
    I_{n_+ n_-}^{\text{virt.}} &= -\frac{\hat{g}_2^2 \delta(\omega)}{4 \pi^2} \left(\frac{\mu}{m_W}\right)^{\!2\epsilon}\! \left(\frac{\nu}{m_W}\right)^{\!\eta} \!e^{\gamma_E \epsilon} \left[\frac{\Gamma\big(\frac{1}{2}-\frac{\eta}{2}\big) \Gamma\big(\epsilon+\frac{\eta}{2}\big)}{2^\eta\, \pi^{\frac{1}{2}} \,\eta}  - \frac{\Gamma\big(\epsilon + \frac{\eta}{2}\big)}{\Gamma\big(1 + \frac{\eta}{2}\big)} \left(\frac{i \,\pi}{2} + \mathcal{O} (\eta)\right) \right] \nonumber \\
&= -\frac{\hat{\alpha}_2}{2 \pi} \delta(\omega) \left[-\frac{1}{\epsilon^2}+ \frac{2}{\epsilon \eta} - \frac{i \pi}{\epsilon}+ \frac{2}{\epsilon} \ln\frac{m_W}{\mu} -\frac{2}{\epsilon} \ln\frac{m_W}{\nu} - \frac{4}{\eta} \ln \frac{m_W}{\mu} \right. \nonumber \\
&\hspace{2.4cm}\left. + \frac{\pi^2}{12} + 2 \pi i \ln\frac{m_W}{\mu}- 2 \ln^2 \frac{m_W}{\mu} + 4 \ln\frac{m_W}{\mu}\ln\frac{m_W}{\nu}\right]. \label{eq:Ivirtnpnm} 
\end{align}
\paragraph{The $v n_+$ and $v n_-$ virtual integrals }\mbox{}\\
The second virtual integral we analyze is the scalar integral which appears in initial-final state soft $W$ exchange. The integrals for the Wilson line combinations $v n_+$ and $v n_-$ give the same result because the virtual part of the soft function is symmetric under the exchange $n_+ \leftrightarrow n_-$. We discuss the integral
\begin{align}
I_{v n_+}^{\text{virt.}} =-i \hat{g}_2^2 \delta(\omega) (v \cdot n_+) \int [dk]\frac{\nu^{\eta}}{[k^2-m^2_W+i \varepsilon][n_+ k-i \varepsilon][v \cdot k-i \varepsilon] |2 k^3|^\eta} \, .
\end{align}
We proceed in a very similar way to the integral $I^{\text{virt.}}_{n_+ n_-}$ above and first do the contour integration in $k^0$. The integral has four poles in the $k^0$ complex plane and only one ($E_k-i \varepsilon$) is situated in the lower half plane. It is therefore easier to close the contour in the lower half plane both for $k^3>0$ and $k^3<0$. By summing the positive and the negative $k^3$ regions and by integrating over $k_\perp$ we obtain
\begin{align}
    I^{\text{virt}}_{v n_+} &= -\frac{\hat{g}_2^2}{8 \pi^2} \delta(\omega) \left(\frac{\mu}{m_W}\right)^{2\epsilon} \left(\frac{\nu}{m_W}\right)^{\eta} e^{\gamma_E \epsilon}\, \frac{\Gamma\big(\frac{1}{2}-\frac{\eta}{2}\big) \Gamma\big(\epsilon+\frac{\eta}{2}\big)}{2^\eta \, \pi^{\frac{1}{2}} \,\eta} \nonumber \\
&= -\frac{\hat{\alpha}_2}{4 \pi}  \delta(\omega)\left[-\frac{1}{\epsilon^2} + \frac{2}{\epsilon \eta} + \frac{2}{\epsilon} \ln \frac{m_W}{\mu} - \frac{2}{\epsilon} \ln \frac{m_W}{\nu}- \frac{4}{\eta} \ln\frac{m_W}{\mu}\right. \nonumber \\
&\hspace{2.4cm}\left. + \frac{\pi^2}{12} +4 \ln \frac{m_W}{\mu}\ln\frac{m_W}{\nu}- 2 \ln^2 \frac{m_W}{\mu}\right] \, . \label{eq:Ivirtnpv}
\end{align}
\paragraph{The $v v$ virtual integral}
Another virtual integral originates from the connection of the two heavy DM Wilson lines. It is defined as
\begin{align}
    I_{vv}^{\text{virt.}} &= -i \hat{g}_2^2 \delta(\omega) (v \cdot v) \int [dk] \frac{1}{[k^2-m_W^2+i\varepsilon][k^0+ i \varepsilon][k^0-i\varepsilon]} \, . \end{align}
We perform the integration in $k^0$ noting that the pinched poles at $k^0=\pm i \varepsilon$ must not be picked up. These poles correspond to the potential region and are already taken into account in the one-loop contribution to the Sommerfeld effect. The integration in $\mathbf{k}$ is then straightforward. The result is 
\begin{align}
I_{vv}^{\text{virt.}}&= - \frac{\hat{\alpha}_2}{2 \pi} \delta(\omega) \left[\frac{1}{\epsilon} + \ln\frac{\mu^2}{m_W^2}\right] \, .\label{eq:Ivirtvv}
\end{align}

In \cite{Baumgart:2017nsr} the integrals $I_{n_+ n_-}^{\text{virt.}},I_{n_+ v}^{\text{virt.}},I_{n_- v}^{\text{virt.}}$ were already computed. We find agreement except for the integral $I_{n_+ n_-}^{\text{virt.}}$ where we find an additional term which results in the imaginary parts of \eqref{eq:Ivirtnpnm}.

\subsection{Real soft integrals}
\label{sec:softreal}

The real emission contribution is extracted by applying the Cutkosky rules to the cut propagators in the previous integrals
\begin{align}
\frac{1}{k^2 - m_{W}^2 + i \varepsilon} \to -2 \pi i  \, \delta(k^2 - m_{W}^2) \, \theta(k^0)\, .
\end{align}
We still have to keep the rapidity regulator to regulate the limit $\omega \to 0$, therefore we introduce star distributions \cite{DeFazio:1999ptt}, see \eqref{eq:stardistdef} 
for the definition. 

\paragraph{The $n_+ n_-$ real integral}\mbox{}\\
We start with the $n_+ n_-$ real emission contribution
\begin{align}
I^{\text{real}}_{n_+ n_-} = (n_+ \cdot n_-)\, \hat{g}_2^2 \int [dk] \frac{\left(- 2 \pi \delta \left(k^2 - m_W^2\right) \theta (k^0)\right) }{(n_+ k)(n_- k)} \delta (\omega - n_- k) \frac{\nu^\eta}{\left|2 k^3 \right|^\eta} \, .
\label{eq:softnpnn}
\end{align}
We first perform the integration in $n_- k$ using $\delta(\omega - n_- k)$, which 
leaves
\begin{align}
I^{\text{real}}_{n_+ n_-} = -\frac{\hat{\alpha}_2 e^{\gamma_E \epsilon}}
{2 \pi^{2-\epsilon}} \,\mu^{2 \epsilon} \nu^\eta \int d n_+k d^{d-2} k_T 
\frac{ \delta(\omega n_+ k - k_T^2 - m_W^2) \theta(\omega+n_+k)}
{\omega n_+k \left|n_+ k - \omega \right|^\eta} \, . 
\end{align}
In performing the $n_+ k$ integral next, the step function can be dropped as it 
only ensures $n_+ k >-\omega$, but $\omega \geq 0$. Since $k_T^2 + m_W^2 >0$, the 
delta-function contributes only for positive $n_+ k$, i.e. the step function does not pose a further restriction. Hence, 
\begin{align}
I^{\text{real}}_{n_+ n_-} = -\frac{\hat{\alpha}_2 e^{\gamma_E \epsilon}}
{ \pi \Gamma(1-\epsilon)} \,\mu^{2 \epsilon}\omega^{\eta-1} \nu^\eta \int_0^{\infty} d k_T \,  \frac{k_T^{1-2 \epsilon}}{k_T^2 + m_W^2}\frac{1}{\left|k_T^2 + m_W^2 - \omega^2\right|^\eta} \, .
\end{align}
We pulled a factor of $\omega^\eta$ into the absolute value, as $\omega\geq0$. The absolute value inside the integral forces us to consider two cases. Either $m_W>\omega$, in which case the absolute value can be dropped, as $k_T,m_W>0$. Or $\omega>m_W$, 
then we split the integrand into an integral from $0$ to $\sqrt{\omega^2 - m_W^2}$ with a factor of $(-1)^\eta$ from the absolute value, and a second integral from $\sqrt{\omega^2-m_W^2}$ to $\infty$, where the absolute value can be dropped.
To make the structure more transparent, we perform the substitution 
$k_T^\prime = k_T/m_W$ and define $\omega^\prime = \omega/m_W$, which 
turns the previous integral into the integral
\begin{align}
I^\text{real}_{n_+ n_-} &= - \frac{\hat{\alpha}_2}{\pi} \left(\frac{\mu^2 e^{\gamma_E}}{m_W^2}\right)^\epsilon \left( \frac{\nu \omega}{m_W^2}\right)^\eta \frac{1}{\omega \Gamma(1-\epsilon)} \int_0^\infty d k_T^\prime \frac{k_T^{\prime 1-2\epsilon}}{\left(k_T^{\prime 2} +1\right)\left|k_T^{\prime 2} +1 - \omega^{\prime 2}\right|^\eta} \, .
\end{align}
over dimensionless quantities. 

We start with the first case $\omega^\prime < 1$. The absolute value can be 
dropped and the integration results in
\begin{align}
    I^{\text{real}}_{n_+ n_-} &= -\frac{\hat{\alpha}_2}{2 \pi} \left(\frac{\mu^2 e^{\gamma_E}}{m_W^2}\right)^{\!\epsilon} \left(\frac{\nu \omega}{m_W^2}\right)^{\!\eta} \frac{1}{\omega \Gamma(1-\epsilon)} \left\{\left(\omega^\prime\right)^{-2 \eta} \Gamma(\epsilon+\eta) \Gamma(1-\epsilon-\eta) \vphantom{\frac{\Gamma}{\Gamma}} \right.\nonumber \\
& \hspace{0.5cm}\left. + \left(1-\omega^{\prime 2}\right)^{1-\epsilon -\eta} \frac{\Gamma(1-\epsilon) \Gamma(\epsilon + \eta -1)}{\Gamma(\eta)} 
\,{}_2F_{1}\left(1,1-\epsilon,2-\epsilon-\eta,1-\omega^{\prime 2}\right)\vphantom{\frac{\Gamma}{\Gamma}}\right\}
\end{align}
with ${}_2F_1$ the hypergeometric function. This is the exact result to all orders in $\epsilon,\eta$. The dimensionless terms inside the curly brackets are finite in the limits $\omega, \eta \to 0$. Therefore the only terms involving $\eta$-poles may come from $\omega^{\eta -1}m_W^{-\eta}=\frac{\delta(\omega)}{\eta}+\left[\frac{1}{\omega}\right]_*^{[m_W]} + \mathcal{O}(\eta)$ in front of the bracket. 
The $\delta(\omega)$ term in this identity requires to expand the expression in the curly brackets up to order $\eta^1$, but allows to set $\omega^{\prime}=0$ in the function arguments. Therefore we can simplify the hypergeometric ${}_2F_1$ function in this case. For the term involving the star distribution, we only keep the $\eta^0$  term in 
the bracket, which is $\omega$ independent. 
Therefore to order $\mathcal{O}\left(\eta,\epsilon\right)$, 
the result can be written as 
\begin{align}
I^{\text{real}}_{n_+ n_-} &= 
- \frac{\hat{\alpha}_2}{2 \pi} \left(\frac{\mu^2 e^{\gamma_E}}{m_W^2}\right)^\epsilon \left(\frac{\nu\, \omega}{m_W^2}\right)^{\!\eta} \frac{\Gamma(\epsilon + \eta)}{\omega \Gamma(1+\eta)} 
+ \mathcal{O}(\eta,\epsilon) \nonumber \\
&= -\frac{\hat{\alpha}_2}{2 \pi}\,
\bigg[\delta(\omega) \,\bigg(-\frac{1}{\epsilon^2}
+\frac{1}{\epsilon\, \eta} + \frac{1}{\eta} \ln\frac{\mu^2}{m_W^2} 
+ \frac{1}{\epsilon}\left(-\ln\frac{\mu^2}{m_W^2} + 
\ln\frac{\nu}{m_W}\right)
\nonumber \\
& \hspace{80pt} + \frac{\pi ^2}{12}-\frac{1}{2} \ln^2\frac{\mu
   ^2}{m_W^2} + \frac{1}{2} \ln\frac{\mu^2}{m_W^2}\ln\frac{\nu^2}{m_W^2}\bigg) \nonumber \\
   &\hspace{42pt} + \left[\frac{1}{\omega}\right]_{*}^{[m_W]} \left(\frac{1}{\epsilon} + \ln\frac{\mu^2}{m_W^2}\right)\bigg] \, . \label{eq:Irealnpnm}
\end{align}

For the above integral we assumed $\omega^\prime < 1$. For the second 
case $\omega^\prime > 1$, we show that the integral can be written in the same form as above. To do so we go back to the integral in $k_T^\prime$
\begin{align}
\int_0^\infty d k_T^\prime \frac{k_T^{\prime 1-2\epsilon}}{\left(k_T^{\prime 2} +1\right)\left|k_T^{\prime 2} +1 - \omega^{\prime 2}\right|^\eta} &= \int_0^{\sqrt{\omega^{\prime 2} -1}} d k_T^\prime \frac{k_T^{\prime 1-2\epsilon}}{\left(k_T^{\prime 2} +1\right)\left(k_T^{\prime 2} +1 - \omega^{\prime 2}\right)^\eta} \nonumber \\
&\hphantom{==}+ \int_{\sqrt{\omega^{\prime 2}-1}}^\infty d k_T^\prime \frac{k_T^{\prime 1-2\epsilon}}{\left(k_T^{\prime 2} +1\right)\left(-k_T^{\prime 2} -1 + \omega^{\prime 2}\right)^\eta} \, .
\end{align}
The individual terms yield
\begin{eqnarray}
&& \int_0^{\sqrt{\omega^{\prime 2} -1}} d k_T^\prime \frac{k_T^{\prime 1-2\epsilon}}{\left(k_T^{\prime 2} +1\right)\left(k_T^{\prime 2} +1 -\omega^{\prime 2}\right)^\eta} 
\nonumber\\
&& \hspace*{1cm} = \, \frac{\left(\omega^{\prime 2} -1\right)^{1-\epsilon-\eta}}{2}\frac{\Gamma(1-\epsilon) \Gamma(1-\eta)}{\Gamma(2-\epsilon-\eta)} 
\, {}_2F_1(1,1-\epsilon,2-\epsilon-\eta,1-\omega^{\prime \,2})\,, 
\nonumber \\
&& \int_{\sqrt{\omega^{\prime 2}-1}}^\infty d k_T^\prime \frac{k_T^{\prime 1-2\epsilon}}{\left(k_T^{\prime 2} +1\right)\left(-k_T^{\prime 2} -1 + \omega^{\prime 2}\right)^\eta} 
\nonumber\\
&& \hspace*{1cm} =\,\frac{\left(\omega^{\prime 2}-1\right)^{-\epsilon} \left(1-\omega^{\prime 2}\right)^{-\eta}}{2} \frac{\Gamma(1-\eta) \Gamma(\epsilon + \eta)}{\Gamma (1+\epsilon)}
\,{}_2F_1\left(1,\epsilon+\eta,1+\epsilon,\frac{1}{1-\omega^{\prime 2}}\right).
\qquad
\end{eqnarray}
The rest of the discussion is analogous to the case $\omega^\prime < 1$. 
To order $\mathcal{O}(\eta,\epsilon)$ 
we find the same result as in that case.

\paragraph{The $ v n_+$ real integral}\mbox{}\\
The $v n_+$ real emission integral is
\begin{align}
I^{\text{real}}_{v n_+} = (v \cdot n_+) \hat{g}_2^2 \int [dk] \frac{\left(-2 \pi \delta(k^2 -m_W^2) \theta(k^0)\right)}{(v \cdot k)(n_+ k)} \delta(\omega - n_- k) \frac{\nu^\eta}{\left|2 k^3\right|^\eta} \, .
\end{align}
We perform the integration in $n_+k, n_- k$ using the two delta-functions 
as for the $n_+ n_-$ case, and obtain 
\begin{align}
I^{\text{real}}_{v n_+} = - \frac{\hat{\alpha}_2}{\pi} \frac{\mu^{2 \epsilon} e^{\epsilon \gamma_E}}{\Gamma(1-\epsilon)} \nu^\eta \omega^{\eta+1} \int_0^\infty d k_T \frac{k_T^{1-2\epsilon}}{k_T^2 + m_W^2} \frac{1}{\omega^2 + k_T^2 + m_W^2}\frac{1}{\left|\omega^2 - k_T^2 -m_W^2\right|^\eta} \, .
\end{align}
Other than for the $n_+ n_-$ integral, the prefactor is now $\omega^{\eta +1}$, which is finite in the limit $\eta,\omega \to 0$, regardless of how the limit is 
taken. Hence, at this point we can set $\eta$ to 0. 
The expression is then a standard integral, that is easily solved. The result reads
\begin{align}
I_{v n_+}^{\text{real}} &= -\frac{\hat{\alpha}_2 e^{\epsilon \gamma_E}}{2 \pi \omega} \mu^{2 \epsilon} \Gamma(\epsilon) \left(m_W^{-2\epsilon} - (m_W^2 + \omega^2)^{-\epsilon}\right)+ \mathcal{O}(\eta) \nonumber \\
                        &= - \frac{\hat{\alpha}_2}{2 \pi} \frac{1}{\omega} \ln \left(\frac{m_W^2+\omega^2}{m_W^2}\right) + \mathcal{O}(\eta,\epsilon)\,, \label{eq:Irealnpv}
\end{align}
and is finite. 
To compare with the other terms, we may also replace $\frac{1}{\omega} \to \left[\frac{1}{\omega}\right]_\star$, as the integral is non-singular as $\omega \to 0$ and hence the star-distribution is equivalent to $\omega^{-1}$.
\paragraph{The $v n_- $ real integral}\mbox{}\\
This integral is related to the $n_+ n_-$ and $v n_+$ integrals. The identity
\begin{align}
    \frac{(n_+ \cdot n_-) }{(n_+ k) (n_- k) } - \frac{(v \cdot n_+)}{(v \cdot k) (n_+  k)}=\frac{(v \cdot n_-)}{(n_-  k) (v \cdot k)} \, ,
\end{align}
ensures that the integral obeys the relation
\begin{align}
    I^{\text{real}}_{vn_-} = I^{\text{real}}_{n_+ n_-} - I^{\text{real}}_{v n_+} \, . \label{eq:Irealnmv}
\end{align}
\paragraph{The $v v$ real integral}\mbox{}\\
The calculation of the $v v$ integral follows the logic of the $v n_+$ integral. We start with the expression
\begin{align}
I^{\text{real}}_{vv} &= (v \cdot v) \,\hat{g}_2^2 \int [d k] \frac{1}{(v \cdot k)^2 }(-2 \pi \delta(k^2 - m_W^2) \theta(k^0)) \delta(\omega -n_- k) \frac{\nu^\eta}{\left|2 k^3 \right|^\eta} \, .
\end{align}
The two delta-functions are used for the $n_+k, n_- k$ integrations. 
This results in the expression
\begin{align}
I^\text{real}_{vv} = - \frac{2 \hat{\alpha}_2}{\pi} \frac{\mu^{2\epsilon} e^{\epsilon \gamma_E} \nu^\eta}{\Gamma(1-\epsilon)} \omega^{\eta +1} \int_0^\infty d k_T \frac{k_T^{1-2\epsilon}}{\left(\omega^2 + k_T^2 + m_W^2\right)^2} \frac{1}{\left|\omega^2 -k_T^2 -m_W^2\right|^\eta } \, .
\end{align}
For the same reasons as for the $v n_+$ integral, we can expand the integrand in $\eta$ and drop the $\mathcal{O}(\eta)$ terms. Hence 
\begin{align}
I^{\text{real}}_{vv} &=- \frac{2 \hat{\alpha}_2}{\pi} \frac{\mu^{2 \epsilon} e^{\epsilon \gamma_E}}{\Gamma(1-\epsilon)}\  \omega \int_0^\infty d k_T \frac{k_T^{1-2\epsilon}}{\left(\omega^2 + k_T^2 + m_W^2\right)^2} + \mathcal{O}(\eta) \nonumber \\
&= - \frac{\hat{\alpha}_2}{\pi} \epsilon \, \Gamma(\epsilon) \mu^{2 \epsilon} e^{\epsilon \gamma_E} \omega \left(\frac{1}{m_W^2 + \omega^2}\right)^{1+\epsilon} + \mathcal{O}(\eta) \nonumber \\
&= - \frac{\hat{\alpha}_2}{\pi} \frac{\omega}{m_W^2+\omega^2} + \mathcal{O}(\eta,\epsilon) \, . \label{eq:Irealvv}
\end{align}
The same result is obtained if we keep the full $\eta$-dependence and expand the hypergeometric functions that arise for the full integrals.

All the real integrals except $I^{\text{real}}_{vv}$ were also computed in \cite{Baumgart:2017nsr}, and we confirm these results.

\subsection{Cut two-loop diagrams}
The integrals now allow us to determine the total discontinuity of a given two-loop diagram, after summing over all cuts. The sum of all cuts is
\begin{align}
\sum_{\text{cuts}}=\text{Disc}(i \mathcal{M}) = -2 \, \text{Im}\, \mathcal{M}\, .
\end{align}

\begin{figure}[t]
	\centering
	\includegraphics[width=0.7\textwidth]{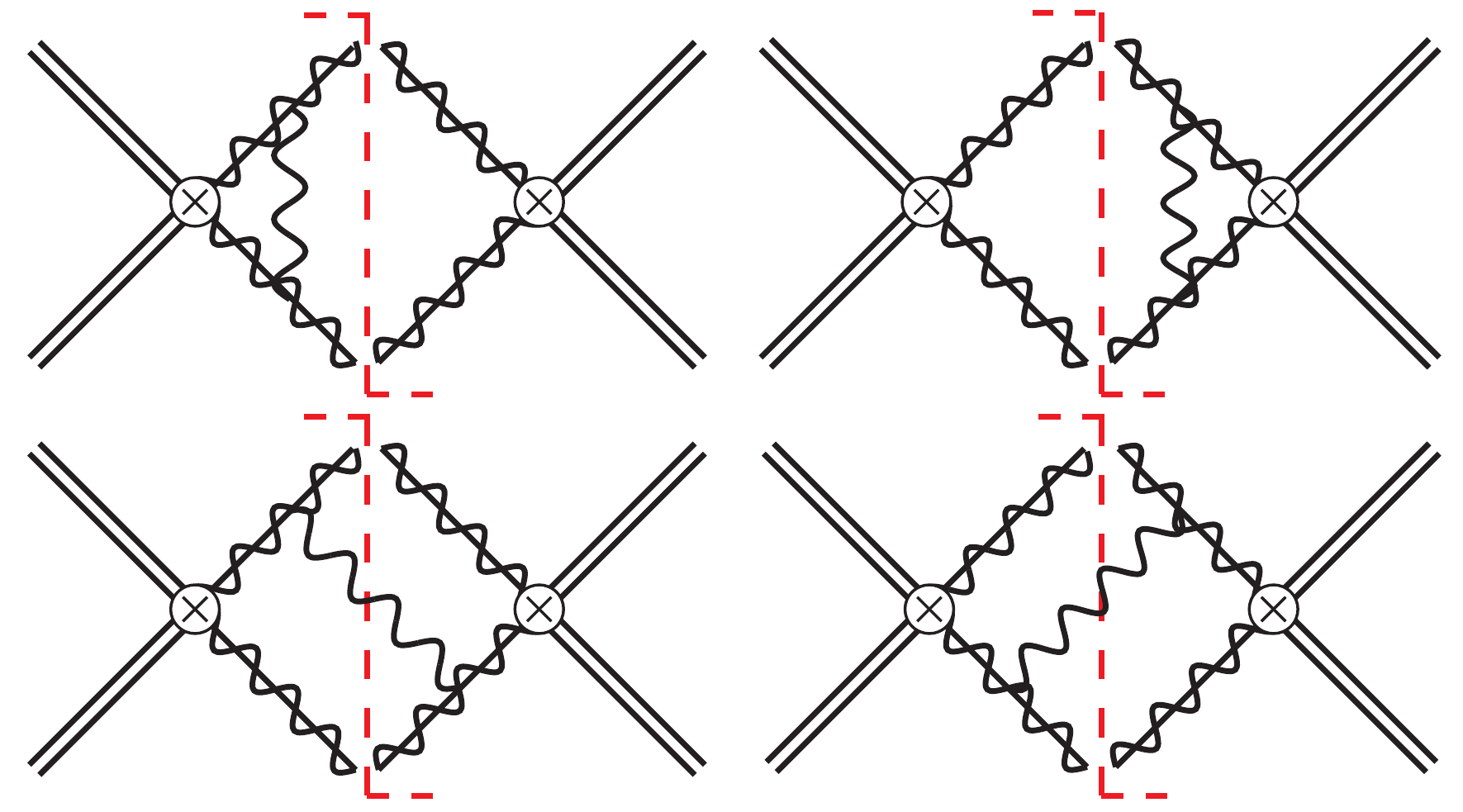}
	\caption{The four possible cuts through the $n_+ n_-$-two-loop diagram.}\label{fig:cutnnbar}
\end{figure}

In Figure~\ref{fig:cutnnbar}, we show the four possible cuts of the $n_+ n_-$-diagram. For the other diagram types, we apply the same procedure. The total discontinuity for the $n_+ n_-$-type-diagrams is\footnote{Note that the choice of scalar integrals in the previous sections implies a relative plus/minus sign between the real and virtual contributions in some of the cut diagrams.} 
\begin{align}
\text{Disc}(i \mathcal{M}_{n_+ n_-}) &= 2\,\text{Re} \left(I_{n_+ n_-}^{\text{real}}-I_{n_+ n_-}^{\text{virt.}}\right) \nonumber \\
&= 
\frac{\hat{\alpha}_2}{\pi}\left[ \delta(\omega) \left(\frac{1}{\epsilon \eta} - \frac{2}{\eta} \ln\frac{m_W}{\mu}- \frac{1}{\epsilon} \ln\frac{m_W}{\nu} + 2\ln\frac{m_W}{\mu}\ln\frac{m_W}{\nu}\right) \right. \nonumber \\
&\hphantom{=\frac{\hat{\alpha}_2}{\pi}\quad}\left.+\left[\frac{1}{\omega}\right]_{*}^{[m_W]} \left( -\frac{1}{\epsilon} + 2 \ln\frac{m_W}{\mu} \right)\right]\, . 
\end{align}
In the $n_+ n_-$-diagrams the real contribution cancels half of the virtual $\eta$ 
rapidity divergence. 

There is more than one $v n_-$ two-loop diagram. We discuss only one of them, as the others differ only via relative overall signs and prefactors:
\begin{align}
\text{Disc}(i \mathcal{M}_{vn_-}) &= I_{vn_-}^\text{virt.} - I_{vn_-}^\text{real} \nonumber \\
&=\frac{\hat{\alpha}_2}{4\pi}\left[ \delta(\omega)  \left(  -\frac{1}{\epsilon^2} + \frac{2}{\epsilon} \ln\frac{m_W}{\mu}+ \frac{\pi^2}{12} - 2\ln^2\frac{m_W}{\mu}\right)\right. \nonumber \\
&\hphantom{=\frac{\hat{\alpha}_2}{4\pi}\quad}\left.+\left[\frac{1}{\omega}\right]_{*}^{[m_W]} \left(\frac{2}{\epsilon} -2\ln\left(\frac{m_W^2 + \omega^2}{m_W^2}\right)-2\ln\frac{m_W^2}{\mu^2} \right)\right] 
\end{align}
For these diagrams the rapidity divergence is completely cancelled. Following the 
same logic, the $v n_+$-diagrams give
\begin{align}
\text{Disc}(i \mathcal{M}_{v n_+}) &= I_{v n_+}^\text{virt.}+I_{v n_+}^\text{real} \nonumber \\
&= - \frac{\hat{\alpha}_2}{4\pi}\,\bigg[\delta(\omega) \left(  - \frac{1}{\epsilon^2} + \frac{2}{\epsilon \eta} - \frac{4}{\eta} \ln\frac{m_W}{\mu} +\frac{2}{\epsilon} \ln\frac{m_W}{\mu}-\frac{2}{\epsilon} \ln\frac{m_W}{\nu} \right. \nonumber \\
&\hphantom{=  + \frac{\hat{\alpha}_2}{4\pi}\delta(\omega) \quad} \left. + \frac{\pi^2}{12} - 2 \ln^2\frac{m_W}{\mu} +4 \ln\frac{m_W}{\mu} \ln\frac{m_W}{\nu} \right) \nonumber \\
&\hphantom{=\frac{\hat{\alpha}_2}{4\pi}\quad}+\left[\frac{1}{\omega}\right]_{*}^{[m_W]} 2  \ln\left(\frac{m_W^2+\omega^2}{m_W^2}\right) \bigg]
\, .
\end{align}
The $\eta$-divergence for this two-loop diagram is the same as for the virtual integrals only, as the corresponding real integral is $\eta$ finite. 

We observe that the left-over rapidity divergences among the two-loop diagrams 
are such that 
\begin{align}
\sum_{\text{virt.}} |_{\eta-\text{div.}} + \sum_{\text{real}} |_{\eta-\text{div.}} = \frac{1}{2}\sum_{\text{virt.}} |_{\eta-\text{div.}} \, .
\end{align}
In the {\em narrow} resolution case $\Eres\sim m_W^2/m_\chi$, the rapidity divergence in the sum of all virtual soft diagrams cancels the rapidity divergence of the photon jet function and the narrow resolution unobserved-jet function. The fact that the intermediate resolution case allows for real soft radiation implies that it has only half the $\eta$-divergence in the soft sector compared to the narrow resolution case. This matches precisely to the fact that the jet function for the unobserved final state 
at intermediate resolution and  hard-collinear 
virtuality $\mathcal{O}(\mchi m_W)$ does not have rapidity divergences anymore.

\subsection{Soft functions in momentum space}
\label{app:soft}

In this Appendix we give the individual components of the index-contracted soft 
function as defined in \eqref{eq:softtwoindexfn} and \eqref{eq:softfndelta}.
For the operator combination $ij=11$ we find
\begin{align}
    &W^{11}_{(00)(00)} (\omega,\mu,\nu) =W^{11}_{(00)(+-)}(\omega,\mu,\nu) =W^{11}_{(+-)(00)}(\omega,\mu,\nu)=W^{11}_{(+-)(+-)} (\omega,\mu,\nu)  \nonumber \\
    &\hspace{1.2cm}= \delta(\omega) + \frac{\hat{\alpha}_2}{4 \pi} \left[\delta(\omega)(-16) \ln\frac{m_W}{\mu}\ln\frac{m_W}{\nu}+\left[\frac{1}{\omega}\right]_*^{\left[m_W\right]}\hspace{-.15cm} (-16) \ln\frac{m_W}{\mu}\right] .
\end{align}
The operator combinations $ij=\{12,21\}$ are given by
\begin{align}
    &W^{12}_{(00)(00)}(\omega,\mu,\nu) = W^{21 *}_{(00)(00)}(\omega,\mu,\nu) \nonumber \\
    &\hspace{1.2cm}=\frac{\hat{\alpha}_2}{4 \pi}\left[ \delta(\omega)\left(8 + 8 \pi i\right) \ln \frac{m_W}{\mu} + \left[\frac{1}{\omega}\right]^{[m_W]}_* 8 \ln \left(\frac{m_W^2 + \omega^2}{m_W^2}\right)\right],  \nonumber \\
    &W^{12}_{(00)(+-)} (\omega,\mu,\nu) = W^{21 *}_{(+-)(00)}(\omega,\mu,\nu) \nonumber \\
    &\hspace{1.2cm}= \delta (\omega) + \frac{\hat{\alpha}_2}{4 \pi}\left[ \delta(\omega)\left( \left(4+4\pi i\right) \ln\frac{\mu}{m_W} -16 \ln\frac{m_W}{\mu} \ln\frac{m_W}{\nu}\right)\right.  \nonumber \\
&\hspace{4.4cm}\left.+ \left[\frac{1}{\omega}\right]^{[m_W]}_* \left(- 4 \ln \left(\frac{m_W^2+\omega^2}{m_W^2}\right)+8\ln\frac{\mu^2}{m_W^2}\right)\right], \nonumber \\
&W^{12}_{(+-)(00)}(\omega,\mu,\nu)= W^{21 *}_{(00)(+-)} (\omega,\mu,\nu) = W^{12}_{(00)(00)} (\omega,\mu,\nu) \,, \nonumber \\ 
&W^{12}_{(+-)(+-)}(\omega,\mu,\nu)= W^{21 *}_{(+-)(+-)}(\omega,\mu,\nu) \nonumber \\
&\hspace{1.2cm}= W^{12}_{(00)(+-)}(\omega,\mu,\nu) + \frac{\hat{\alpha}_2}{4 \pi} \left[\frac{1}{\omega}\right]^{[m_W]}_* (-2) \ln \left(\frac{m_W^2+ \omega^2}{m_W^2}\right)\,.
\end{align}
Finally, the operator combination $ij=22$ is
\begin{align}
&W^{22}_{(00)(00)} (\omega,\mu,\nu)= \frac{\hat{\alpha}_2}{4 \pi} \left[\frac{1}{\omega}\right]^{[m_W]}_* \left(8 \ln \left(\frac{m_W^2 + \omega^2}{m_W^2}\right) - 8\frac{\omega^2}{m_W^2 + \omega^2}\right) \nonumber \,, \\
&W^{22}_{(00)(+-)} (\omega,\mu,\nu)= W^{22*}_{(+-),(00)} (\omega,\mu,\nu) \nonumber \\
&\hspace{1.2cm}= \frac{\hat{\alpha}_2}{4 \pi}\left[ \delta(\omega)\left(8 - 8 \pi i\right) \ln \frac{m_W}{\mu}+ \left[\frac{1}{\omega}\right]^{[m_W]}_* \left(4 \ln \left(\frac{m_W^2 + \omega^2}{m_W^2}\right) + 4\frac{\omega^2}{m_W^2 + \omega^2}\right)\right] \nonumber \\ 
&W^{22}_{(+-)(+-)} (\omega,\mu,\nu) = \delta(\omega) + \frac{\hat{\alpha}_2}{4 \pi}\left[ \delta(\omega) \left( - 8 \ln \frac{m_W}{\mu}-16 \ln \frac{m_W}{\mu}\ln \frac{m_W}{\nu}\right) \right. \nonumber \\
&\hspace{3cm}\left.+ \left[\frac{1}{\omega}\right]^{[m_W]}_* \left(-6 \ln \left(\frac{m_W^2+\omega^2}{m_W^2}\right) -2 \frac{\omega^2}{m_W^2 + \omega^2} + 8 \ln\frac{\mu^2}{m_W^2}\right)\right] \, ,
\end{align}

\subsection{Expressions for the resummed soft coefficients $\hat{W}^{ij}_{IJ}$}
\label{app:softPrime}

We collect here the inverse Laplace-transformed soft coefficients $\hat{W}$ as discussed in Section \ref{sec:intressoftfn} and defined in \eqref{eq:softrescoeff}. 
We also make use of the inverse Laplace transform $F(\omega)$ which was defined in \eqref{eq:invTrans2}. For the operator combination $ij=11$, the $\hat{W}$ coefficients are given by
\begin{align}
\hat{W}^{11}_{(00)(00)}(\omega,\mu_s,\nu)&=\hat{W}^{ 11}_{(00)(+-)}(\omega,\mu_s,\nu)=\hat{W}^{11}_{(+-)(00)}(\omega,\mu_s,\nu)=\hat{W}^{ 11}_{(+-)(+-)}(\omega,\mu_s,\nu) \nonumber \\
                                         &=\left(1+\frac{\hat{\alpha}_2}{4 \pi} \left(-16\right) \ln \frac{m_W}{\mu_s} \,\partial_\eta\right) \frac{e^{-\gamma_E \eta}}{\Gamma(\eta)} \frac{1}{\omega} \left(\frac{\omega}{\nu}\right)^\eta \, .
\end{align}
We note that here $\eta$ is defined as in \eqref{eq:etadef} and should not be confused with the rapidity regulator. For the operator combination $ij=12$, the results read
\begin{align}
\hat{W}^{12}_{(00)(00)}(\omega,\mu_s,\nu)&=\hat{W}^{12}_{(+-)(00)} (\omega,\mu_s,\nu)\nonumber \\
&= \frac{\hat{\alpha}_2}{4 \pi} \left[\left(8+8\pi i\right) \ln \frac{m_W}{\mu_s}\right] \frac{e^{-\gamma_E \eta}}{\Gamma(\eta)} \frac{1}{\omega} \left(\frac{\omega}{\nu}\right)^\eta + \frac{\hat{\alpha}_2}{4 \pi} \left[8 F(\omega)\right] \, , 
\nonumber\\
\hat{W}^{12}_{(00)(+-)}(\omega,\mu_s,\nu)&=\left[1+\frac{\hat{\alpha}_2}{4 \pi} \left(\left(-16 \ln \frac{m_W}{\mu_s}\, \partial_\eta\right) - \left(4+4\pi i\right)\ln \frac{m_W}{\mu_s}\right)\right] \frac{e^{-\gamma_E \eta}}{\Gamma(\eta)}\frac{1}{\omega} \left(\frac{\omega}{\nu}\right)^\eta \nonumber \, 
\nonumber\\
                                         &\hspace{.5cm}+ \frac{\hat{\alpha}_2}{4 \pi} \left[-4 F(\omega)\right]\, , \nonumber\\
\hat{W}^{12}_{(+-)(+-)}(\omega,\mu_s,\nu)&=\hat{W}^{12}_{(00)(+-)}(\omega,\mu_s,\nu) + \frac{\hat{\alpha}_2}{4 \pi} \left[-2 F(\omega)\right] \, ,
\end{align}
and for $ij=21$, 
\begin{align}
\hat{W}^{21}_{(00)(00)}(\omega,\mu_s,\nu)&=\hat{W}^{12 *}_{(00)(00)}(\omega,\mu_s,\nu) \nonumber\\
\hat{W}^{21}_{(00)(+-)}(\omega,\mu_s,\nu)&=\hat{W}^{12 *}_{(+-)(00)}(\omega,\mu_s,\nu) \nonumber\\
\hat{W}^{21}_{(+-)(00)}(\omega,\mu_s,\nu)&=\hat{W}^{12 *}_{(00)(+-)}(\omega,\mu_s,\nu) \nonumber\\
\hat{W}^{21}_{(+-)(+-)}(\omega,\mu_s,\nu)&=\hat{W}^{12 *}_{(+-)(+-)}(\omega,\mu_s,\nu) \, .
\end{align}
Finally, for the operator combination $ij=22$, we have the inverse Laplace-transformed soft coefficients
\begin{align}
\hat{W}^{22}_{(00)(00)}(\omega,\mu_s,\nu) &= \frac{\hat{\alpha}_2}{4 \pi} \left[8 F(\omega) - 8 P(\omega)\right] \, , \nonumber\\
\hat{W}^{22}_{(00)(+-)}(\omega,\mu_s,\nu)&=\hat{W}^{22 *}_{(+-)(00)} (\omega,\mu_s,\nu)\nonumber \\
&= \left[ \frac{\hat{\alpha}_2}{4 \pi} \left(8-8\pi i\right) \ln \frac{m_W}{\mu_s}\right] \frac{e^{-\gamma_E \eta}}{\Gamma(\eta)} \frac{1}{\omega} \left(\frac{\omega }{\nu}\right)^\eta + \frac{\hat{\alpha}_2}{4 \pi} \left[4 F(\omega) + 4 P(\omega)\right]\, , \nonumber\\
\hat{W}^{22}_{(+-)(+-)}(\omega,\mu_s,\nu)&= \left[1+\frac{\hat{\alpha}_2}{4 \pi} \left(\left(-16 \ln \frac{m_W}{\mu_s} \,\partial_\eta\right) - 8 \ln \frac{m_W}{\mu_s}\right)\right]\frac{e^{-\gamma_E \eta}}{\Gamma(\eta)} \frac{1}{\omega} \left(\frac{\omega }{\nu}\right)^\eta \nonumber \, \\
&\hspace{.5cm}+ \frac{\hat{\alpha}_2}{4 \pi} \left[-6 F(\omega) -2 P(\omega)\right]\, .
\end{align}

\section{RG and RRG invariance for the narrow resolution}
\label{app:RGEconsistency}
In this Appendix, we provide the RG and RRG equations of the unobserved-jet function and the soft coefficients $D^i$ ($i=1,2$)\footnote{The following does not 
depend on the indices $I,33$ of $D^i_{I,33}$, which are therefore dropped 
to simplifiy the notation.} of the narrow resolution case \cite{Beneke:2018ssm}. Together with the results for $Z_\gamma$ from Section \ref{sec:photonjet}, we can then show the RG and RRG invariance in the narrow resolution case. For the soft coefficients, the RG and RRG equations read
\begin{align}
\frac{d}{d \ln \mu} D^i \left(\mu,\nu\right) &= \bm{\gamma}^{\mu}_{D, ij} D^j\left(\mu,\nu\right) \,,\\
\frac{d}{d \ln \nu} D^i\left(\mu,\nu\right) &= \bm{\gamma}^{\nu}_{D} D^i\left(\mu,\nu\right) \,, 
\end{align}
where the one-loop anomalous dimensions are given by
\begin{align}
\bm{\gamma}^{\nu\,(0)}_D &= 16 \ln \frac{m_W}{\mu} \mathbf{1}_2  \label{eq:anomdimnarSoftcoeffnu}\,, \\
\bm{\gamma}^{\mu\,(0)}_D &= \begin{pmatrix}
    {\displaystyle 16 \ln \frac{\mu}{\nu}+ 8 \pi i 
    \quad}& 0 \\[0.2cm]
    c_2(j)(-4+4 \pi i)\;\; \;\;& {\displaystyle 16\ln \frac{\mu}{\nu}+ (12-4\pi i)}
\end{pmatrix} \label{eq:anomdimnarSoftcoeffmu}\,.
\end{align}

For the unobserved-jet function, whose explicit results are collected in Appendix \ref{app:recjetfnnarrow}, the RG and RRG equations are given by
\begin{align}
    \frac{d}{d \ln \nu} J^{33}\left(p^2,\mu,\nu\right) &= \gamma^{\nu}_{J_{\text{rec}}} J^{33}\left(p^2,\mu,\nu\right) \,, \\
    \frac{d}{d \ln \mu} J^{33}\left(p^2,\mu,\nu\right) &= \gamma^{\mu}_{J_{\text{rec}}} J^{33}\left(p^2,\mu,\nu\right) \,,
\end{align}
and the one-loop anomalous dimensions read 
\begin{align}
\gamma^{\nu\,(0)}_{J_{\text{rec}}} &= 16\ln \frac{\mu}{m_W} , 
\label{eq:anomdimnarJetnu}\\
\gamma^{\mu\,(0)}_{J_{\text{rec}}} &= 16 \ln \frac{\nu}{2 m_\chi} + \frac{19}{3}
\label{eq:anomdimnarJetmu}\,.
\end{align}
Analogous to the discussion in Section~\ref{sec:RGI} for the intermediate resolution case, the independence of \eqref{eq:SIJGIJ} and \eqref{eq:factformulanarrow} on the scales $\mu,\nu$ implies consistency relations among the anomalous dimensions:
\begin{align}
\gamma^{\nu}_{Z_\gamma} \mathbf{1}_2 + \gamma^{\nu}_{J_{\text{rec}}} \mathbf{1}_2 + \gamma^{\nu}_{D} + \gamma^{\nu *}_{D} = 0 \, , \\
\mathbf{\Gamma} + \mathbf{\Gamma}^{*} + \bm{\gamma}^{\mu}_{D} + \bm{\gamma}^{\mu *}_{D} + \gamma^{\mu}_{Z_\gamma} \mathbf{1}_2 + \gamma^{\mu}_{J_{\text{rec}}} \mathbf{1}_2 = 0 \, .
\end{align}
Using the expressions for the anomalous dimensions from the RG and RRG equations for the hard function \eqref{eq:anDimWilson}, the photon jet function \eqref{eq:photnAnDimVirt}, \eqref{eq:photnAnDimRap}, the unobserved-jet function \eqref{eq:anomdimnarJetnu}, \eqref{eq:anomdimnarJetmu} and the soft coefficients of the narrow resolution case \eqref{eq:anomdimnarSoftcoeffnu}, \eqref{eq:anomdimnarSoftcoeffmu}, we can explicitly check that these consistency constraints are satisfied.


\section{Complete NNLO expansions}
\label{app:logexpansions}

In Section \ref{sec:logexpansions} we showed the coefficients $c_{(+-)(+-)}^{(n,2n)}$, $c_{(+-)(+-)}^{(n,2n-1)}$ and $c_{(+-)(+-)}^{(n,2n-2)}$ defined in \eqref{eq:logexpand} for $n\leq2$. In this Appendix, we list the remaining coefficients relevant to the understanding of the logarithmic structure of NLL' resummation at NNLO, that is, $c_{IJ}^{(n,m)}$ with $0 \le n \le 2$ and $0 \le m \le 2n$ and $I,J \in \{(00),(+-)\}$. 

Additionally, this Appendix provides more details on the computation of the resummed cross section, as well as the fixed-order expansions. We introduce the following abbreviations (partially already given in Section \ref{sec:logexpansions} but repeated here for completeness):
\begin{eqnarray}
L \equiv \ln\frac{4m_\chi^2}{m_W^2} \,, \quad  l_{\mu_h} &\equiv& \ln\frac{\mu_h^2}{4m_\chi^2} \,, \quad l_{\mu_j} \equiv \ln\frac{\mu_j^2}{2m_\chi m_W} \,, \quad l_{\mu_s} \equiv \ln\frac{\mu_s^2}{m_W^2} \,, \\
l_{\mu} \equiv \ln\frac{\mu^2}{m_W^2} \,, \quad l_{\nu_h} &\equiv& \ln\frac{\nu_h^2}{4m_\chi^2} \,, \quad l_{\nu_s} \equiv \ln\frac{\nu_s^2}{m_W^2} \,, \quad l_R \equiv \ln x_\gamma \,, \\
\kappa_R = \kappa_R(x_\gamma) &\equiv& \frac{1}{2} \ln (1 + x_\gamma^2) \,, \\
\lambda_R = \lambda_R(x_\gamma) &\equiv& -\frac{1}{2} \text{Li}_2 (-x_\gamma^2) \,, \\
\varphi_{f_R} = \varphi_{f_R}(x_\gamma) &\equiv& \int_0^{x_\gamma}\frac{\dif y}{y}[f_R(x_\gamma-y)-f_R(x_\gamma)] \,, \label{eq:varphi}\\
\vartheta_{f_R} = \vartheta_{f_R}(x_\gamma) &\equiv& \int_0^{x_\gamma} \dif y\frac{\ln( y)}{y}[f_R(x_\gamma-y)-f_R(x_\gamma)] \,,
\end{eqnarray}
where
\begin{eqnarray}
x_\gamma \equiv \frac{2 E^\gamma_{\text{res}}}{m_W} \,.
\end{eqnarray}

\subsection{Narrow resolution coefficients}

\subsubsection{\boldmath$I,J=(00),(00)$}
The fixed-order annihilation cross section starts at the two-loop 
order, and the two-loop coefficient exhibits at most two logarithms. 
Hence $c_{(00)(00)}^{{\rm nrw}(n,m)} = 0$ for $n\leq 2$, except 
for
\begin{align}
    c_{(00)(00)}^{{\rm nrw}(2,2)} &= 1 + \pi^2 \nonumber\\
c_{(00)(00)}^{{\rm nrw}(2,1)} &= 4 - \frac{\pi^2}{2} \nonumber\\
c_{(00)(00)}^{{\rm nrw}(2,0)} &= 4 - \pi^2 + \frac{\pi^4}{16}
\end{align}

\subsubsection{\boldmath$I,J=(+-),(00)$}
The fixed-order annihilation cross section starts at the one-loop 
order. Hence $c_{(+-)(00)}^{{\rm nrw}(0,0)} = 0$, and 
\begin{align}
    c_{(+-)(00)}^{{\rm nrw}(1,2)} &= 0 \nonumber\\
c_{(+-)(00)}^{{\rm nrw}(1,1)} &= -1 - i\pi \nonumber\\
c_{(+-)(00)}^{{\rm nrw}(1,0)} &= -2 + \frac{\pi^2}{4} \nonumber\\
c_{(+-)(00)}^{{\rm nrw}(2,4)} &= 0 \nonumber\\
c_{(+-)(00)}^{{\rm nrw}(2,3)} &= 1 + i\pi \nonumber\\
c_{(+-)(00)}^{{\rm nrw}(2,2)} &= \frac{55}{48} - \frac{53 i\pi }{48} \nonumber\\
c_{(+-)(00)}^{{\rm nrw}(2,1)} &= \left[\left(\frac{11}{12} + \frac{11i\pi}{12}\right)\hat{s}_W^2 -\frac{19}{12} - \frac{19i\pi}{12}\right] l_{\mu} + \frac{55}{12} + \frac{55i\pi}{18} - \frac{55\pi^2}{96} - \frac{11i\pi^3}{24} \nonumber\\
&\quad-\left(\frac{1}{4} + \frac{i\pi}{4}\right)\left(z_\gamma + j(E^\gamma_{\text{res}})\right) \nonumber\\
c_{(+-)(00)}^{{\rm nrw}(2,0)} &= \left(\frac{1}{2}+\frac{i \pi }{2}\right) l_{\mu_h}^3+\left(-\frac{31}{48}+\frac{5 i \pi }{48}+\frac{3 \pi ^2}{8}\right) l_{\mu_h}^2+\left(\frac{31}{12}+\frac{19 i \pi }{18}-\frac{13 \pi ^2}{32}-\frac{7 i \pi ^3}{24}\right) l_{\mu_h} \nonumber\\
&\quad+\left(\frac{1}{2}+\frac{i \pi }{2}\right) l_{\mu_s}^3+ \left[(-1-i \pi ) l_{\nu_s}+\frac{\pi^2}{4}+\frac{43 i \pi }{48}+\frac{55}{48}\right]l_{\mu_s}^2 \nonumber\\
&\quad+\left(\frac{35 i\pi }{18}-\frac{\pi ^2}{12}-\frac{i \pi ^3}{4}\right) l_{\mu_s} + \left[\left(\frac{11}{6}-\frac{11 \pi^2}{48}\right) \hat{s}_W^2+\frac{19 \pi ^2}{48}-\frac{19}{6}\right]l_{\mu} \nonumber\\
&\quad+\frac{7\pi ^4}{96}-\frac{4 \pi ^2}{3}+6 +\left(\frac{\pi ^2}{16}-\frac{1}{2}\right) \left(z_\gamma + j(E_{\text{res}}^\gamma)\right)
\end{align}
The coefficients for the index combination $I,J=(00),(+-)$ are obtained by taking the complex conjugate of the coefficients given in this section, i.e. $c_{(00)(+-)}^{(n,m)} = (c_{(+-)(00)}^{(n,m)})^*$.

\subsubsection{\boldmath$I,J=(+-),(+-)$}
\begin{align}
    c_{(+-)(+-)}^{{\rm nrw}(0,0)} &= \hat{s}_W^2+ \hat{c}_W^2\Theta\left(\Eres-\frac{m_Z^2}{4m_\chi}\right) \nonumber\\
c_{(+-)(+-)}^{{\rm nrw}(1,2)} &= -1 \nonumber\\
c_{(+-)(+-)}^{{\rm nrw}(1,1)} &= 1 \nonumber\\
c_{(+-)(+-)}^{{\rm nrw}(1,0)} &= \left(\frac{19}{24}-\frac{11}{12}\hat{s}_W^2\right)l_{\mu} - 6 + \frac{3\pi^2}{4} + \frac{1}{4}\left(z_\gamma + j(E^\gamma_{\text{res}})\right) \nonumber\\
c_{(+-)(+-)}^{{\rm nrw}(2,4)} &= \frac{1}{2} \nonumber\\
c_{(+-)(+-)}^{{\rm nrw}(2,3)} &= -\frac{53}{72} \nonumber\\
c_{(+-)(+-)}^{{\rm nrw}(2,2)} &= \left(-\frac{19}{12}+\frac{11 \hat{s}_W^2}{12}\right)l_{\mu}-\frac{13 \pi^2}{12}+\frac{671}{144} -\frac{1}{4}\left(z_\gamma + j(E^\gamma_{\text{res}})\right) \nonumber\\
c_{(+-)(+-)}^{{\rm nrw}(2,1)} &= \frac{19}{24}l_{\mu_s}^2 + \left(\frac{35}{9}-\frac{\pi^2}{3}\right)l_{\mu_s} +\left(\frac{19}{12}-\frac{11}{12}\hat{s}_W^2\right)l_{\mu} \nonumber\\
&\quad+ \frac{3}{4} -\frac{65\pi^2}{288} -\frac{\beta_{1,\text{SU}(2)}}{8} + \frac{1}{4}\left(z_\gamma + j(E^\gamma_{\text{res}})\right) \nonumber\\
c_{(+-)(+-)}^{{\rm nrw}(2,0)} &= -\frac{1}{4} l_{\mu_h}^4-\frac{17}{72}l_{\mu_h}^3+\left(\frac{25 \pi ^2}{24}-\frac{203}{144}\right) l_{\mu_h}^2 + \left(-\frac{\beta_{1,\text{SU}(2)}}{8}-\frac{149 \pi^2}{288}+\frac{15}{4}\right) l_{\mu_h} \nonumber\\
&\quad-\frac{1}{4}l_{\mu_s}^4+ \left(l_{\nu_s}-\frac{55}{72}\right)l_{\mu_s}^3 + l_{\mu_s}^2\bigg[-l_{\nu_h}^2+\left(\frac{11 \hat{s}_W^2}{12}-\frac{19}{24}\right) l_{\nu_h}-l_{\nu_s}^2+l_{\nu_s} \nonumber\\
&\quad-\frac{121 \hat{s}_W^4}{144}+\pi^2-\frac{2141}{576}\bigg] +l_{\mu_s} \bigg[\left(-\frac{j(E^\gamma_{\text{res}})}{4}-\frac{z_{\gamma }}{4}\right)l_{\nu_h} -\frac{\pi^2}{6} l_{\nu_s} \nonumber\\
&\quad+\frac{31 \pi^2}{144} +\left(\frac{1}{16}-\frac{\hat{s}_W^2}{16}\right)\beta_{1,\text{SU}(2)}-\frac{\hat{s}_W^4}{16 \hat{c}_W^2}\beta_{1,Y} \nonumber\\
&\quad+\left(\frac{11 \hat{s}_W^2}{48}-\frac{19}{96}\right) z_{\gamma } -\frac{19 j(E^\gamma_{\text{res}})}{48} \bigg]+l_{\mu}^2
\left(\frac{121 \hat{s}_W^4}{144}-\frac{209 \hat{s}_W^2}{288}+\frac{361}{576}\right) \nonumber\\
&\quad+l_{\mu} \bigg[\left(\frac{11}{2}-\frac{11 \pi^2}{16}\right)
\hat{s}_W^2+\frac{19 \pi^2}{16}-\frac{19}{2} +
\left(\frac{\hat{s}_W^2}{16}+\frac{1}{16}\right)\beta_{1,\text{SU}(2)}+\frac{\hat{s}_W^4}{16 \hat{c}_W^2}\beta_{1,Y} \nonumber\\
&\quad+\left(\frac{19}{48}-\frac{11 \hat{s}_W^2}{48}\right) \left(z_{\gamma } +j(E^\gamma_{\text{res}})\right) \bigg] + \left(-\frac{3}{2}+\frac{3\pi^2}{16}+\frac{z_{\gamma}}{16}\right) j(E^\gamma_{\text{res}}) \nonumber\\
&\quad+9-\frac{7 \pi^2}{4}+\frac{37 \pi ^4}{576}+\left(-\frac{3}{2}+\frac{3 \pi ^2}{16}\right) z_{\gamma}
\end{align}

\subsection{Intermediate resolution coefficients}
\subsubsection{\boldmath$I,J=(00),(00)$}
\begin{align}
    c_{(00)(00)}^{{\rm int}(0,0)} &= 0 \nonumber\\
c_{(00)(00)}^{{\rm int}(1,2)} &= 
\;c_{(00)(00)}^{{\rm int}(1,1)} =0 \nonumber\\
c_{(00)(00)}^{{\rm int}(1,0)} &= 2\lambda_R -2\kappa_R \nonumber\\
c_{(00)(00)}^{{\rm int}(2,4)} &= c_{(00)(00)}^{{\rm int}(2,3)} =0 \nonumber\\
c_{(00)(00)}^{{\rm int}(2,2)} &= 1 + \pi^2 -\frac{3\lambda_R}{2} +\frac{3\kappa_R}{2} \nonumber\\
c_{(00)(00)}^{{\rm int}(2,1)} &= 2l_R\left(\lambda_R - \kappa_R\right) + 4 - \frac{\pi^2}{2} + \frac{29}{24} \lambda_R -\frac{125}{24} \kappa_R + 2\left(\varphi_{\lambda_R} -\varphi_{\kappa_R}\right) \nonumber\\
c_{(00)(00)}^{{\rm int}(2,0)} &= \left(-1-\pi ^2\right) l_{\mu_s}^2 +l_{\mu_s} \bigg(2 l_{\nu_s} (\lambda_R -\kappa_R) -4l_R(\lambda_R -\kappa_R)  -\frac{43 \lambda _R}{12} +\frac{91 \kappa_R}{12} \nonumber\\
&\quad-4 (\varphi_{\lambda_R} - \varphi_{\kappa_R})\bigg)
+l_{\mu}(\lambda_R - \kappa_R) \left(\frac{19}{6}-\frac{11}{6}
\hat{s}_W^2\right) +2 l_R^2 (\lambda_R - \kappa_R) \nonumber \\
&\quad+l_R \left(-\frac{19}{12}(\lambda_R - \kappa_R) +4 (\varphi_{\lambda_R} -\varphi_{\kappa_R})\right) +4 -\pi ^2 +\frac{\pi^4}{16} +\frac{z_{\gamma }}{2}(\lambda_R - \kappa_R) \nonumber\\
&\quad+\left(-\frac{73}{9} + \frac{5 \pi^2}{6}\right) \lambda _R +\left(\frac{1}{9} + \frac{\pi^2}{6}\right) \kappa_R-\frac{19}{12}(\varphi_{\lambda_R} - \varphi_{\kappa_R})+4 (\vartheta_{\lambda_R} - \vartheta_{\kappa_R})
\end{align}

\subsubsection{\boldmath$I,J=(+-),(00)$}
\begin{align}
    c_{(+-)(00)}^{{\rm int}(0,0)} &= 0 \nonumber\\
c_{(+-)(00)}^{{\rm int}(1,2)} &= 0 \nonumber\\
c_{(+-)(00)}^{{\rm int}(1,1)} &= -1-i\pi \nonumber\\
c_{(+-)(00)}^{{\rm int}(1,0)} &= -2 + \frac{\pi^2}{4} + \lambda_R +\kappa_R \nonumber\\
c_{(+-)(00)}^{{\rm int}(2,4)} &= 0 \nonumber\\
c_{(+-)(00)}^{{\rm int}(2,3)} &= \frac{3}{4} + \frac{3i\pi}{4} \nonumber\\
c_{(+-)(00)}^{{\rm int}(2,2)} &= \left(-1-i\pi\right)l_R + \frac{25}{24} - \frac{17i\pi}{24} +\frac{\pi^2}{16} -\frac{3}{4}(\lambda_R +\kappa_R) \nonumber\\
c_{(+-)(00)}^{{\rm int}(2,1)} &= l_{\mu}\left[- \frac{19}{12} - \frac{19i\pi}{12} + \left(\frac{11}{12} + \frac{11i\pi}{12}\right)\hat{s}_W^2 \right] -\left(1+i\pi\right)l_R^2 \nonumber\\ 
&\quad+\left(-\frac{29}{24} + \frac{19i\pi}{24} +\frac{\pi^2}{4} +\lambda_R +\kappa_R\right) l_R +\frac{247}{72} +\frac{10i\pi}{9} -\frac{65\pi^2}{192} - \frac{i\pi^3}{8} \nonumber\\
&\quad- \left(\frac{1}{4} + \frac{i\pi}{4}\right)z_\gamma + \left(\frac{77}{48} + 3i\pi\right)\lambda_R +\frac{125}{48} \kappa_R + \varphi_{\lambda_R} + \varphi_{\kappa_R} \nonumber\\
c_{(+-)(00)}^{{\rm int}(2,0)} &= \left(\frac{1}{2} + \frac{i\pi}{2}\right) l_{\mu_h}^3 + \left(-\frac{31}{48} + \frac{5i\pi}{48} + \frac{3\pi^2}{8}\right) l_{\mu_h}^2 + \left(\frac{31}{12} + \frac{19i\pi}{18} -\frac{13\pi^2}{32} -\frac{7i\pi^3}{24}\right) l_{\mu_h} \nonumber\\
&\quad+ \left[-\left(1+i\pi\right)l_{\nu_s} + \left(2 + 2i\pi\right)l_R + \frac{79}{48} +\frac{91i\pi}{48} -\frac{\pi^2}{4} \right] l_{\mu_s}^2 \nonumber\\
&\quad+ \bigg[l_{\nu_s}\left(\lambda_R +\kappa_R\right) -2l_R\left(\lambda_R +\kappa_R\right) + \frac{35i\pi}{18} -\frac{i\pi^3}{6} + \left(-\frac{67}{24} -3i\pi \right)\lambda_R \nonumber \\
&\quad -\frac{91}{24}\kappa_R -2\left(\varphi_{\lambda_R} +\varphi_{\kappa_R}\right) \bigg] l_{\mu_s} + \bigg[-\frac{19}{6} +\frac{19\pi^2}{48} + \left(\frac{11}{6} - \frac{11\pi^2}{48}\right)\hat{s}_W^2 \nonumber\\
&\quad+ \left(\frac{19}{12} - \frac{11}{12}\hat{s}_W^2\right)\left(\lambda_R +\kappa_R\right) \bigg]l_{\mu} +\left(-2 + \frac{\pi^2}{4} + \lambda_R +\kappa_R\right) l_R^2 \nonumber\\
&\quad+ \left(\frac{19}{12} - \frac{19\pi^2}{96} -\frac{19}{24}\left(\lambda_R +\kappa_R\right) + 2\left(\varphi_{\lambda_R} +\varphi_{\kappa_R}\right)\right) l_R +\frac{19}{9} -\frac{13\pi^2}{72} - \frac{\pi^4}{96} \nonumber\\
&\quad+ \left(-\frac{1}{2} + \frac{\pi^2}{16}\right)z_\gamma + \frac{z_\gamma}{4}\left(\lambda_R +\kappa_R\right) + \left(-\frac{37}{18} + \frac{\pi^2}{6}\right) \lambda_R + \left(-\frac{1}{18} - \frac{\pi^2}{12}\right) \kappa_R \nonumber\\
&\quad- \frac{19}{24}\left(\varphi_{\lambda_R} +\varphi_{\kappa_R}\right) + 2(\vartheta_{\lambda_R} +\vartheta_{\kappa_R})
\end{align}
The coefficients for the index combination $I,J=(00),(+-)$ are obtained by taking the complex conjugate of the coefficients given in this section, i.e. $c_{(00)(+-)}^{(n,m)} = (c_{(+-)(00)}^{(n,m)})^*$.

\subsubsection{\boldmath$I,J=(+-),(+-)$}
\begin{align}
    c_{(+-)(+-)}^{{\rm int}(0,0)} &= 1 \nonumber\\
c_{(+-)(+-)}^{{\rm int}(1,2)} &= -\frac{3}{4} \nonumber\\
c_{(+-)(+-)}^{{\rm int}(1,1)} &= l_R + \frac{29}{48} \nonumber\\
c_{(+-)(+-)}^{{\rm int}(1,0)} &= \left(\frac{19}{24} - \frac{11}{12}\hat{s}_W^2\right)l_{\mu} + l_R^2 - \frac{19}{24}l_R - \frac{73}{18} +\frac{5\pi^2}{12}  + \frac{z_\gamma}{4} - \frac{3}{2}\lambda_R -\frac{1}{2}\kappa_R \nonumber\\
c_{(+-)(+-)}^{{\rm int}(2,4)} &= \frac{9}{32} \nonumber\\
c_{(+-)(+-)}^{{\rm int}(2,3)} &= -\frac{3}{4}l_R-\frac{2}{9} \nonumber\\
c_{(+-)(+-)}^{{\rm int}(2,2)} &=  \left(-\frac{19}{16} + \frac{11}{16} \hat{s}_W^2\right) l_{\mu} -\frac{1}{4}l_R^2 + l_R + \frac{4489}{2304} -\frac{37\pi^2}{48} - \frac{3}{16}z_\gamma + \frac{9}{8}\lambda_R +\frac{3}{8}\kappa_R \nonumber\\
c_{(+-)(+-)}^{{\rm int}(2,1)} &= \frac{19}{48} l_{\mu_s}^2 + \left[\left(\frac{19}{12} - \frac{11}{12} \hat{s}_W^2\right)l_R + \frac{551}{576} - \frac{319}{576}\hat{s}_W^2\right]l_{\mu} + l_R^3 - \frac{7}{12}l_R^2 \nonumber\\
&\quad+ \left( - \frac{437}{192} - \frac{\pi^2}{12} + \frac{1}{4} z_\gamma  -\frac{3}{2}\lambda_R -\frac{1}{2}\kappa_R\right)l_R + \frac{1525}{432} - \frac{\beta_{1,\text{SU}(2)}}{8} - \frac{227\pi^2}{576} \nonumber\\
&\quad+2 \zeta(3)+ \frac{29}{192}z_\gamma -\frac{29}{32}\lambda_R -\frac{125}{96}\kappa_R  -\frac{3}{2}\varphi_{\lambda_R} -\frac{1}{2}\varphi_{\kappa_R} \nonumber\\
c_{(+-)(+-)}^{{\rm int}(2,0)} &= -\frac{1}{4} l_{\mu_h}^4 - \frac{17}{72} l_{\mu_h}^3 + \left(-\frac{203}{144} + \frac{25\pi^2}{24}\right)l_{\mu_h}^2 + \left(\frac{15}{4} - \frac{149\pi^2}{288} - \frac{\beta_{1,\text{SU}(2)}}{8}\right)l_{\mu_h} \nonumber\\
&\quad-\frac{1}{2}l_{\mu_j}^4 + \left(2l_R-\frac{19}{18}\right) l_{\mu_j}^3 + \left(-3l_R^2 + \frac{19}{6}l_R - \frac{289}{64} +\pi^2\right) l_{\mu_j}^2 \nonumber\\
&\quad+ l_{\mu_j}\left[ 2l_R^3 - \frac{19}{6}l_R^2 + \left(\frac{289}{32} - 2\pi^2\right)l_R - \frac{665}{216} + \frac{19\pi^2}{18} +4\zeta(3) \right]  \nonumber\\
&\quad+ l_{\mu_s}^2\bigg[-\frac{1}{2}l_{\nu_h}^2 + \frac{11}{12}\hat{s}_W^2 l_{\nu_h} -\frac{1}{2}l_{\nu_s}^2 + \left(2l_R+1\right)l_{\nu_s} -2l_R^2 -\frac{67}{24}l_R -\frac{67}{48}\nonumber\\
&\quad +\frac{5\pi^2}{6} -\frac{121}{144}\hat{s}_W^4 \bigg] +l_{\mu_s}\bigg[\left(-\frac{35}{18} + \frac{\pi^2}{6} - \frac{1}{4} z_\gamma\right)l_{\nu_h} + \left(\frac{35}{18} - \frac{\pi^2}{6} - \frac{3}{2}\lambda_R -\frac{1}{2}\kappa_R\right)l_{\nu_s} \nonumber\\
&\quad+ \left(-\frac{35}{9} + \frac{\pi^2}{3} + 3\lambda_R +\kappa_R\right)l_R -\frac{1}{16}\frac{\hat{s}_W^4}{\hat{c}_W^2}\beta_{1,Y} + \frac{1}{16}(1-\hat{s}_W^2)\beta_{1,\text{SU}(2)} \nonumber\\
&\quad +\left(-\frac{19}{96} + \frac{11}{48}\hat{s}_W^2\right)z_\gamma +\frac{43}{16}\lambda_R +\frac{91}{48}\kappa_R+ 3\varphi_{\lambda_R} +\varphi_{\kappa_R} \bigg] \nonumber\\
&\quad+ l_{\mu}^2\left(\frac{361}{576} - \frac{209}{288}\hat{s}_W^2 + \frac{121}{144}\hat{s}_W^4 \right) + l_{\mu}\bigg[\left(\frac{19}{12} - \frac{11}{12}\hat{s}_W^2\right) l_R^2 \nonumber\\
&\quad+\left(-\frac{361}{288} + \frac{209}{288}\hat{s}_W^2\right) l_R -\frac{1387}{216} + \frac{1}{16}\beta_{1,\text{SU}(2)} + \frac{95\pi^2}{144} \nonumber\\
&\quad+ \left(\frac{803}{216} + \frac{1}{16}\beta_{1,\text{SU}(2)} - \frac{55\pi^2}{144}\right) \hat{s}_W^2 + \frac{1}{16}\frac{\hat{s}_W^4}{\hat{c}_W^2}\beta_{1,Y} + \left(\frac{19}{48} - \frac{11}{48}\hat{s}_W^2\right)z_\gamma \nonumber\\
&\quad+ \left(-\frac{19}{8} + \frac{11}{8}\hat{s}_W^2\right) \left(\lambda_R +\frac{1}{3}\kappa_R\right) \bigg] + l_R^2\left( -6 + \frac{11\pi^2}{12} + \frac{1}{4} z_\gamma -\frac{3}{2}\lambda_R -\frac{1}{2}\kappa_R\right) \nonumber\\
&\quad+ l_R\left(+\frac{19}{4} -\frac{209\pi^2}{288} - \frac{19}{96} z_\gamma +\frac{19}{16}\left(\lambda_R +\frac{1}{3}\kappa_R\right) -3\varphi_{\lambda_R} -\varphi_{\kappa_R} \right) -\frac{8}{3} + \frac{439\pi^2}{216} \nonumber\\
&\quad - \frac{143\pi^4}{576} + \left(-\frac{73}{72} + \frac{5\pi^2}{48}\right) z_\gamma - \frac{3}{8} z_\gamma \left(\lambda_R +\frac{1}{3}\kappa_R\right) + \left(\frac{73}{12} - \frac{5\pi^2}{8}\right) \lambda_R \nonumber \\
&\quad +\left(\frac{1}{36} + \frac{\pi^2}{24}\right) \kappa_R +\frac{19}{16} \left(\varphi_{\lambda_R} +\frac{1}{3}\varphi_{\kappa_R}\right) - 3\left(\vartheta_{\lambda_R} +\frac{1}{3}\vartheta_{\kappa_R}\right)
\end{align}

\subsection{Further input}
In obtaining the coefficients just listed, we made use of a series of expressions and properties which we summarize below.

\subsubsection{Running couplings at two loops}
At one-loop the running of the SU(2) coupling $\hat{\alpha_2}$ is determined from $\beta_{\text{SU(2)}}^{\rm 1-loop}(\hat\alpha_2)=-\beta_{0,\text{SU}(2)}\frac{\hat\alpha^2_2}{2\pi}$ while the running of  the Weinberg angle $\sin^2\theta_W(\mu)\equiv\hat{s}_W^2(\mu)$ follows from its definition in terms of the SU(2) and U(1) hypercharge gauge couplings,
\begin{equation}
\hat{s}_W^2(\mu)=\frac{\hat\alpha_1(\mu)}{\hat\alpha_1(\mu)+\hat\alpha_2(\mu)} \ .
\end{equation}
At two loops, contributions from other SM couplings affect the running of the EW gauge couplings. However, we can neglect the $\mu$-dependence of the other couplings in such terms 
in the beta-functions, as for example, in $\hat\alpha_2(\mu)\alpha_3(\mu)$,  since it would be relevant only at the NNNLO.  
Equivalently, to NNLO we can let the couplings run as 
\begin{equation}
\hat\alpha_2(\mu)=\hat\alpha_2(\mu_0)+\frac{\hat\alpha^2_2(\mu_0)}{4\pi}\beta_{0,\text{SU}(2)}\ln\frac{\mu_0^2}{\mu^2}+\frac{\hat\alpha^3_2(\mu_0)}{16\pi^2}\left(\beta_{1,\text{SU}(2)}\ln\frac{\mu_0^2}{\mu^2}+\beta_{0,\text{SU}(2)}^2\ln^2\frac{\mu_0^2}{\mu^2}\right)+\ldots\ ,
\end{equation}
where, in the SM,
\begin{eqnarray}
&& \beta_{0,\text{SU}(2)} =\frac{19}6\,, \quad
\beta_{1,\text{SU}(2)} =
\displaystyle -\frac{35}6-\frac32\frac{\hat{s}_W^2}{\hat c_W^2}-12\frac{\hat{\alpha}_3}{\hat\alpha_2}+\frac32\frac{y_t^2}{4\pi\hat\alpha_2} \,,
\\
&& \beta_{0,Y} = -\frac{41}{6} \,,\quad\;\;
\beta_{1,Y} = -\frac{199}{18}-\frac92\frac{\hat{c}_W^2}{\hat{s}_W^2}-\frac{44}3\frac{\hat{\alpha}_3}{\hat\alpha_1}+\frac{17}6\frac{y_t^2}{4\pi\hat\alpha_1}\,.
\end{eqnarray}
The coupling constant ratios in $\beta_{1,\text{SU}(2)}$ and $\beta_{1,Y}$ are treated as constants for the expansion to the 
two-loop order. The two-loop running of $\hat{s}_W^2$ is given by
\begin{eqnarray}
\label{eq:sWrunning}
\hat{s}_W^2(\mu) &=& \hat{s}_W^2(\mu_0)
+\frac{\hat\alpha_2(\mu_0)\hat{s}_W^2(\mu_0)}{4\pi}
\,\big[-\beta_{0,\text{SU}(2)}+(\beta_{0,\text{SU}(2)}+\beta_{0,Y})\hat{s}_W^2(\mu_0)\big]\ln\frac{\mu_0^2}{\mu^2} 
\nonumber\\
&& +\, \frac{\hat\alpha^2_2(\mu_0)\hat{s}_W^2(\mu_0)}{16\pi^2}\left[\frac1{\hat{c}_W^2(\mu_0)}\left(\hat{s}_W^4(\mu_0)\beta_{1,Y}-\hat c_W^4(\mu_0)\beta_{1,\text{SU}(2)}\right)\ln\frac{\mu_0^2}{\mu^2}\,+\right.
\nonumber\\
&&\left.+\,(\beta_{0,\text{SU}(2)}+\beta_{0,Y})\hat{s}_W^2\left(\hat{s}_W^2(\mu_0)\beta_{0,Y}-\hat{c}_W^2(\mu_0)\beta_{0,\text{SU}(2)}\right)\ln^2\frac{\mu_0^2}{\mu^2}\right]+\ldots\ 
\end{eqnarray}

\subsubsection{Identities for the star distributions}

The identities that are given here are useful for the fixed-order expansions of the resummed cross sections. They also need to be applied when checking the pole and scale cancellations of the individual cross sections:
\begin{equation}
\left[\frac1x\right]_*^{[a]}=\left[\frac1x\right]_*^{[b]}-
\log\frac{a}b\,\delta(x)\,,
\label{eq:id1}
\end{equation}
\begin{equation}
\left[\frac{\ln\frac{x}{a}}x\right]_*^{[a]}=\left[\frac{\ln\frac{x}{b}}x\right]_*^{[b]}+\ln\frac{b}{a}\left[\frac1x\right]_*^{[\sqrt{bc}]}+\frac12
\ln\frac{b}a\ln\frac{c}a\,\delta(x) \, .
\label{eq:id2}
\end{equation}

\subsubsection{Convolutions}

The convolutions provided in this section are used for both the implementation of the resummed annihilation cross sections to compute numerical results, as well as for the fixed-order expansion. 
Defining 
\begin{equation}
\label{eq:convol}
f(\omega)\otimes g(p^2) 
\equiv \int_0^{\frac{p^2}{2\mchi}} d\omega \,
f(\omega)\,g(p^2-2\mchi\omega)\big|_{p^2=4\mchi\Eres} ,
\end{equation}
we have
\begin{eqnarray}
\int_0^{\Eres}d E_\gamma\,\delta(\omega)\otimes\delta(p^2) &=& \frac1{4m_\chi}\label{eq:convInt1} \, , \\
\int_0^{\Eres}d E_\gamma\left[\frac1{\omega}\ln\left(1+\frac{\omega^2}{m_W^2}\right)\otimes\delta(p^2)\right]&=& \frac1{4m_\chi}\lambda_R(x_\gamma)\label{eq:convInt2} \, , \\
\int_0^{\Eres}d E_\gamma\left[\frac{1}{\omega}\right]_{*}^{[\nu_s]}\otimes\delta(p^2) &= & \frac1{4m_\chi}\ln\left(\frac{m_W}{\nu_s}x_\gamma\right) \label{eq:convInt3} \, , \\
\int_0^{\Eres}d E_\gamma\left[\frac{\ln\frac{\omega}{\nu_s}}{\omega}\right]_{*}^{[\nu_s]}\otimes\delta(p^2) &= & \frac1{8m_\chi}\ln^2\left(\frac{m_W}{\nu_s}x_\gamma\right) \label{eq:convInt4} \, , 
\end{eqnarray}
\begin{eqnarray}
\int_0^{\Eres}d E_\gamma\,\delta(\omega)\otimes\left[\frac{1}{p^2}\right]_{*}^{[\mu_j^2]} &=& \frac1{4m_\chi}\ln\left(\frac{2m_\chi m_W}{\mu_j^2}x_\gamma\right)\label{eq:convInt5} \, , \\
\int_0^{\Eres}d E_\gamma\left\{\frac1{\omega}\ln\left(1+\frac{\omega^2}{m_W^2}\right)\otimes\left[\frac{1}{p^2}\right]_{*}^{[\mu_j^2]} \right\}&=& \frac1{4m_\chi}\left[\lambda_R(x_\gamma)\ln\left(\frac{2m_\chi m_W}{\mu_j^2}x_\gamma\right)+\right.\nonumber\\
{} &{}&{}\hspace{30pt}+\varphi_R(x_\gamma)\bigg{]}\label{eq:convInt6} \, ,
\end{eqnarray}
\begin{eqnarray}
\int_0^{\Eres}d E_\gamma\left[\frac{1}{\omega}\right]_{*}^{[\nu_s]}\otimes\left[\frac{1}{p^2}\right]_{*}^{[\mu_j^2]}  &= & \frac1{4m_\chi}\left[\ln\left(\frac{m_W}{\nu_s}x_\gamma\right)\ln\left(\frac{2m_\chi m_W}{\mu_j^2}x_\gamma\right)-\frac{\pi^2}{6}\right]\,,\qquad 
\label{eq:convInt7} \\
\int_0^{\Eres}d E_\gamma\,\delta(\omega)\otimes\left[\frac{\ln\frac{p^2}{\mu_j^2}}{p^2}\right]_{*}^{[\mu_j^2]} &=& \frac1{8m_\chi}\ln^2\left(\frac{2m_\chi m_W}{\mu_j^2}x_\gamma\right)\label{eq:convInt8} \, , 
\end{eqnarray}
\begin{eqnarray}
\int_0^{\Eres}d E_\gamma\left\{\frac1{\omega}\ln\left(1+\frac{\omega^2}{m_W^2}\right)\otimes\left[\frac{\ln\frac{p^2}{\mu_j^2}}{p^2}\right]_{*}^{[\mu_j^2]}\right\}&=& \frac1{4m_\chi}\left[\frac12\lambda_R(x_\gamma)\ln^2\left(\frac{2m_\chi m_W}{\mu_j^2}x_\gamma\right)+\right.\nonumber\\
{}&{}&{}\hspace{-10pt}+\varphi_R(x_\gamma)\ln\frac{2m_\chi m_W}{\mu_j^2}+\vartheta_R(x_\gamma)\bigg{]}\label{eq:convInt9} \, , \\
\int_0^{\Eres}d E_\gamma\left[\frac{1}{\omega}\right]_{*}^{[\nu_s]}\otimes\left[\frac{\ln\frac{p^2}{\mu_j^2}}{p^2}\right]_{*}^{[\mu_j^2]}&= & \frac1{4m_\chi}\left[\frac12\ln\left(\frac{m_W}{\nu_s}x_\gamma\right)\ln^2\left(\frac{2m_\chi m_W}{\mu_j^2}x_\gamma\right)\right.\nonumber\\
{} &{}&{}\hspace{8pt}-\frac{\pi^2}{6}\ln\left(\frac{2m_\chi m_W}{\mu_j^2}x_\gamma\right)+\zeta(3)\bigg{]} \label{eq:convInt10} \, ,\quad
\\
\int_0^{\Eres}d E_\gamma\,\delta(\omega)\otimes\left[\frac{\ln^2\frac{p^2}{\mu_j^2}}{p^2}\right]_{*}^{[\mu_j^2]} &=& \frac1{12m_\chi}\ln^3\left(\frac{2m_\chi m_W}{\mu_j^2}x_\gamma\right)\label{eq:convInt11} \, .
\end{eqnarray}

\providecommand{\href}[2]{#2}\begingroup\raggedright\endgroup


\end{document}